%% file: apacic2_hepph.tex
\documentclass[12pt,a4]{article}
\usepackage{graphicx}
\usepackage{amsmath}

\usepackage{pstricks}
\usepackage{rotating}

\input{mysettings}

\begin{document}
\psset{linewidth=1pt}
\psset{unit=1pt}
%\begin{fmffile}{figs}
%\input{figures}
%\begin{frontmatter}
\title{\apa\ 2.0\\ A PArton Cascade In C++}

\author{F. Krauss\thanks{email: krauss@theory.phy.tu-dresden.de},
  A. Sch\"alicke\setcounter{footnote}{3}\thanks{email: dreas@theory.phy.tu-dresden.de}, 
  G. Soff\setcounter{footnote}{1}\thanks{deceased} \\
\small\em Institut f\"ur Theoretische Physik,
   TU Dresden, 01062 Dresden, Germany}

%\ead{}
%\ead{dreas@theory.phy.tu-dresden.de}

\maketitle
\begin{abstract}
The new version of the parton shower module \apa\ for the \sherpa\
event generator framework is presented. It incorporates some features,
that are specific for the consistent merging with multi-particle
matrix elements at tree-level. This publication also includes some exemplary results
and a short description of the upgraded class structure of \apa, 
{\tt version 2.0}. 
\end{abstract}
%\begin{keyword}
%\end{keyword}
%\end{frontmatter}

\tableofcontents

\section{Introduction}

\noindent
Event generators for the simulation of high-energetic reactions of
particles play an important role in understanding
inclusive and exclusive final states at collider experiments. In order
to live up to the complicated task of a complete simulation of such
processes, event generators decompose the full reaction into a
sequence of individual steps, usually divided by a
characteristic scale. Typically, the starting point for the simulation
of an individual event is the selection of the hardest process in it,
commonly this is denoted as the signal process and described at the
parton level. The emerging partons are then passed to the parton
shower. This part of the simulation describes additional multiple
parton emission. In modelling this Bremsstrahlung-like radiation,
certain approximations are made, that retain the leading soft or
collinear contributions and allow the description of complex
radiation patterns through individual parton splittings at decreasing
scales. When the splitting scale reaches a cut-off value, typically of
the order of a few $\Lambda_{\rm QCD}$, the parton shower stops. At
these scales, the realm of perturbative QCD is left, and the regime of
soft QCD is reached. There, the final translation of the parton to
primordial hadrons is modelled through phenomenological
approaches. If the primordial hadrons are unstable, they are
further decayed until only stable hadrons are left. 

\noindent 
In such a framework, the parton shower provides the link between
perturbatively calculable differential cross sections at the parton
level and models for their transition to observable hadrons with
phenomenological parameters, which need to be tuned to data. The
inclusion of the parton shower does not only create more realistic
high-multiplicity parton final states from low parton multiplicities
available through corresponding exact matrix elements, it also reduces
the average distance of the partons in momentum space down to a
fixed size and thus ensures that the hadronisation
parameter tunes are rather independent of the hard process in
question. It is this aspect of the parton shower that renders it
indispensable for a meaningful simulation with any predictive power.
This can then be used in order to calculate hadronisation corrections
or to judge detector effects in certain processes. In most cases,
results from parton shower Monte Carlos have been in astonishing
agreement with data, giving rise to the confidence that simulation
tools can also be used to predict signals and backgrounds for the
current and the next round of collider experiments. 

\noindent
Although, for more than two decades, parton showers have been widely
used in multi-purpose event generators, details of their specific
implementation traditionally depend on certain choices and, due to recent
developments, on the specific form the hard matrix elements generate
the initial parton configurations. Traditional multi-purpose event
generators, like {\tt Pythia} \cite{Sjostrand:1993yb,Sjostrand:2001yu}
or {\tt Herwig} \cite{Corcella:2000bw,Corcella:2002jc} usually have a 
$2\to 2$ process at leading order (LO) as the signal process. There,
after defining the corresponding, process-dependent starting
conditions for the shower, the latter is allowed to evolve freely. Only
recently, this treatment has been consistently extended to next-to
leading order (NLO) precision for $2\to 2$ processes in the 
{\tt MC@NLO} framework \cite{Frixione:2002bd,Frixione:2004wy}, which 
uses the machinery of {\tt Herwig}. There, the NLO calculation is
modified in order to match the requirements imposed by the specific
form of the parton shower. Another approach, going beyond $2\to 2$
processes, has been taken in {\tt SHERPA} \cite{Gleisberg:2003xi};
there, a fully automated consistent merging of $2\to n$ processes at
the tree-level with the parton shower according to the formalism of
\cite{Catani:2001cc,Krauss:2002up} has been implemented and tested
\cite{Krauss:2004bs}\footnote{
  A similar approach for the merging of matrix elements with
  the dipole cascade \cite{Lonnblad:1992tz} has been taken in
  \cite{Lonnblad:2002sy}; for some specific processes the formalism
  has been used to merge multi-jet matrix elements with {\tt Pythia}
  and {\tt Herwig} \cite{Mrenna:2003if}. A slightly different method
  \cite{Mangano:2001xp} has been advocated by the authors of {\tt Alpgen}
  \cite{Mangano:2002ea}, who merge their matrix elements with the
  parton shower of {\tt Herwig}.}.
For the merging with such multi-leg tree-level matrix elements, the
parton shower has to be supplemented with the determination of more
involved starting conditions and with additional constraints on the
phase space of the emitted particles. 

\noindent 
In this paper the specific realisation and implementation of the
parton shower in {\tt SHERPA} will be presented. A first version of
this module, \apa, {\tt version 1.0}, has been published some time ago
in \cite{Kuhn:2000dk}; it covered the parton shower in the final
state only, including some algorithms for the merging with the matrix
elements provided by \ame\ \cite{Krauss:2001iv,Schalicke:2002ck}. In
this first version, many of the steering and service classes were still
located within \apa; as time went by, \apa\ transformed from a
stand-alone code to a mere module of the full framework {\tt SHERPA}.
In its present state, {\tt version 2.0}, which is discussed in the
following, \apa\ thus includes the parton shower in the initial and
final state and an improved handling of algorithms necessary for the
parton shower aspects of the merging procedure\footnote{
  However, it should be noted that situations like
  deep-inelastic scattering are still beyond the reach of \apa\ and,
  thus, of {\tt SHERPA}.}. 
The outline of this paper is as follows: After briefly reviewing the
parton shower formalism in Sec.\ \ref{PS_Sec}, the ideas underlying 
its particular implementation in \apa\ will be discussed in Sec.\
\ref{Impl_Sec}. There, special focus will be on those aspects, which
are specific for the merging procedure with matrix elements. In
Sec.\ \ref{Results_Sec} exemplary results will be presented, which
compare the parton shower results of \apa\ with analytical approaches
and with data. Details of the implementation, including a short
description of the class structure of \apa\ will be given in the
appendix, App.\ \ref{app:program}.  
 
\section{Parton shower\label{PS_Sec}}

\noindent
This section comprises a brief overview of the parton shower formalism
and the corresponding algorithms implemented in \apa. There, the
sequence of parton emissions in the shower evolution is organised by
virtuality as the ordering parameter. In this respect, the algorithm
presented here is closely related to the implementation in 
{\tt Pythia} \cite{Sjostrand:1993yb,Sjostrand:2001yu}\footnote{
  Note that recently, a reformulation of {\tt Pythia}s parton
  shower has been presented \cite{Sjostrand:2004ef}, which employs
  transverse momentum as ordering parameter. It has been implemented
  into the recent version of {\tt Pythia} \cite{Sjostrand:2003wg}. 
  In \cite{Gieseke:2003rz} an alternative evolution variable, also
  related to transverse momentum, has been discussed.  It has already
  been implemented into {\tt Herwig++} \cite{Gieseke:2003hm}.}.   
However, the \apa\ version differs in some details like the treatment 
of massive particles and particular scale choices for the evaluation 
of coupling constants and parton density functions (PDFs). In addition,
it incorporates some unique features that facilitate the merging
of matrix elements with the parton shower according to the
formalism of \cite{Catani:2001cc,Krauss:2002up}.

\subsection*{Basics of parton showering}

\noindent
The parton shower evolution relies on the fact that parton
emission processes become singular in the soft or collinear limit.
When the available phase space is cut accordingly, these singularities
translate into large logarithms, which can be resummed according
to the DGLAP evolution
\cite{Dokshitzer:1977sg,Gribov:1972ri,Lipatov:1974qm,Altarelli:1977zs}.
By taking into account the leading logarithms only, the parton shower
picture reduces complex radiation patterns of multiple parton
emissions to chains of individual independent parton splittings. They
are organised in a probabilistic manner by an ordering parameter,
usually some quantity like the virtual mass of the decaying parton or
the transverse momentum of the decay products. Suitable constraints on
this ordering parameter avoid singular regions of phase space. The
probability for no parton splitting to occur between two scales
$t_0<t_1$ is encoded in the Sudakov form factor. It is given by 
\begin{equation}\label{eq:Sudakov}
  \Delta_{a}(t_0,t_1) = 
  \exp \left\{ - \intl_{t_0}^{t_1} \frac{\dt}{t} \intl \dz 
    \frac{\alpha_S(p_{\perp})}{2\pi} \sum\limits_{b,c} P_{a\to bc}(z)
  \right\} \;,
\end{equation}
where $p_\perp$ is the transverse momentum. It can be written as a
function of the scale $t$ and the energy splitting variable
$z$. $P_{a\to bc}(z)$ denotes the splitting function for the branching
$a\to bc$. A complete list of splitting functions implemented in \apa\
can be found in App.\ \ref{app:splittings}. Because of its
interpretation as no-splitting probability between two scales, ratios
of two Sudakov form factors like 
\begin{equation}\label{eq:TL_No_Branch}
  P_{\rm no branch} = 
  \frac{\Delta(t_0,t_1)}{\Delta(t_0,t)}
\end{equation}
yield the probability for no emission between $t_1$ and $t$, which
could be resolved at the scale $t_0$. Given a hard starting scale $t_1$,
this allows the scale for the actual branching $t$ to be generated by
equating a random number $R$ with this ratio and solving for
$t$. With a second random number then the splitting variable $z$ can
be selected. It is distributed according to 
\begin{equation}\label{eq:TL_Z}
  P_{\rm split} =
  \frac{\alpha_s(p_\perp)}{2\pi}\,P(z)\,.
\end{equation}
This probabilistic interpretation allows the formulation of the
parton shower as a Markov-chain of independent $1\to2$-branchings,
where the scale $t_1$, at which the former splitting occurs, sets the upper
limit for the subsequent branching. Organising the parton shower in
such a way results in what is known as ``forward evolution''. This
type of evolution is employed for parton showers in the final state,
i.e.\ for the ``time-\-like parton shower'', where the resulting parton
ensemble is not subject of any other constraint.  

\noindent 
For the initial state, however, i.e.\ for space-\-like parton evolution,
the situation changes drastically. This is because standard forward
evolution of a parton ensemble from some comparably low fixed hadronic
scale $Q_0^2<0$, of the order of a few $\Lambda_{\rm QCD}$, to the
fixed scale of hard interaction, $Q^2\ll Q_0^2$, distributed
according to the appropriate matrix elements, would be highly
inefficient. Therefore it is more convenient to start with specifying
the scale $Q^2$ of the core process and its momentum fractions $x_1$
and $x_2$, and subsequently evolve ``backwards'' to the partons
assumed to be resolved from the incoming hadrons. The correct way to
perform this backward evolution has been introduced in
\cite{Sjostrand:1985xi,Marchesini:1987cf}; it boils down to modifying
the Sudakov form factors by dividing by the appropriate PDF at the
given scale and at the relevant Bjorken-$x$. The resulting
no-branching probabilities are then given by\footnote{
  A more efficient, non-Markovian, evolution algorithm is currently 
  under investigation \cite{Jadach:2004talk}.}  
\begin{equation}\label{eq:SL_No_Branch}
  P_{\rm no branch} = 
  \frac{f(x,\,t)}{f(x,\,t_1)}\cdot\frac{\Delta(t_0,t_1)}{\Delta(t_0,t)}\,.
\end{equation}
Again, equating this probability with a random number and solving for
$t$ defines the previous branching scale. This time, however, the
corresponding splitting variable $z$ is distributed according to
\begin{equation}\label{eq:SL_Z}
  P_{\rm split} =
  \frac{\alpha_s(p_\perp)}{2\pi}
  \frac{P(z)}{z}f\left(\frac{x}{z},\,t\right)\,,
\end{equation}
where $x$ is the momentum fraction of the decaying parton, and $z=x/x_1$
with $x_1$ being the momentum fraction of the resulting parton from
which the backward step started.  
  
\noindent
In both time-\-like and space-\-like evolution, the notion of soft colour
coherence \cite{Marchesini:1987cf,Mueller:1981ex,Ermolaev:1981cm,Dokshitzer:1982fh,Marchesini:1983bm}
plays a crucial role. It results in an angular ordering of subsequent
branchings, which can be implemented directly through a corresponding
choice of the evolution variable. This has been done in {\tt Herwig} 
\cite{Corcella:2000bw,Corcella:2002jc}, where a suitable angular
variable has been chosen. Alternatively, colour coherence can 
be implemented by choosing transverse momentum as ordering parameter;
this choice has been made in {\tt Ariadne} \cite{Lonnblad:1992tz}
which bases its multiple emission treatment on splitting colour
dipoles. Recently, two new formalisms to incorporate ordering according
to transverse momentum into the more conventional parton picture have
been presented in \cite{Sjostrand:2004ef,Gieseke:2003rz}. In its most
trivial version, however, angular ordering can be imposed through a
direct veto on increasing opening angles. This is how it has been
implemented in {\tt Pythia}, where virtuality is the evolution
variable. Of course, such a way of implementing colour coherence is by
far not as sophisticated as the methods above; nevertheless this
method has also been chosen in the framework of \apa.  

\subsection*{Variables in \apa}

\noindent
As indicated above, the parton shower in \apa\ is organised in terms
of virtuality, i.e.\ in terms of the virtual mass of the decaying
particle, and the splitting variable $z$ is interpreted as the energy 
fraction of one decay product (daughter) w.r.t.\ the decaying 
parton (mother).\\[2mm]
\begin{enumerate}
\item\underline{Time-like evolution:}\\[2mm]
  \noindent
  For parton branchings $a\to bc$ in the final state, the evolution
  variable $t$ is given by  
  \begin{equation}
    t = t_a = p_a^2 = (p_b+p_c)^2\,.
  \end{equation}
  The splitting variable $z$ is defined as energy fraction, i.e. 
  \begin{equation}\label{eq:fs_zvar_def}
    z = \frac{E_b}{E_a}\,,
  \end{equation}
 taken in the rest frame of the complete final state parton shower\footnote{
    In \apa, the time-like evolution of a given parton ensemble is
    performed in its rest frame. After the shower terminates, the
    resulting partonic final state is fully reconstructed and boosted
    into the relevant (lab) frame.}. 
  For the respective transverse momentum, which serves as an argument in
  the running coupling constant and as a low-energy cut-off, a
  definition following \cite{Catani:1991hj} has been chosen. There,
  \begin{equation}
    k_\perp^2 = 2\,\mbox{\rm Min}\{E_b^2,\,E_c^2\}(1-\cos\theta_{bc})\,,
  \end{equation}
  which in terms of $t_a$ and $z$ translates into
  \begin{equation}\label{eq:k_perp}
    k_\perp^2 = 
    \mbox{\rm Min}\left\{\frac{z}{1-z},\,\frac{1-z}{z}\right\}t_a\,. 
  \end{equation}
  There is, however, some residual freedom in the exact choice of
  scales, which can be altered by a (logarithmically) small factor.
  In \apa, this freedom is used to {\bf define} the transverse
  momentum as 
  \begin{equation}
    p_\perp^2 = \frac14\,
    \mbox{\rm Min}\left\{\frac{z}{1-z},\,\frac{1-z}{z}\right\}t_a\,.
  \end{equation}
  Demanding that all $p_\perp$ are larger than some minimal cut-off,
  $p_\perp^{(0)}$, immediately poses a constraint on the allowed
  $(t_a,\,z)$-range for individual branchings, namely
  \begin{equation}
    t_a>t_0=4{p_\perp^{(0)}}^2\;\;\;\mbox{\rm and}\;\;\; 
    z\in\left[\frac{t_0}{t_a+t_0},\,\frac{t_a}{t_a+t_0}\right]\,.
  \end{equation}
  In addition, the transverse momentum must be positive. Using the
  definition of the splitting variable, Eq.\ \eqref{eq:fs_zvar_def},
  yields a kinematical condition on $z$, namely
  \begin{equation}\label{eq:fs_kinematical_zcondition}
    z\in\left[\frac12\left(1-\sqrt{\frac{t_a}{E_a^2}}\right)\,,
      \frac12\left(1+\sqrt{\frac{t_a}{E_a^2}}\right)\right]\,.
  \end{equation}
  For the explicit angular veto employed to model coherence effects, the
  following approximation for the angle is used  
  \begin{equation}\label{eq:TL_Angle_Est}
    \theta_{bc} \approx \sqrt{\frac{t_a}{z(1-z)E_a^2}}\,.
  \end{equation}
  It stems from $t_a=2z(1-z)E_a^2\,(1-\cos\theta_{bc})$. This
  approximation is valid for small $\theta_{bc}$, but it can conveniently
  be used for large angle emission too, since the actual value is
  irrelevant for its role as ordering parameter.

  \noindent
  Having split parton $a$ into $b$ and $c$, the two offsprings are
  massless first. They gain a virtual mass by continuing the parton
  shower evolution for another step with starting scale $t_a$. Of
  course, the sum of their virtual masses cannot be larger than the
  virtual mass of $a$, i.e.\ 
  \begin{equation}\label{eq:mass_sum_condition}
    \sqrt{t_b}+\sqrt{t_c}<\sqrt{t_a}\,.
  \end{equation}
  This, together with the angular ordering constraint, sets some
  additional limits on the splitting variables. However, there
  is still an issue to be resolved: When $z$ was defined, it was
  implicitly assumed that both outgoing partons are on their mass-shell;
  this is not true any longer, which implies that the branching
  kinematics of $a\to bc$ has to be redefined. In \apa, this is achieved
  through a redefinition of the energy splitting variable $z$, also quite
  along the lines of what happens inside {\tt Pythia}. In both codes, 
  the modified splitting variable reads
  \begin{equation}\label{eq:TL_Shuffle_Z}
    \tilde{z} = 
    \left(z-\frac12\right)
    \frac{\lambda(t_a,\,t_b,\,t_c)}{t_a}+\frac{t_a+t_b-t_c}{2t_a}\,,
  \end{equation}
  where 
  \begin{equation}
    \lambda(a,b,c)=\sqrt{(a-b-c)^2-4bc}\,.
  \end{equation}
  This modification constitutes a simple linear transformation
  from the massless to the massive domain, which is always possible as
  long as the condition given by Eq.\ \eqref{eq:mass_sum_condition} is
  satisfied.

  \noindent
  If one or more of the partons $a$, $b$, and $c$ are massive, the
  definitions above change as follows: 
  \begin{itemize} 
    \item  The evolution variable is now defined as 
    $\tilde {t}_a = t_a - m^2_a$. This in addition to the usage of
    mass terms in the slitting kernels (cf.\ App.\
    \ref{app:splittings}), leads to a slightly modified Sudakov form
    factor. 
  \item The argument of the running coupling is now given by the
    ``complete'' $k_\perp$ formula, i.e.\ it is defined through
    \begin{equation}
      p_\perp^2 = \frac{\mbox{\rm Min}\{E_b^2,\,E_c^2\}}{2}(1-\cos\theta_{bc})\,,
    \end{equation}
    which assumes on-shell, but not necessarily massless daughters. 
  \end{itemize}
  The other conditions introduced above are, in principle, not
  altered, provided that the kinematical variable $t_a$ is used rather
  than the evolution variable $\tilde{t}_a$. Consequently, the
  kinematical $z$-domain is still given by Eq.\
  \eqref{eq:fs_kinematical_zcondition}, and $z$ is translated into the
  massive case by Eq. \eqref{eq:TL_Shuffle_Z}. This leads to a
  reinterpretation of the splitting variable, which is necessary,
  since the splitting functions are calculated with $z$ being the
  light-cone momentum fraction.

\item\underline{Space-like evolution:}\\[2mm]
  \noindent
  For the parton shower in the initial state, i.e.\ space-\-like showers,
  the evolution proceeds backwards. The ordering parameter here is the
  virtual mass of the respective initial state parton. Therefore in
  branchings $b\to ac$ the scale is given by
  \begin{equation}
    t = t_a = p_a^2 <0\,,  
  \end{equation}
  and, wherever it is needed, the absolute value is taken. The
  definition of the splitting variable is a little bit more tricky.
  In order to ensure four-momentum conservation in the reconstruction
  of the showering kinematics, the two shower branches of the two
  incoming partons are coupled. For the two incoming particles,
  labelled by $a$ and $\tilde a$, the Sudakov form factors, cf.\ Eq.\
  (\ref{eq:SL_No_Branch}), are used to choose their virtual mass, $t_a$
  and $t_{\tilde a}$. The parton with larger off-shellness is selected
  for the reconstruction of the corresponding backward step. Assume
  that $t_a<t_{\tilde a}$, and, hence, that $a$ is selected. Then $z$
  is defined through
  \begin{equation}
    z = \frac{(p_a+p_{\tilde a})^2}{(p_b+p_{\tilde a})^2}\,,
  \end{equation}
  clearly a Lorentz-invariant measure. This implies that, step by
  step, the c.m.\ energy squared in the parton system is enhanced by a
  factor $1/z$. In the same way, the Bjorken-$x$ of partons $a$ and
  $b$ are related by
  \begin{equation}
    x_b = x_a/z\,,
  \end{equation}
  where the first Bjorken-$x$ of the partons entering the hard process
  are taken directly from the matrix element evaluation. In this
  respect, at each step of the space-\-like shower evolution, the current
  pair of partons is oriented along the beam axis, with Bjorken-$x$ as
  if they were massless. It is this Bjorken-$x$ and a scale
  $Q^2=p_\perp^2$ that constitute the parameters for the calculation
  of the PDF at the corresponding step. In the c.m.\ system of partons
  $b$ and $\tilde a$, the transverse momentum of $a$ and $c$ is given
  by    
  \begin{equation}
    k_\perp^2 = -(1-z)\,t_a\,.
  \end{equation}
  Following a similar reasoning as in the final state treatment, the
  transverse momentum entering the running coupling as well as the PDF
  evaluation is {\bf defined} through 
  \begin{equation}
    p_\perp^2 = -\frac{1-z}{4}\,t_a\,.
  \end{equation}
  
  \noindent
  Having thus constructed the backward step $b\to ac$ leading to the
  parton $b$, it is clear that $c$ may experience a final state shower
  evolution. Its starting scale, by default, is given by $t_a$. For
  kinematic reasons, however, the actual scale that is chosen in the
  shower evolution must also satisfy
  \begin{equation}\label{eq:MaxT}
    t_c <t_{c,{\rm max}} = 
    \frac{\lambda_{a\tilde a}\lambda_{b\tilde a}-
          s_{a\tilde a}s_{b\tilde a}}{2 t_{\tilde a}} + 
        t_a + t_{\tilde a}
  \end{equation}    
  with
  \begin{align}    
    s'_{a\tilde a} & = s_{a\tilde a} - t_a - t_{\tilde a} \;,   & 
    s'_{b\tilde a} & = s_{b\tilde a} - t_b - t_{\tilde a} \\
    \lambda_{a\tilde a} &= \lambda(s'_{a\tilde a},t_a,t_{\tilde a}) & 
    \lambda_{b\tilde a} &= \lambda(s'_{b\tilde a},t_b,t_{\tilde a}) 
  \end{align}
  where $s_{a\tilde a}=(p_a+p_{\tilde a})^2$ and 
  $s_{b\tilde a}=(p_b+p_{\tilde a})^2 = s_{a\tilde a}/z$. 
  Similar to the approximation of Eq.\ (\ref{eq:TL_Angle_Est}), the
  opening angle of this splitting can be estimated by
  \begin{equation}\label{eq:SL_Angle_Est}
    \theta_{ac} \approx \sqrt{\frac{t_a}{(1-z)\,E_a^2}}\,.
  \end{equation}
  \noindent
  Knowing $t_b$ and $t_c$ as well as the four-momenta $p_a$ and
  $p_{\tilde a}$, the four-momenta $p_b$ and $p_c$ are constructed
  explicitely in the c.m.\ frame of  $p_a$ and $p_{\tilde a}$. 
  In this system, energies and momenta are fixed according to
  \begin{align}
    E_c &= \frac{s_{b{\tilde a}}+  t_b + t_a - t_c}
    {2\sqrt{s_{a\tilde a}}} \\
    p_{{\rm z},c} &= \frac{s_{b\tilde a} - 2 E_{\tilde a}E_c}
    {2 p_{{\rm z},a}} \\
    p_{\perp c} &=  \sqrt{E_c^2 - p_{{\rm z},c}^2 - t_c}\,,
  \end{align}
  where the transverse momentum is distributed uniformly in its
  azimuthal angle. Furthermore, 
  \begin{align}
    p_b &= p_a + p_c\,.
  \end{align}
  Subsequently, the system is boosted into the rest frame of partons
  $b$ and $\tilde a$ and rotated such that their momenta point along
  the beam axis.
\end{enumerate}

\subsection*{Colour treatment}

\noindent
When the parton shower terminates, the resulting multi-parton ensemble
needs to be transformed into hadrons. Due to a lack of quantitative
understanding of non-perturbative physics, this is achieved through
phenomenological models. These models have some different underlying
physics assumptions, highlighted by the examples of the string model
\cite{Artru:1974hr,Artru:1979ye,Bowler:1981sb,Andersson:1983jt,Sjostrand:1984ic,Andersson:1998tv}
stressing the role of colour coherence \cite{Azimov:1986sf}, and of 
cluster fragmentation models
\cite{Gottschalk:1982yt,Webber:1983if,Webber:1986mc,Winter:2003tt,Gieseke:2003hm}
that are closer to the independent fragmentation approach \cite{Field:1977fa}
and the concept of pre-confinement \cite{Amati:1979fg}. However, they
have in common that they rely on having as input a parton ensemble
with a  well-defined colour structure in the $N_c\to\infty$ limit.
This necessitates that the parton shower must distribute colours in
this limit. In most cases, like, e.g.\, in a splitting $q\to qg$ the
colour structure is unambiguously defined, ensuring that a well-defined
colour structure at the beginning of the parton shower evolution can
be mapped onto a well-defined colour structure at its end. However,
there are cases with ambiguities, namely in splittings of the type
$g\to gg$. In this case, there are two ways of locally distributing
colour. Denoting a colour state by its colour triplet and anti-triplet
labels in the $N_c\to\infty$ limit, $(a,\,\bar b)$, these two ways in
the splitting $g(1)\to g(2)g(3)$ can be visualised as
\begin{equation} 
(a,\,\bar b) \to (a,\,\bar c) + (c,\,\bar b)\;\;\;\mbox{\rm or}\;\;\;
(a,\,\bar b) \to (c,\,\bar b) + (a,\,\bar c)\,.
\end{equation} 
In \apa, the choice is made in the following way: In the shower
evolution, the decaying gluon $1$ was produced in another parton
branching process, where the other decay product is colour-connected
to the gluon. For both gluons $2$ and $3$, the transverse momentum
w.r.t.\ this colour partner is determined according to Eq.\
(\ref{eq:k_perp}). The choice is then made such that the gluon with
the smaller relative $k_\perp$ is colour connected with this partner
parton.

\subsection*{Initialisation of the parton shower}

\noindent
Within the {\tt SHERPA} framework, an algorithm along the lines of 
\cite{Catani:2001cc,Krauss:2002up} for the consistent merging of
tree-level matrix elements for multi-particle production with the
parton shower of \apa\ has been implemented \cite{InPrep}. The key
idea of this algorithm is to separate the phase space for parton
emission into the hard region of jet production accounted for by
suitable tree-level matrix elements and the softer region of jet
evolution covered by the parton shower. Then, extra weights are
applied on the former and vetoes on the latter, such that the overall
dependence on the separation cut is minimal. The separation is
achieved through a $k_\perp$  measure
\cite{Catani:1991hj,Catani:1992zp,Catani:1993hr}. The weight attached
to the matrix elements is constructed employing Sudakov form factors,
thereby taking into account those terms that would appear in a
corresponding parton shower evolution. Therefore, a ``pseudo parton
shower history'' is reconstructed by clustering the initial and
final state particles from the tree-level matrix element according to
the $k_\perp$ algorithm. This procedure provides the scales necessary
for the evaluation of the weight, namely the nodal values of the
different $k_\perp$, where two jets have been merged into one.  

\noindent
Within \apa, the reconstructed ``shower history'' is utilised for the
determination of initial conditions for the shower evolution. Then,
starting from the hard $2\to 2$ core process, all partons obtain as
starting scale for their shower evolution the nodal virtuality, where
they emerge for the first time. It should be stressed at this point
that there is some residual mismatch in the parton shower variables
used in \apa, namely virtuality $t=p^2$, and the scales ($k_\perp$
measures) used in the Sudakov form factors employed in the reweighting
of the matrix elements. This point will be discussed in more detail in
the following section.  

\noindent
For the scale of the four partons entering or leaving the $2\to 2$
process, there are a few options, defining the respective choice of
scale $t_{\rm start}$:  
\begin{itemize}
\item 
  Processes of the type $e^+e^-\to q\bar q$:\;\;\; $t_{\rm start} =
  \hat s = (p_q+p_{\bar q})^2$.
\item 
  Drell-Yan type (e.g.\ $q\bar q\to e^+e^-$):\;\;\; $t_{\rm start} = 
  \hat s = (p_q+p_{\bar q})^2$.
\item 
  ``Drell-Yan+jet''-type (e.g.\ $q\bar q\to W^{(*)}g$):\;\;\; 
  For the incoming partons the choice is 
  $t_{\rm start} = \hat s = (p_q+p_{\bar q})^2$. For the final
  state parton, however, the start scale for the shower evolution is
  different, namely the nodal value where it has been produced,
  usually $\hat t$ or $\hat u$.
\item 
  QCD processes: 
  In this case, different colour flows are usually competing.
  The winner is selected according to the respective
  contribution, which is related to a specific propagator structure
  ($\hat s$-, $\hat t$-, or $\hat u$-channel). The choice of starting
  scale follows the selection of the colour structure; in most cases
  therefore the starting scale $t_{\rm start}$ is the minimum of 
  $\hat t$ and $\hat u$.
\end{itemize}
In order to account for colour coherence, a maximal allowed branching
angle $\theta_{\rm crit}$ has to be determined for each parton. Using
the knowledge of the colour connections of the $2\to 2$ process,
$\theta_{\rm crit}$ is defined as the angle between the parton in
question and the parton colour connected to it. In case of gluons 
the choice of this angle is, of course, ambiguous; there, the
maximum of both possible values is taken.

\noindent
Having fixed the starting scales of the hardest partons, the starting
scales for the evolution of softer partons are easy to obtain. In
principle, there are only two ways, in which softer partons may
emerge, which lead to a slightly different treatment.
\begin{itemize}
\item 
  Branchings in the final state:\\
  There the harder, i.e.\ the more energetic, of the two offsprings
  inherits as starting scale and angle the values of the decaying
  parton, whereas the parameters of the softer offspring are taken
  directly from the node. In other words, the initial virtual mass
  equals the invariant mass of the pair, and the starting angle is
  given by its opening angle.
\item 
  Branchings in the initial state:\\
  There, in analogy to the treatment of the space-like shower
  evolution, the virtual mass and the starting angle of the time-like
  offspring are given by the virtual mass of the initial state
  parton that branches ``backwards'' and the respective opening angle
  w.r.t.\ the corresponding beam. The ``new'' initial state parton
  inherits the starting scale and angle of the line pointing towards
  the ``core'' process.
\end{itemize}

\subsection*{Merging issues}

\noindent
Having fixed the starting scales, the parton shower can be evolved.
Due to the merging prescription, emissions inside the parton shower
have to be vetoed, if they result in the production of partons with
transverse momenta $k_\perp$ larger than a given jet resolution
scale, $p_\perp^{\rm cut}$. Parton showers attached to matrix elements
with the highest multiplicity must live up for the production of
eventual extra jets. In this case, the veto scale is not the jet
resolution scale but rather given by the smallest $k_\perp$ of the
partons stemming from the matrix element. 

\noindent
However, in the parton evolution of such multi-parton final states,
there is a last subtlety. It is connected to the fact that, through
showering, the partons may acquire a virtual mass different from their
on-shell mass used in the matrix element. This recoil has to be
compensated for, if possible in such a way that the other partons
stemming from the matrix element are least perturbed. This is solved
in the following way: The merging procedure results in the
reconstruction of a pseudo parton shower history of the parton
ensemble. Thus, to each parton produced in the matrix element a
partner can be assigned such that both emerge in a branching of the
pseudo parton shower history. If one of them or both acquire a mass,
the respective partner takes care of the recoil. The strategy employed
for this is very similar to the one used when the initially massless
partons in the showering acquire a mass; in both cases the energy
splitting variable is shifted through a linear transformation, for the
shower case, cf.\ Eq.\ (\ref{eq:TL_Shuffle_Z}). In the initial
branchings of the shower initiators, there are two cases to be
considered: 
\begin{itemize}
\item 
  Both partners experience a parton shower evolution. Then, the
  reconstructed energy splitting variable of the branching where they
  have been produced, is shifted according to 
  Eq.\ (\ref{eq:TL_Shuffle_Z}).
\item 
  In the branch $a\to bc$, only one partner ($b$) experiences a parton
  shower evolution, the mass of its partner is ($c$) fixed, since it
  can be thought of as an internal line of the matrix element - it
  branches according to the pseudo parton shower history. Then $z$
  is shifted according to 
  \begin{equation}\label{eq:TL_Shuffle_Z_ME}
    \tilde{z} =
    \left(z-\frac{t_a+t^{(0)}_b-t^{(0)}_c}{2 t_a} \right) 
    \frac{\lambda(t_a,t_b,t_c)}{\lambda(t_a,t^{(0)}_b,t^{(0)}_c)} 
    + \frac{t_a+t_b-t_c}{2 t_a}
    \;,
  \end{equation}
  where $t^{(0)}_i$ denote the matrix element masses and $t_i$ the new
  virtualities. This modification results in a change of the opening
  angle $\theta_{bc}$, leading to a modified momentum even for the
  particle with the mass unchanged. In order to comply with
  four-momentum conservation, its offsprings have to be adjusted,
  too. This transformation can easily be achieved through a rotation
  followed by a boost along the direction of particle $c$.
\end{itemize}

\noindent
Within \apa, an additional veto on ``losing'' a jet produced in the
matrix element is introduced. Losing a jet through the shower is
possible due to a mismatch of the quantities responsible for jet
definition ($k_\perp$) and jet evolution (virtual mass). This leads to
recoils imposed by the shower evolution resulting in a change of
$k_\perp$, while the virtuality of the internal line in question is
preserved. Therefore, a check on the number of jets is performed after
the shower evolution is finished. This is done by demanding that the -
now off-shell - shower seeds are still separated in $k_\perp$. If this
is not the case the event is rejected, and a new event is generated
from the beginning, i.e.\ with a new kinematical situation but the
same flavour constellation of the jets.

\newpage

\section{Implementation\label{Impl_Sec}}

\noindent
This section focuses on the basic strategies according to which the
formalism discussed in the previous section, Sec.\ \ref{PS_Sec}, is
implemented in \apa. A more detailed reference to all individual
classes can be found in App.\ \ref{app:program}.

\noindent
The basic unit for the realisation of the parton shower in
terms of a computer program is the individual branching of a single
parton. Connecting such branchings automatically leads to the Markov
structure of the full emission pattern. In other words, the task of a
parton shower program is to fill and to connect such individual
branchings. It is the strength of an object-oriented programming
language like {\tt C++} that it is well suited to map the underlying
physical pictures onto program code by using an appropriate class
structure.  

\subsection*{Representation of the parton shower}

\begin{figure}[bt]
\centerline{\includegraphics[width=6cm]{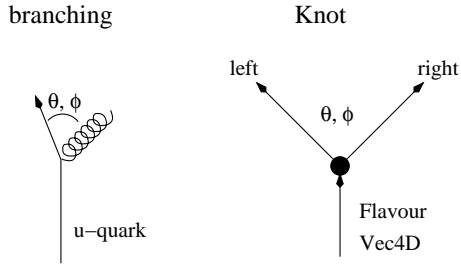}}
\caption{\label{fig:map_knot} 
  The basic building blocks of shower emission are binary branchings,
  represented by {\tt Knot}s. }
\end{figure}

\noindent
Following the reasoning above, the basic structure representing a
single branching is a {\tt Knot}, reflecting the $a\to bc$ binary
decay structure inherent to the parton shower. A {\tt Knot} carries 
information on the incoming particle $a$, encoded in the class 
{\tt Particle}, and on the {\tt Knot}, where it originates from. 
When it is filled, i.e.\ when the decay is specified, it also
yields the two offsprings and their respective decay {\tt Knot}s. 
In addition, a number of other quantities are stored in a {\tt Knot},
namely
\begin{itemize}
\item the scale $t_a$,
\item the energy splitting parameter $z$, 
\item the squared energy, $E_a^2$, 
\item and its minimal virtuality $t_a^{(0)}$, 
\item the respective opening angle in the approximation of
  Eqs.\ (\ref{eq:TL_Angle_Est}, \ref{eq:SL_Angle_Est}) $\theta_{bc}$,
\item the azimuthal angle $\phi$, 
\item and the Bjorken-$x$ of the particle $x_a$ (if needed).
\end{itemize}

\begin{figure}[t]
\centerline{\includegraphics[width=14cm]{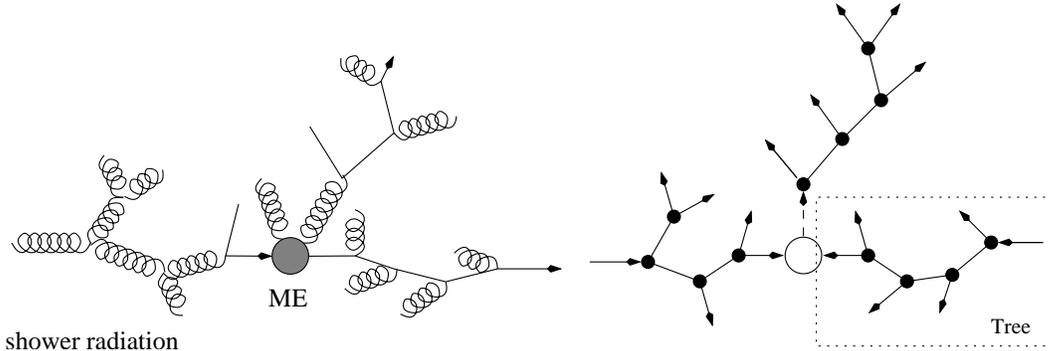}}
\caption{\label{fig:map_tree} 
  Sketch of the mapping between radiation processes and the
  corresponding classes. The full radiation pattern is identified as a
  chain of $1\to 2$ processes, a Markov chain, which translates into
  the class {\tt Tree}. A {\tt Tree} in turn is realised as a list of
  linked {\tt Knot}s. The shower evolution of an event is represented
  by three {\tt Tree}s, one for the final state shower, and two for
  the initial state shower.} 
\end{figure}
\noindent
These {\tt Knots} are then linked in terms of a (binary) Markov chain,
represented in the class {\tt Tree}. Each {\tt Tree} contains a
pointer to its first {\tt Knot}, the root {\tt Knot}\footnote{
  Since a parton shower history is reconstructed in the
  merging procedure, this root {\tt Knot} may reflect the already fully
  defined splitting of an internal line of some matrix element.}. 
Starting from this root {\tt Knot}, all other {\tt Knot}s are then
accessible by successively following the pointer inside the 
{\tt Knot}s, thereby spanning the full {\tt Tree} structure.  
Each event consist of three {\tt Tree}s, as indicated in Fig.\ 
\ref{fig:map_tree}; one {\tt Tree} gives rise to the complete final 
state shower, eventually with a dummy particle as root {\tt Knot}. The
other two {\tt Tree}s represent the initial state showers from both
sides, therefore they have one of the two particles entering the
hardest subprocess as their respective root {\tt Knot}. 
Since the full parton shower evolution is represented by these 
{\tt Tree}s, they provide routines to create and delete individual 
{\tt Knot}s, or to boost and rotate the full structure. The latter 
option is relevant for the transparent implementation of the initial 
state parton shower.

\subsection*{Filling the {\tt Tree}s}

\begin{figure}[t]
\centerline{\includegraphics[width=13cm]{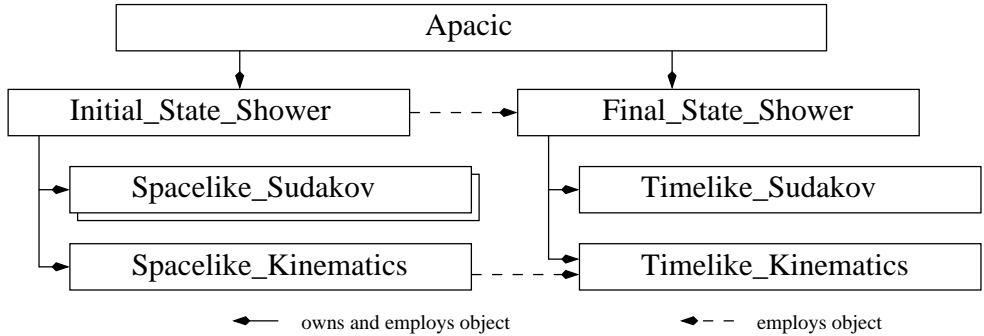}}
\caption{\label{fig:apa_classes} Relation of the main classes of \apa.}
\end{figure}

\noindent
The {\tt Tree}s are now to be filled by the corresponding parton
shower. In \apa, two classes are responsible for this, namely the 
{\tt Final\_\-State\_\-Shower} and the {\tt Initial\_\-State\_\-Shower}, 
who have access to one or two {\tt Tree}s, respectively. Since the
latter, the {\tt Initial\_\-State\_\-Shower}, produces particles that
may undergo a time-like shower evolution, it also has a pointer to the
former, the {\tt Final\_\-State\_\-Shower}. Apart from switches
steering the specific way the shower works \footnote{
  There are some options for specific questions made available in
  \apa, that go beyond the standard settings discussed above.}, 
however, both showers need information from Sudakov form factors
and they have to construct the respective time-\-like or space-\-like
kinematics. These necessities are encoded in corresponding classes,
named {\tt Timelike\_\-Sudakov} or {\tt Spacelike\_\-Sudakov}, and 
{\tt Timelike\_\-Kinematics} or {\tt Spacelike\_\-Kinematics}. It is
clear that the {\tt Spacelike\_\-Kinematics} must use methods from its
time-\-like counterpart, hence it has a pointer to 
{\tt Timelike\_\-Kinematics}. Since the two {\tt Kinematics}-classes are
responsible for jet vetoes, they have a pointer to a {\tt Jet\_Finder}
class made available through the {\tt SHERPA} framework and some flags
steering its proper usage. An overview over the basic relations
between the main classes of \apa\ is depicted in Fig. \ref{fig:apa_classes}.

\noindent
In contrast to the {\tt Kinematics} classes, the two {\tt Sudakov}
classes are a bit more intricate. Since most of their actions are
related to the selection of splitting functions and their usage, both
are derived from the class {\tt Splitting\_\-Group}, and contain
a {\tt Splitting\_\-Group} for each flavour, the shower can handle.
%
%\noindent
Then, each {\tt Splitting\_\-Group} builds a wrapper around all
splitting functions $P_{a\to bc}(z)$ for all branchings allowed for a
certain incoming particle $a$. As an example, consider the case of a
gluon. It may undergo either a $g\to gg$ or a $g\to q\bar q$
branching, each of which is represented by its corresponding 
{\tt Splitting\_\-Function}. Consequently, the class
{\tt Splitting\_\-Group} responsible for the gluon branching contains one 
{\tt Splitting\_\-Function} for the gluon final state and one for each
quark flavour. Any individual {\tt Splitting\_\-Function} incorporates
information like its incoming and outgoing flavours, an estimate for
its integral over $z$, and methods to extract $z$ distributed
according to $P_{a\to bc}(z)$. In addition, {\tt Splitting\_\-Group}s
allow to select a splitting mode and the corresponding flavours. 

\noindent
The two {\tt Sudakov} classes construct all such {\tt Splitting\_\-Group}s
for physical branchings. When a parton, encoded in {\tt Knot} is to be
split, the respective Sudakov form factor implementation
selects the physically relevant {\tt Splitting\_\-Group}, i.e.\ the one
with the flavour of the parton as incoming or outgoing particle, and uses
its estimated $z$ integral to select a corresponding $t$ \footnote{ 
  Other information needed for this selection, like the start scale
  $t_{\rm start}$ and the maximally allowed angle $\theta_{\rm crit}$
  are obtained from the {\tt Knot}.
}. 
Then vetos are applied, either in the procedure of the hit-or-miss
method used in the {\tt Sudakov} classes, or in the framework of the
merging procedure to reject unwanted jets to be produced in the parton
shower. The latter test is performed with the help of the
corresponding {\tt Kinematics} class. It should be noted that the 
{\tt Spacelike\_\-Sudakov} classes contain a PDF due to the backward
evolution described above; in view of the possibility to have two
different beam particles with different PDFs the class
{\tt Initial\_\-State\_\-Shower} has two of them.

\section{Results\label{Results_Sec}}

\noindent
In this section some results will be presented that validate the
implementation of the parton shower algorithms, discussed in the
previous sections. 

\subsection*{Comparison with analytic Sudakov form factors}

\noindent
First of all, the parton shower of \apa\ will be confronted with known 
analytical results dealing with the resummation of large logarithms
emerging in multiple parton emission. Such results are available
mainly for $e^+e^-$ annihilation into hadrons. There, analytical 
calculations describe the relative rates for different jet
multiplicities in the $k_\perp$ algorithm. For the massless case,
results can be found in \cite{Catani:1991hj}, massive quarks are dealt
with in \cite{Krauss:2003cr}. All results presented here relate to 
$e^+e^-$ annihilations at $E_{\rm c.m.}=91.2$ GeV, i.e.\ to the
LEP 1 energy.  

\begin{figure}
  \begin{center}
    \includegraphics[width=6.8cm]{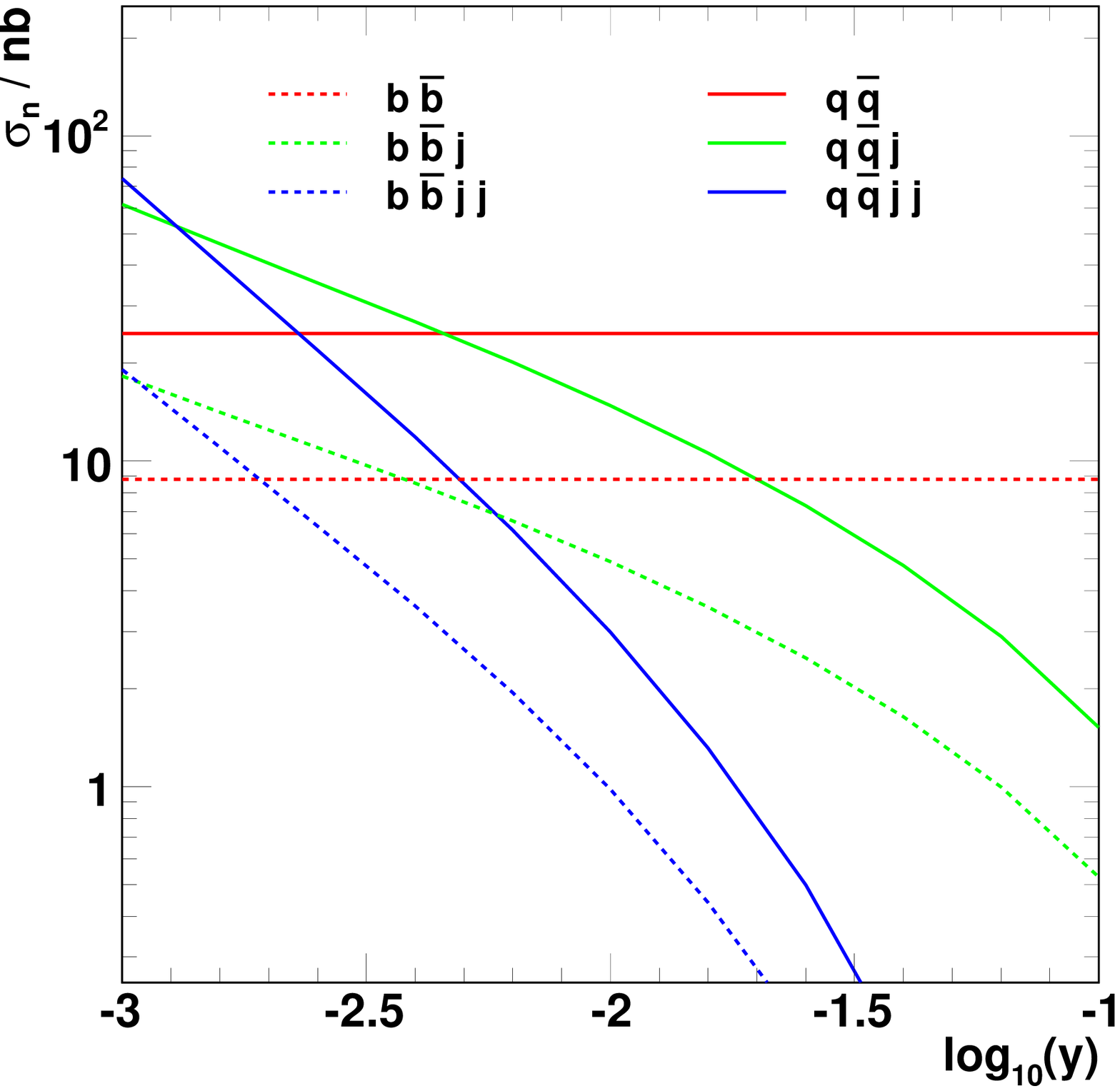}
    \includegraphics[width=6.8cm]{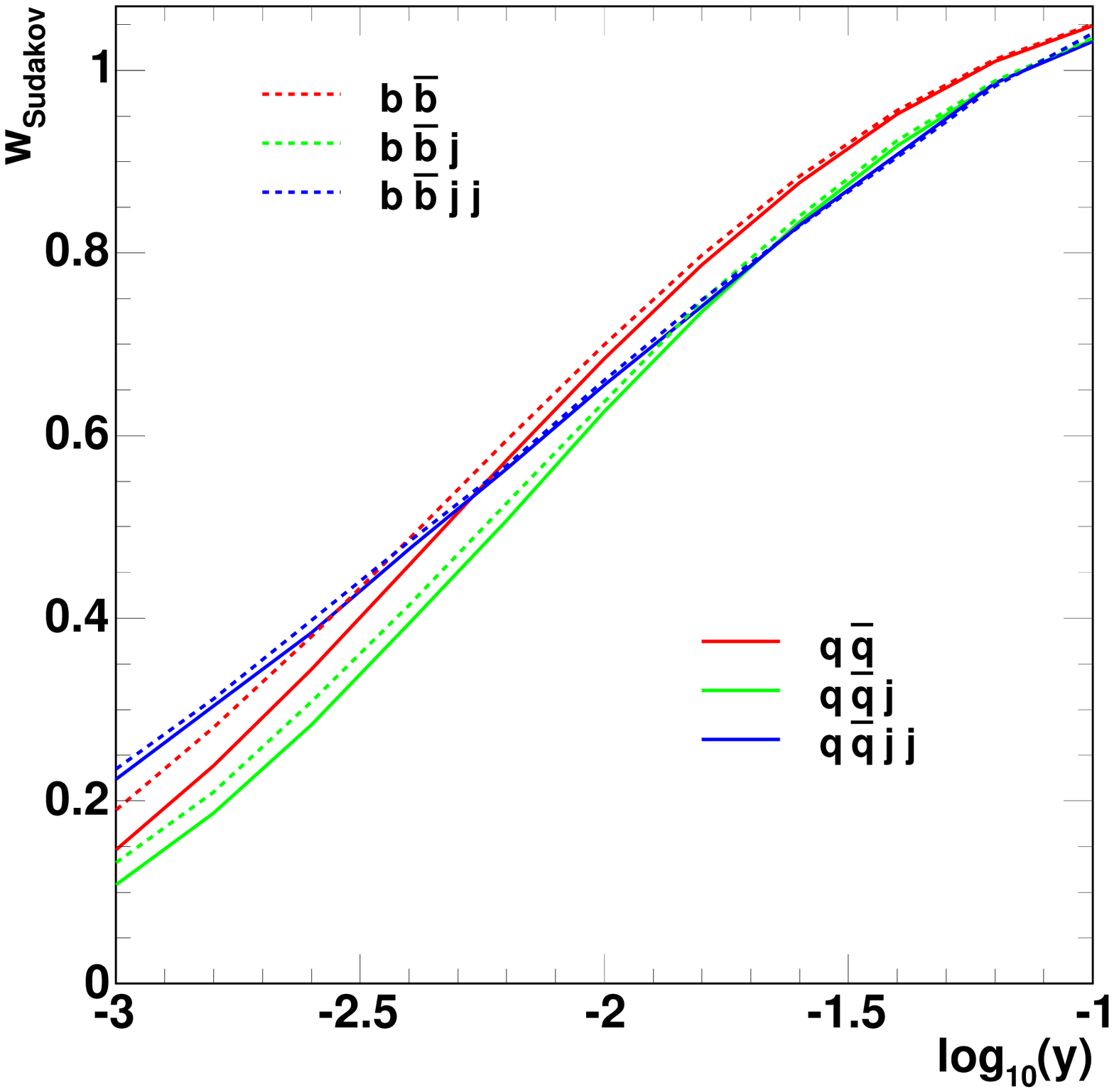}
    \caption{\label{fig:xsec_sud} Cross section for the production of
      up to four jets (left) and the corresponding Sudakov weight that
      will be attached in the merging procedure (right) vs.\ the jet
      resolution $y$. Two/three/four jet topologies are depicted in
      red/green/blue, massless jet-configurations are shown with solid
      lines, jet configurations involving two massive $b$ quarks
      with $m_b=4.5$ GeV are shown with dashed lines.}
  \end{center}
\end{figure}

\begin{figure}
  \begin{center}
    \includegraphics[width=6.8cm]{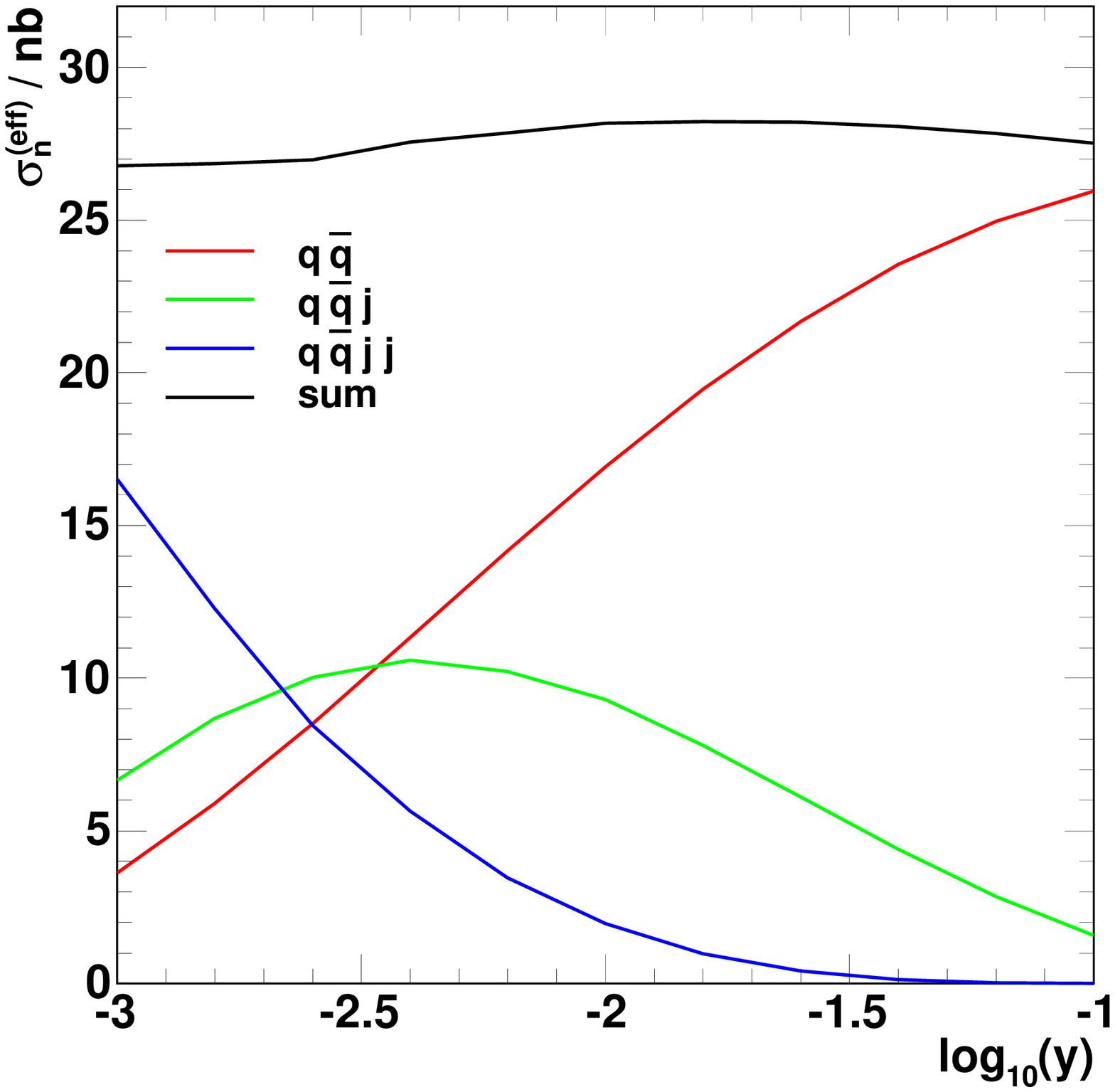}
    \includegraphics[width=6.8cm]{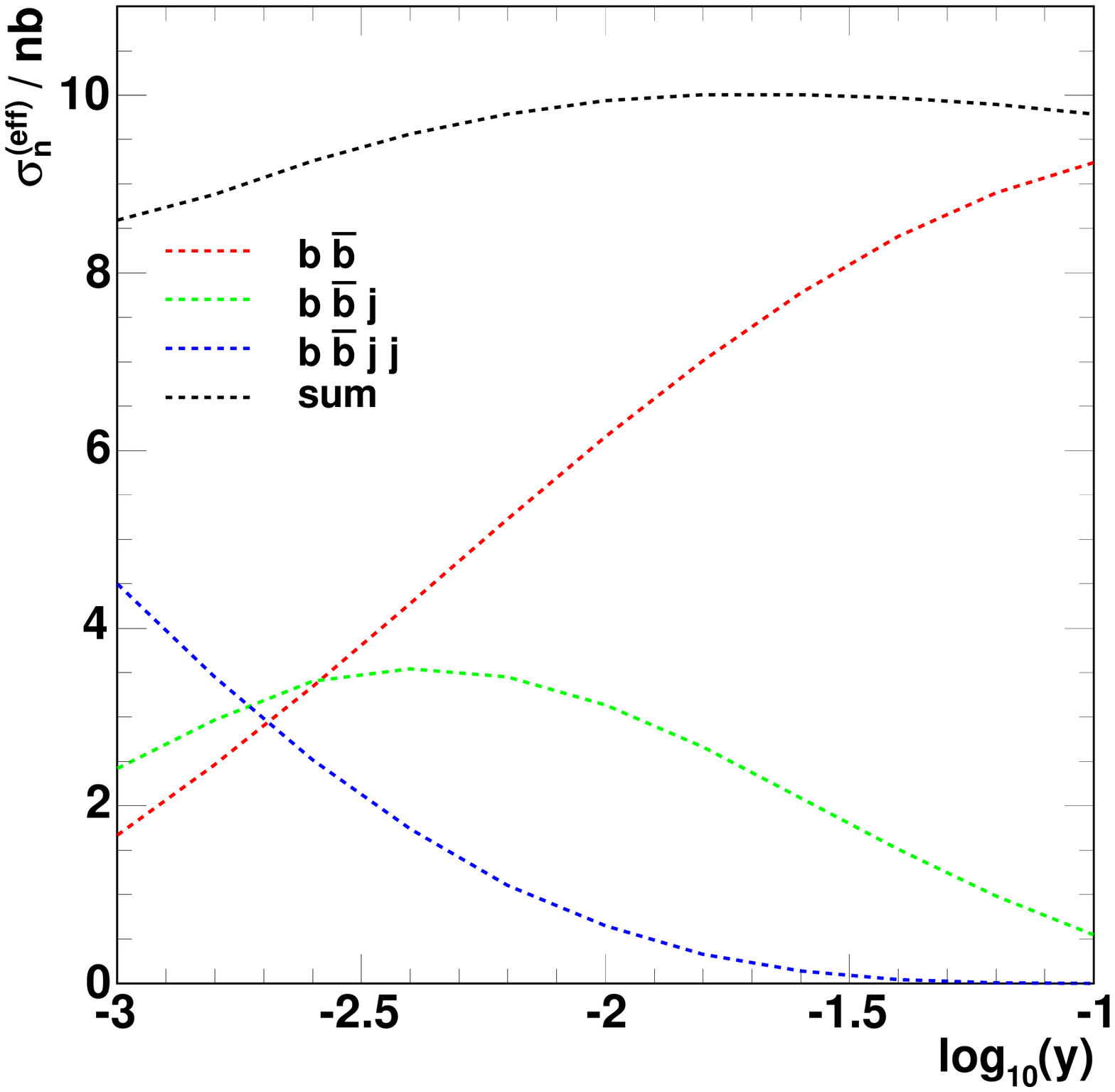}
    \caption{\label{fig:xsec_times_sud} Cross section times Sudakov
      weight vs.\ the jet resolution $y$ for each jet configuration 
      and the sum of them. The left plot shows massless jets, the
      right one displays jet configurations with two massive $b$
      quarks.} 
  \end{center}
\end{figure}

\begin{figure}
  \begin{center}
    \includegraphics[width=8.0cm]{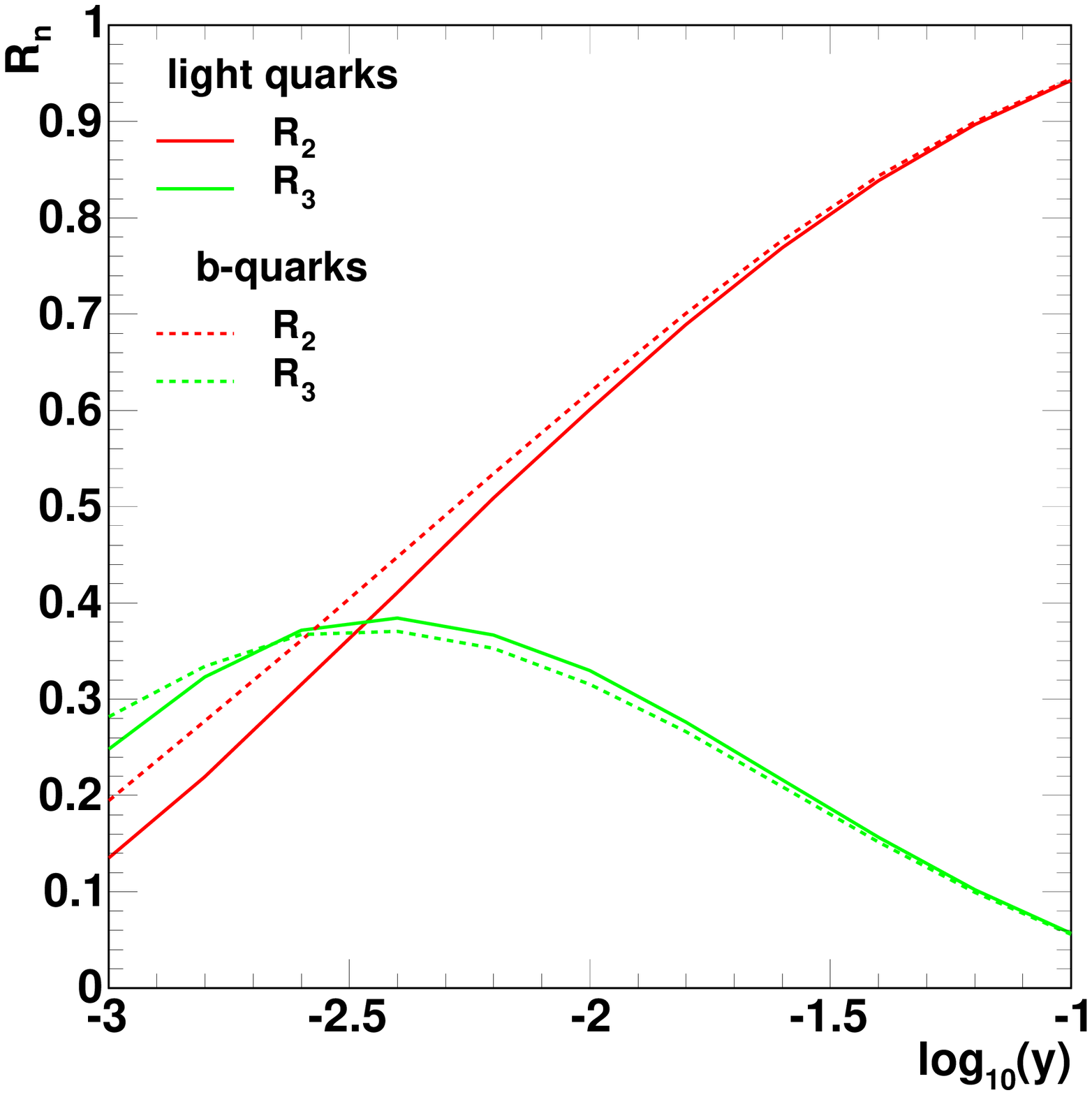}
    \caption{\label{fig:jetrates} Jetrates for two- and three jets
      from the combined matrix element plus Sudakov weight.}
  \end{center}
\end{figure}

\noindent 
To gain some first idea about the behaviour of the merged cross
sections and the effect of the Sudakov form factors, consider Fig.\
\ref{fig:xsec_sud}. There, tree-level cross sections for the
production of up to four jets, out of which two may be $b$-jets 
(left) are contrasted with the weight applied to them in the merging
procedure (right). The jets are defined through a Durham jet measure
with resolution parameter $y$, which represents the minimal relative
transverse momentum of two separated jets. Already here, it becomes
apparent that the massive $b$-quarks are suppressed w.r.t.\ massless
ones in the cross section, whereas the weights are nearly identical
for massless and massive jets. In Fig.\ \ref{fig:xsec_times_sud} the
merging has been performed; in the left plot the resulting jet cross
sections and their sum are shown for the massless case, the right plot
exhibits the corresponding cross sections when two $b$ quarks are
involved. It is surprising how stable the total cross sections are
w.r.t.\ the jet definition cut. In Fig.\ \ref{fig:jetrates} the
resulting two- and three-\-jet rates are displayed. Clearly, due to
the ``dead cone'' effect the $b$-quarks tend to radiate less, leading
to an enhanced two jet rate compared to light quarks. For three jets,
the situation is slightly different; for small value of $y$ it is the
four-\-jet rate that starts dominating. In this region therefore the
three-\-jet rate for massive quarks starts to exceed the massless one,
which tends to have a larger fraction of four-\-jet events.

\begin{figure}
  \begin{center}
    \includegraphics[width=6.8cm]{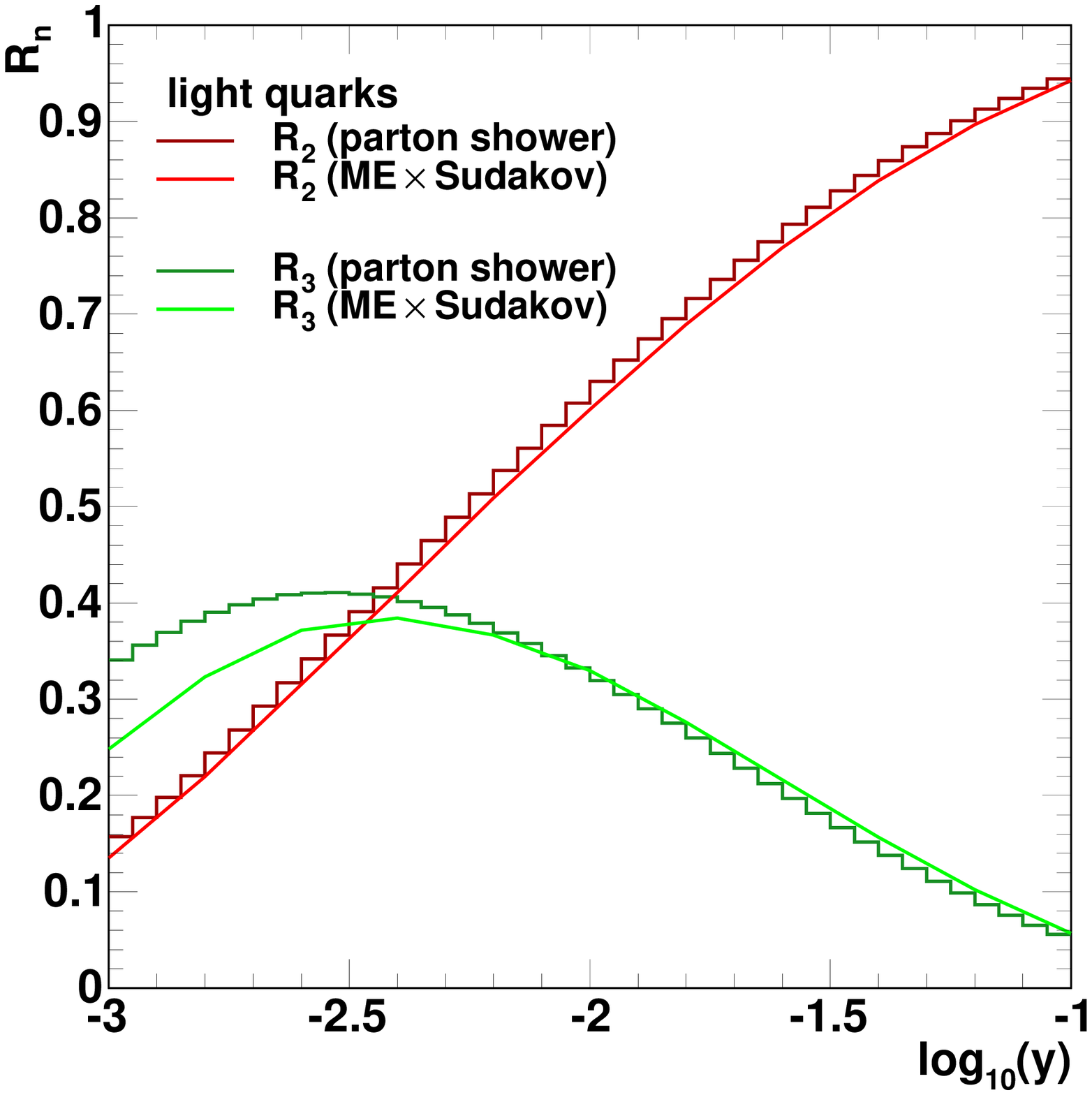}
    \includegraphics[width=6.8cm]{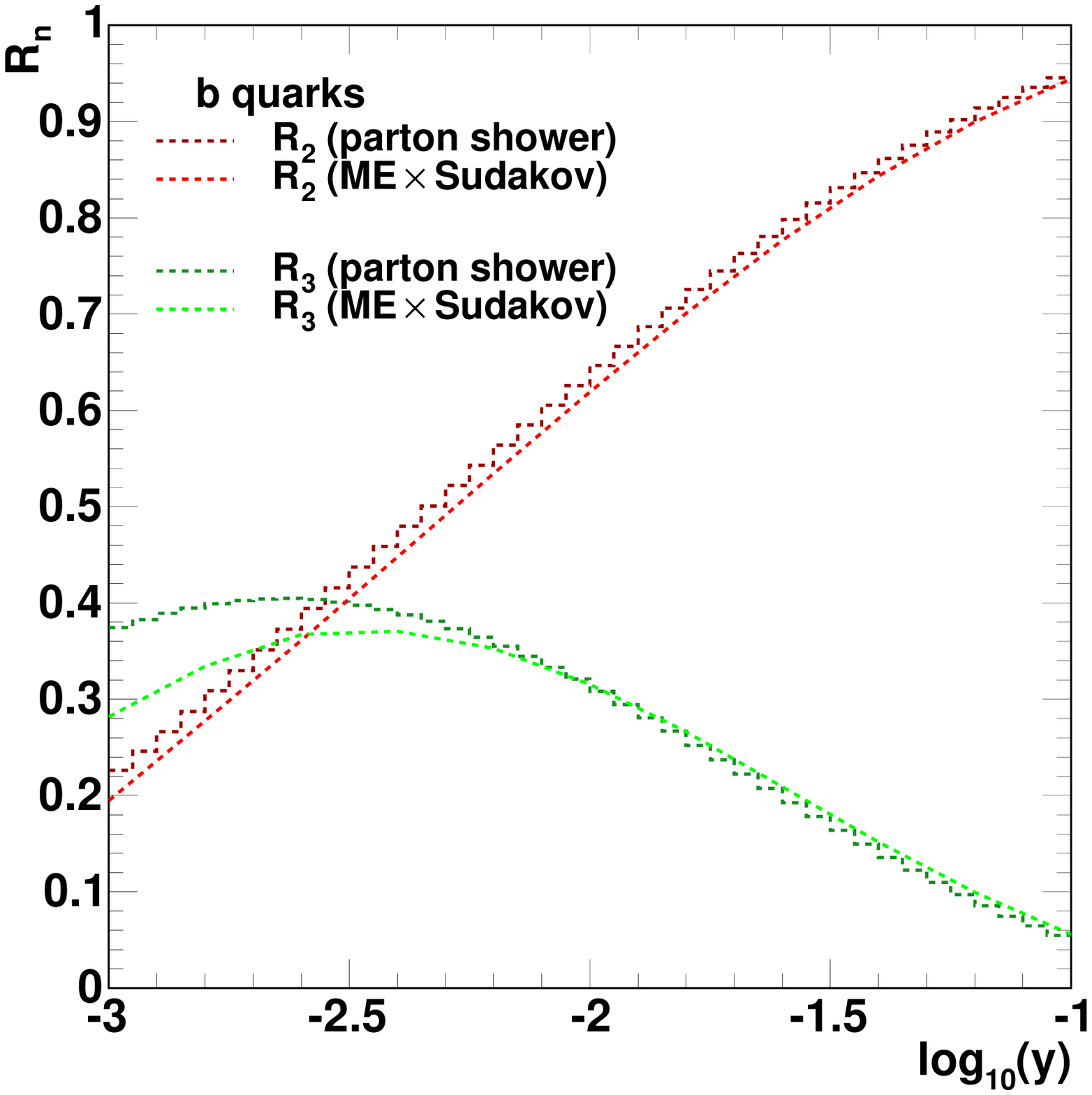}
    \caption{\label{fig:jetrates_vs_shower} Analytical results for the
      jetrates from Fig.\ \ref{fig:jetrates} vs.\ the results obtained
      from the parton shower.}
  \end{center}
\end{figure}

\begin{figure}
  \begin{center}
    \includegraphics[width=8.0cm]{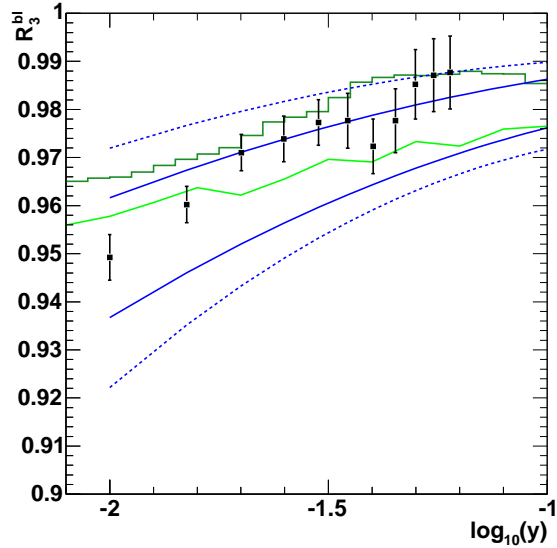}
    \caption{\label{fig:blrate} Heavy-to-light three-jet rate ratio 
      $R_3^{bl}=R_3^{(b)}/R_3^{(uds)}$ in dependence on the Durham jet
      resolution $y$. Data (black points) from a Delphi measurement
      \cite{Abreu:1997ey} are shown together with the {\tt SHERPA} prediction
      (the dark green histogram corresponds to the shower result and the 
      light green curve is obtained by combining matrix elements 
      with Sudakov weights), and with an analytic calculation 
      \cite{Rodrigo:1997gy} (blue lines, dashed=LO, solid=NLO) for 
      two different values of $m_b$, $m_b=$3, 5 GeV.
    }
  \end{center}
\end{figure}

\noindent
In Fig.\ \ref{fig:jetrates_vs_shower}, the results of the parton
shower as implemented in \apa\ are confronted with these analytical
jetrates, which basically represent a resummed tree-level
calculation. The agreement over a wide range of phase space, down to
small values of $y\le 0.004$ is very good. Only for very small values
of $y$ the three-jet rate as obtained from the parton shower starts to
overshoot the analytical result, which is due to the increased
importance of four-jet events in this region. Finally, Fig.\
\ref{fig:blrate}, exhibits the ratio of three-jet rates with 
$b$-quarks and with light quarks ($u$, $d$, and $s$) only. There,
data from the Delphi collaboration \cite{Abreu:1997ey} are contrasted 
with a full next-to leading order calculation \cite{Rodrigo:1997gy}
with varying $b$ quark mass, with the pure parton shower result as
obtained by \apa\ and with the result of the combined matrix element 
plus Sudakov weight. The data clearly prefer a lighter $b$ quark with
$m_b\approx 3$ GeV in the NLO calculation
\footnote{This result is consistent with the value obtained
  by evolving the measurement at the $\Upsilon$ resonance up to the
  $Z$ boson mass, cf.\ \cite{Abreu:1997ey}. }.
In addition, the parton shower obviously describes the data
surprisingly well; for the merged matrix element plus Sudakov weight
result, a running $b$ quark mass seems to be also appropriate. 

\clearpage
\subsection*{Comparison of shower and hadron level: Hadronisation corrections}

\noindent
Before further comparing results of the parton shower with
experimental data, it must be stressed that many of such comparisons
are feasible and meaningful on the hadron level only. This implies
that the partons emerging from the shower must be hadronised with some
phenomenological model. In the case of {\tt SHERPA}, the choice at the
moment is to employ the Lund string model for it. This model has a
number of parameters to be tuned to data. The tuning of Monte Carlo
event generators to data is an intricate procedure, involving an
optimisation in a multi-\-dimensional phase space of parameters.  
These parameters may be perturbative (like, for instance,
$\alpha_s$), or non-\-perturbative (like, for example, the string
tension in the Lund model), or they could characterise the transition
between the perturbative and the non-\-perturbative regime (e.g.\  the
parton shower cut-off). For further details on such a tuning
procedure, the reader is referred to \cite{Abreu:1996na,Hamacher:1995df}. 
The parton shower of \apa\ in its versions 1.0 and 2.0 together with
the Lund string model implemented in {\tt Pythia} has been tuned in
\cite{Uwediss} and \cite{Hendrikdipl}, respectively.  

\begin{figure}
\begin{picture}(400,280)
\put(-15,140){\includegraphics[width=8.cm]{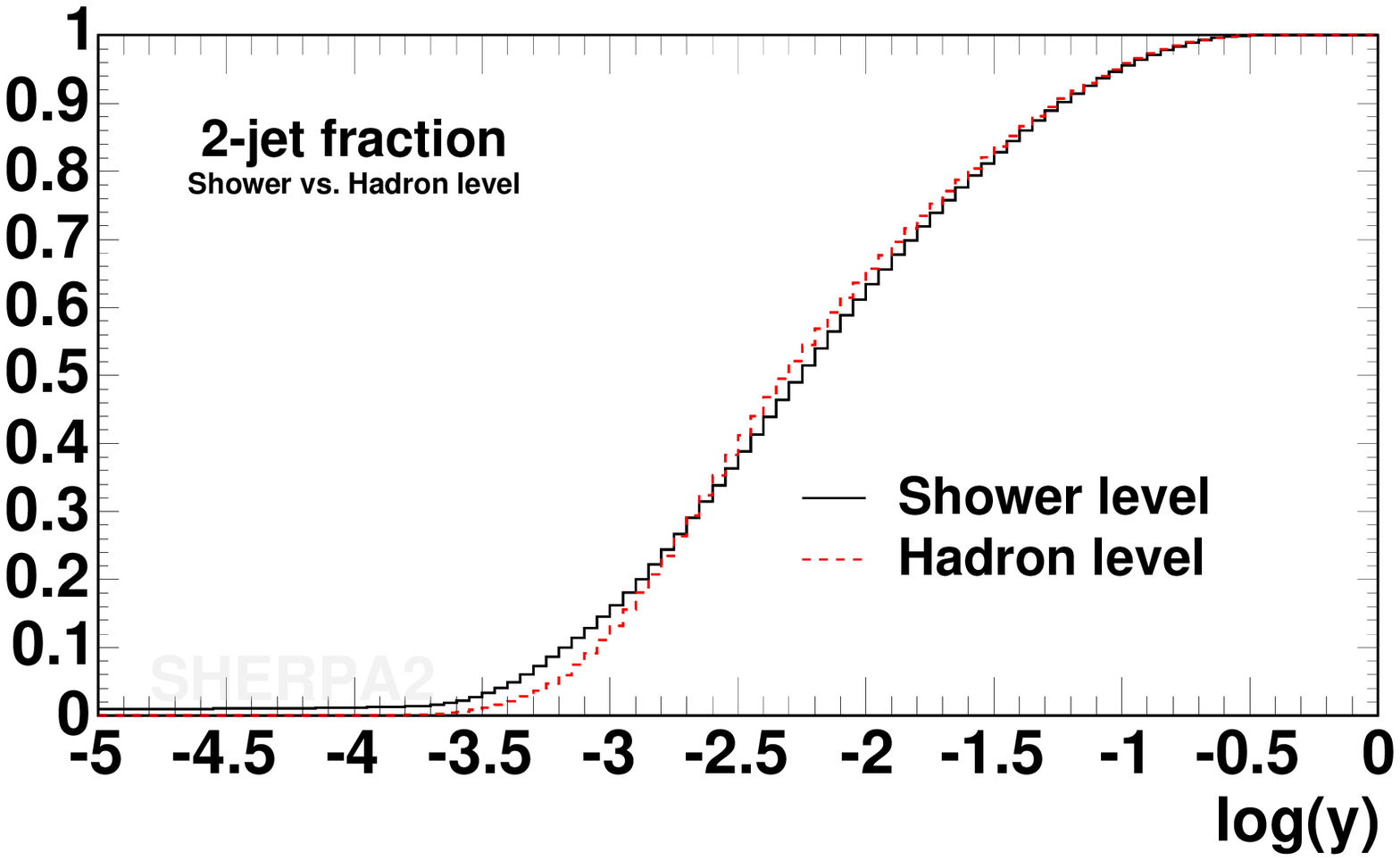}}
\put(200,140){\includegraphics[width=8.cm]{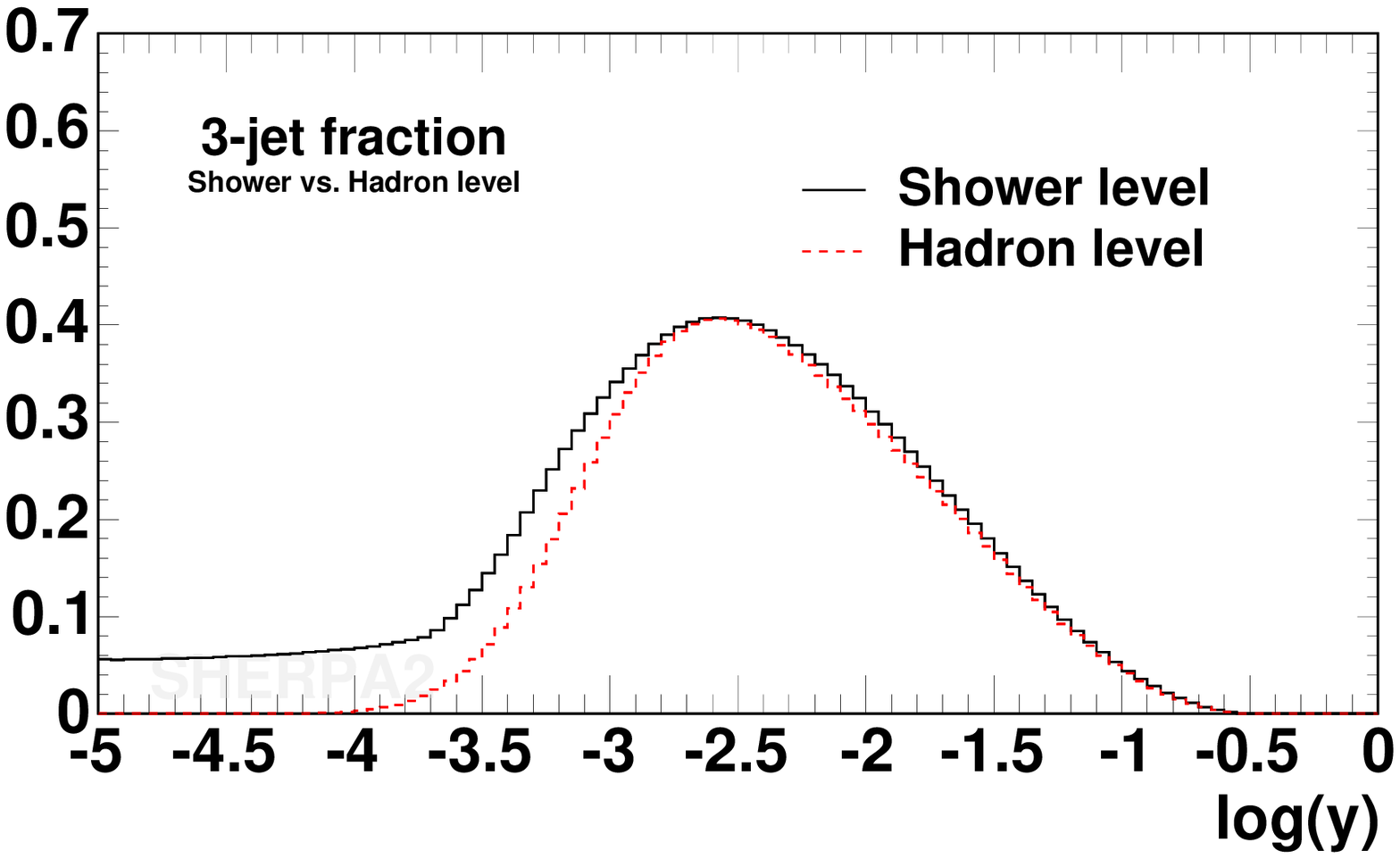}}
\put(-15,0){\includegraphics[width=8.cm]{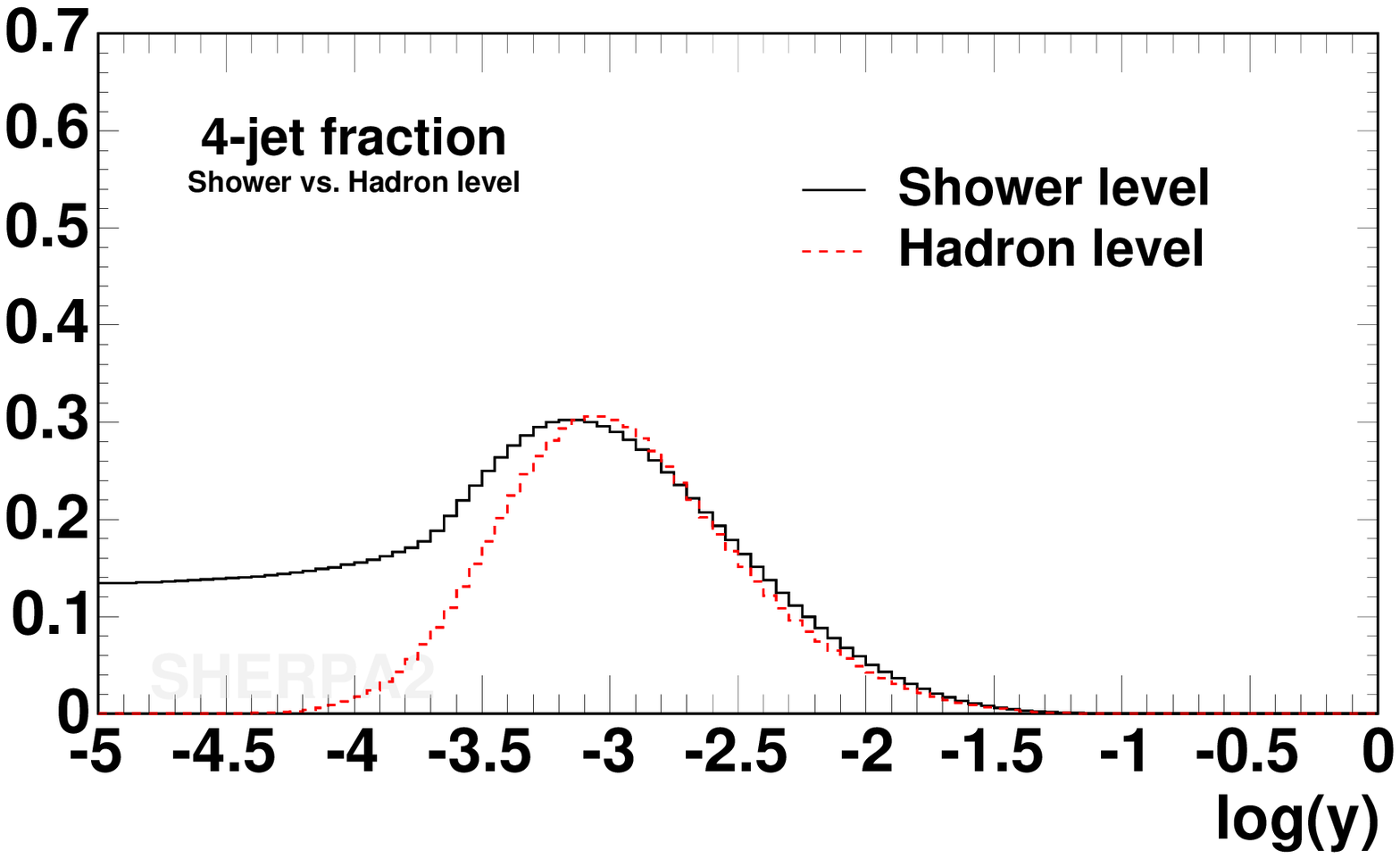}}
\put(200,0){\includegraphics[width=8.cm]{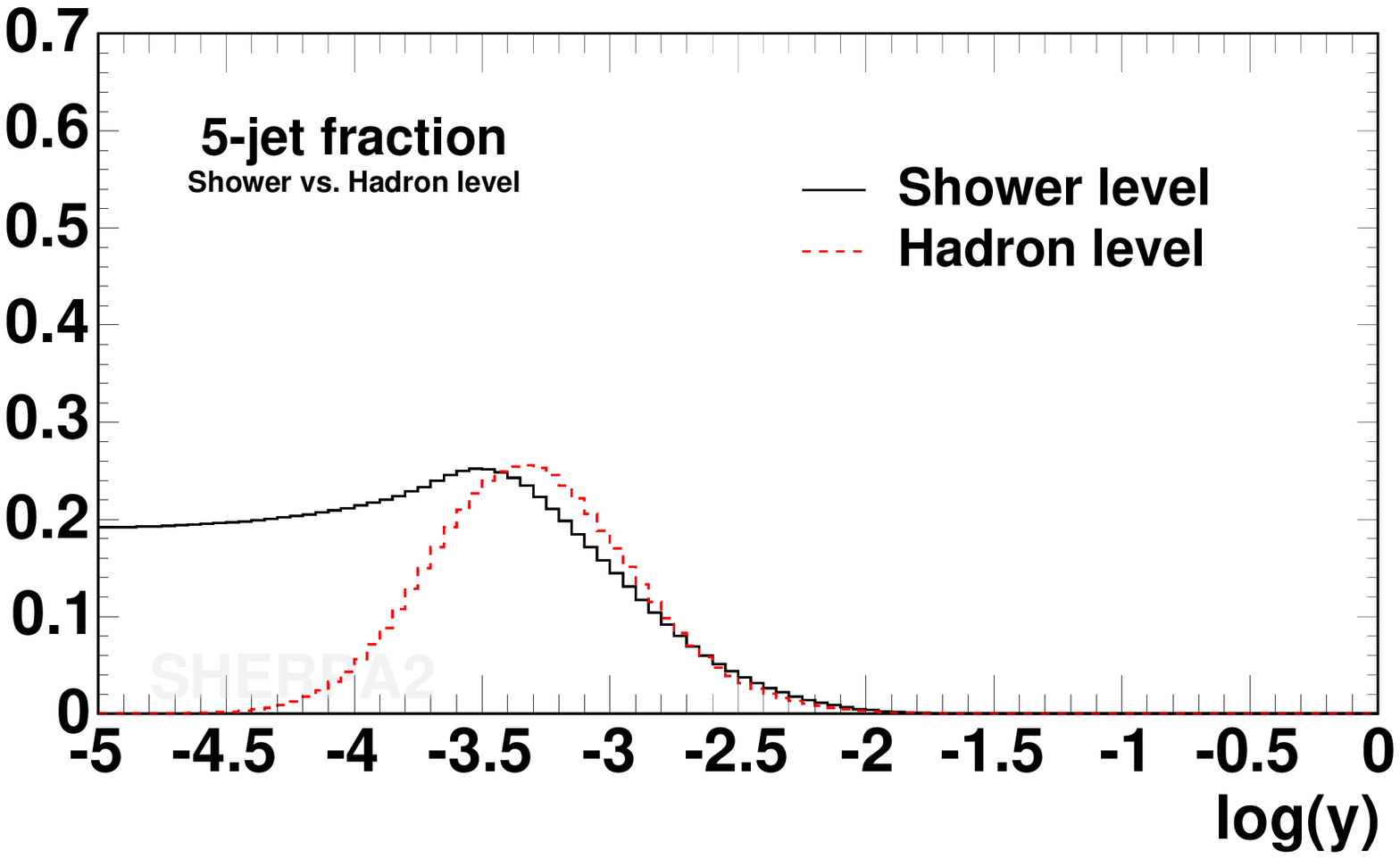}}
%\put(93,0){\includegraphics[width=8.cm]{Analysed_KtJetrates_1_jet_1_1_4.dat_3_had2.eps}}
\end{picture}
\caption{\label{fig:lep_durham_had} Durham jet rates at LEP 1. The
  shower level result (solid lines) of {\tt SHERPA} is contrasted with its result after
  hadronisation  (dashed lines).}
\end{figure}

\begin{figure}
\begin{picture}(400,180)
\put(-15,-60){\includegraphics[width=8.0cm]{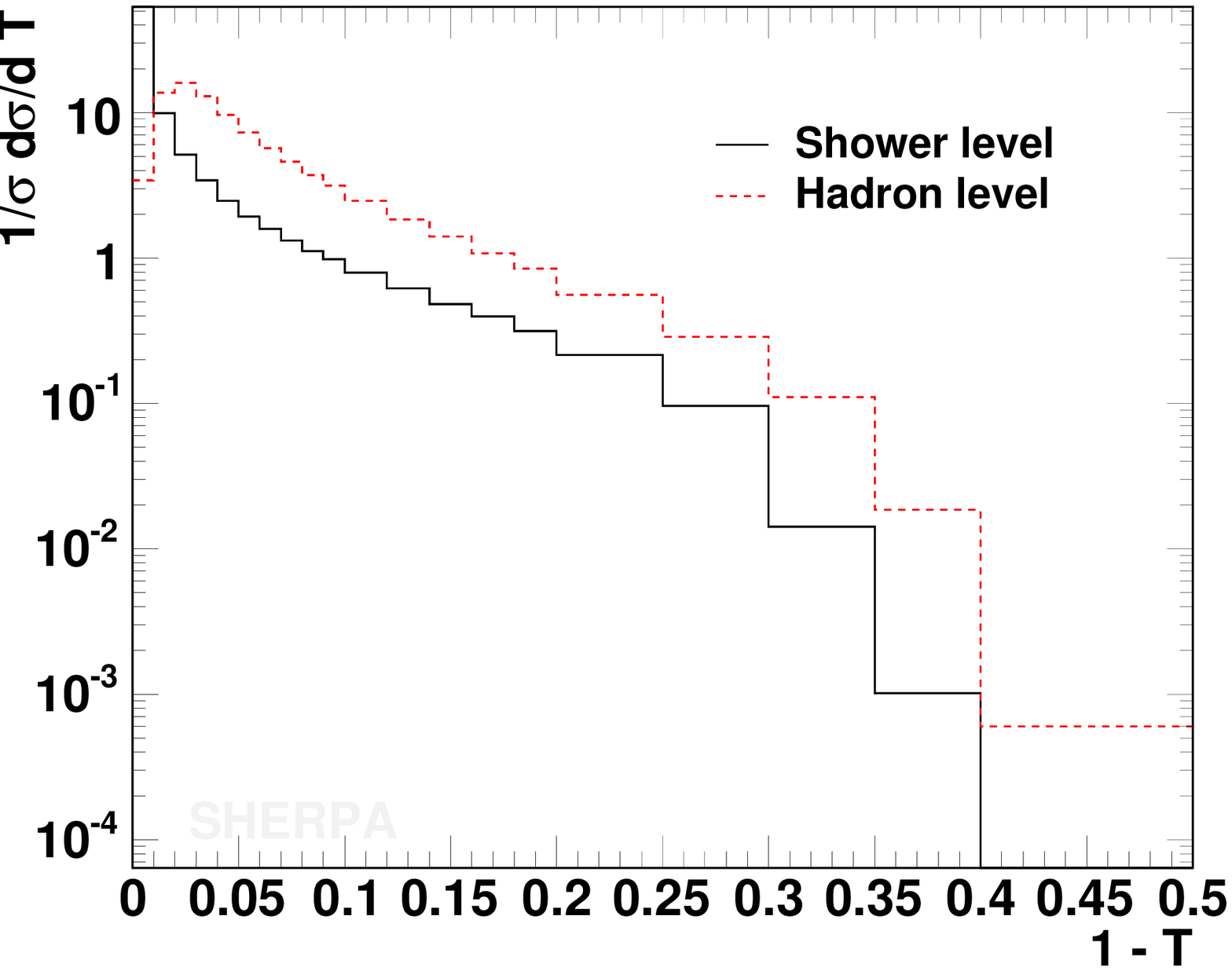}}
\put(200,-60){\includegraphics[width=8.0cm]{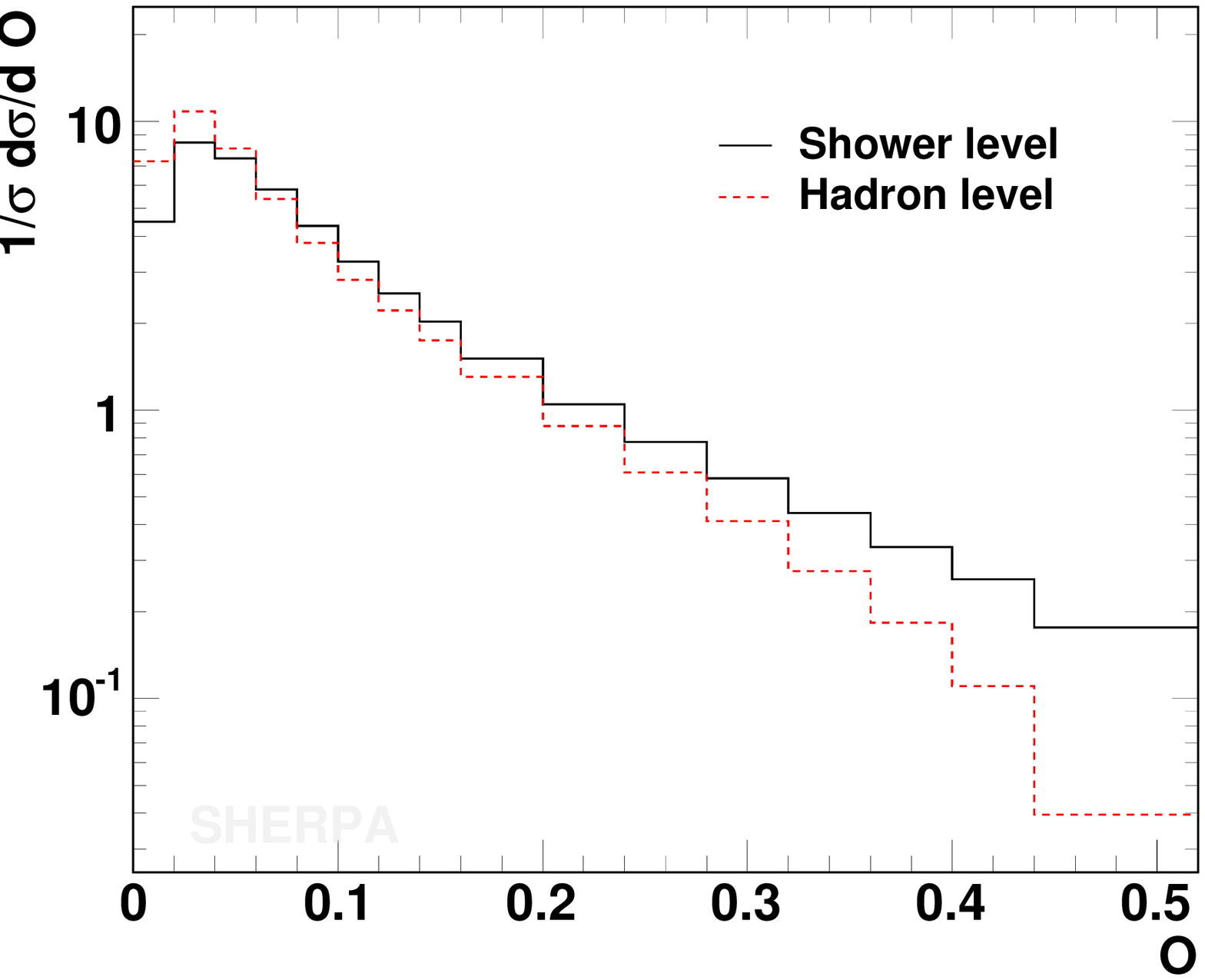}}
\end{picture}
\caption{\label{fig:lep_thrust_had} Thrust and Oblateness at LEP 1. The
  shower level result (solid lines) of {\tt SHERPA} is contrasted with its
  result after hadronisation (dashed lines).}
\end{figure}

\noindent
Hadronisation and the tuning of phenomenological models, however,
induces some source of systematic error in any simulation that needs
to be investigated. In Fig.\ \ref{fig:lep_durham_had} the parton
shower results for total jet rates in the Durham scheme for 2-5 jets
are confronted with results after hadronisation. Evidently, both
results coincide on a level of 10\% or better down to jet resolutions
of $y\approx 0.001$. This corresponds to relative transverse momenta
of the order of $3$ GeV, a kinematical regime, where decays of
$b$-hadrons start to matter. This also implies that down to such low
values parton shower results can be compared directly to data. In
Fig.\ \ref{fig:lep_thrust_had}, two event shape variables, thrust and
oblateness, are studied. Again, results before and after hadronisation
are compared. Pictorially speaking, thrust is a measure for how
``jetty'' an event is. A thrust of 1, for example, describes a perfect,
``pencil-\-like'' two-jet event, where both jets are oriented
back-\-to-\-back and have no transverse spread. Clearly, such a
configuration is severely suppressed at the hadron level, since it
corresponds to events with few, highly collinear particles only. Thus,
in the $T=1$ bin, hadronisation corrections are large. Over a wide
range of thrust values away from such extreme configuration, however,
the shapes of the results before and after hadronisation are nearly
identical; hadronisation corrections in this region therefore a just a
constant that can be taken directly from data. In this region of, say,
$0.05\le T\le 0.6$, the parton shower results can be confronted
directly with data. Similar reasoning holds true not only for the other
exemplary observable, oblateness, but for a plethora of observables. 

\clearpage
\subsection*{Comparison with experimental data from LEP}

\begin{figure}
\begin{picture}(400,240)
\put(-15,0){\includegraphics[width=8.0cm]{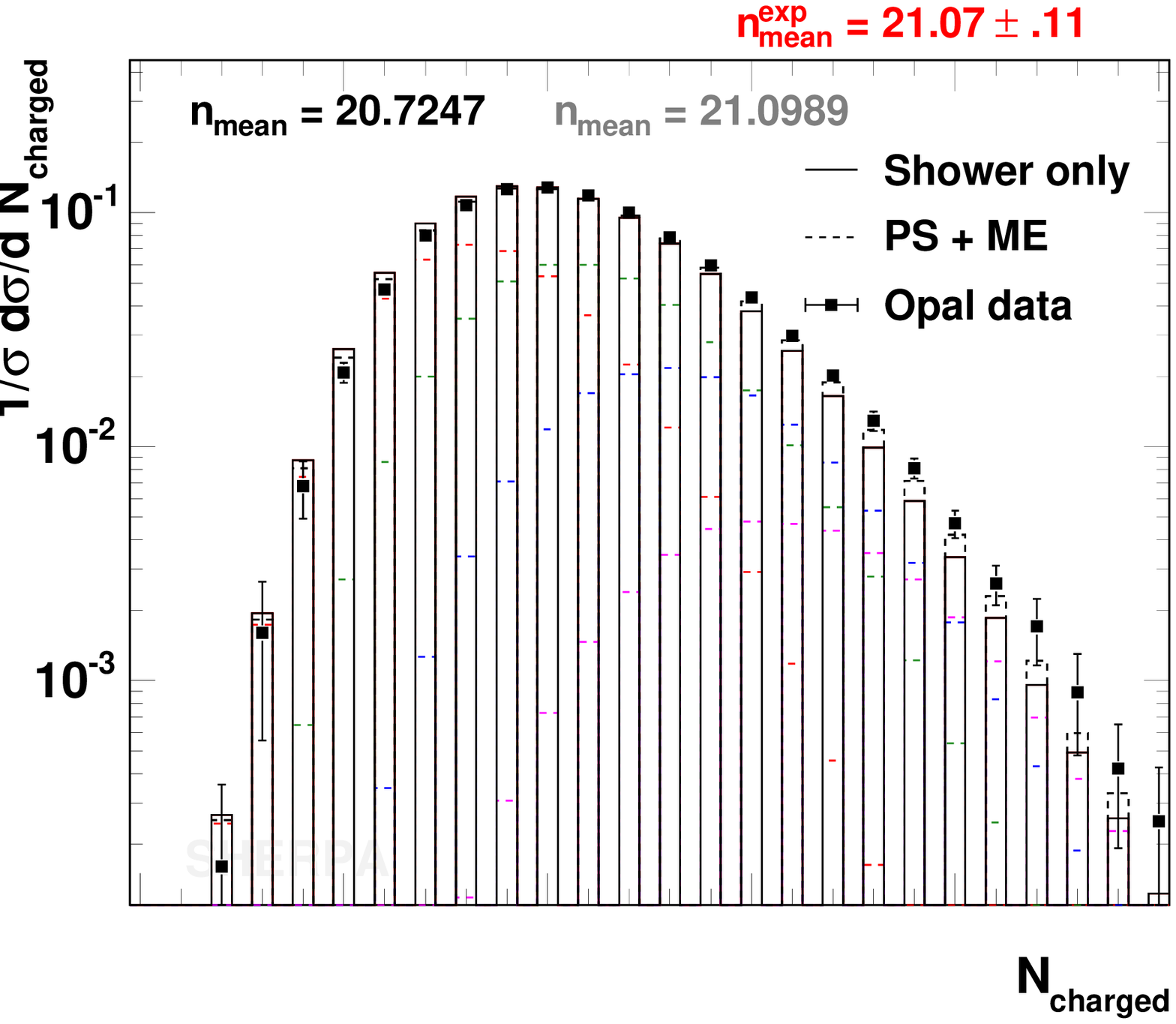}}
\put(-15,0){\includegraphics[width=8.0cm]{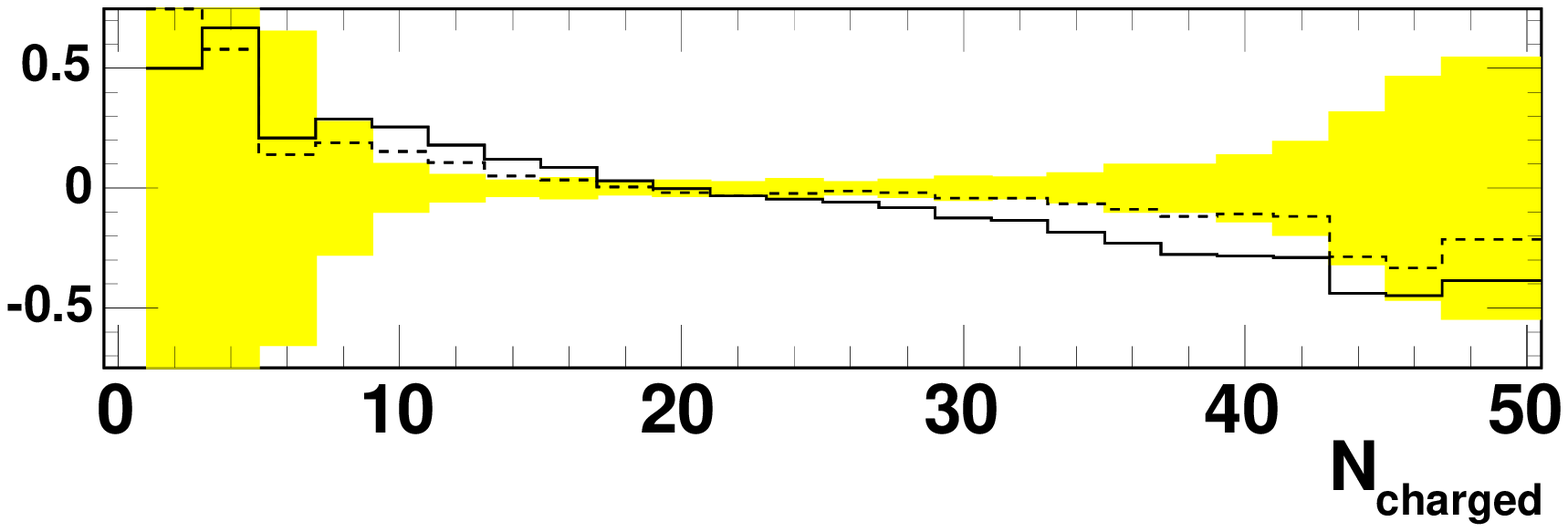}}
\put(200,0){\includegraphics[width=8.0cm]{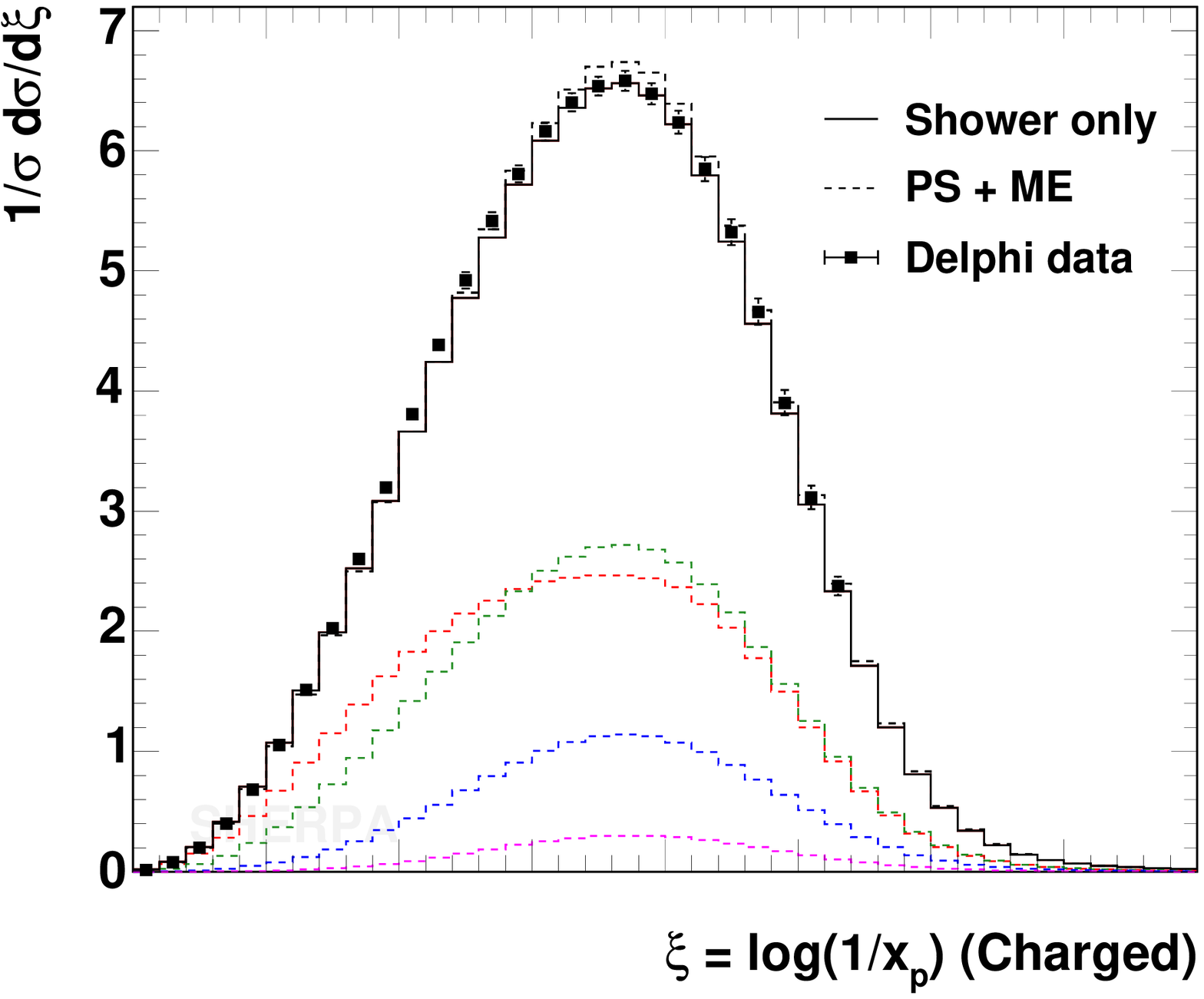}}
\put(200,0){\includegraphics[width=8.0cm]{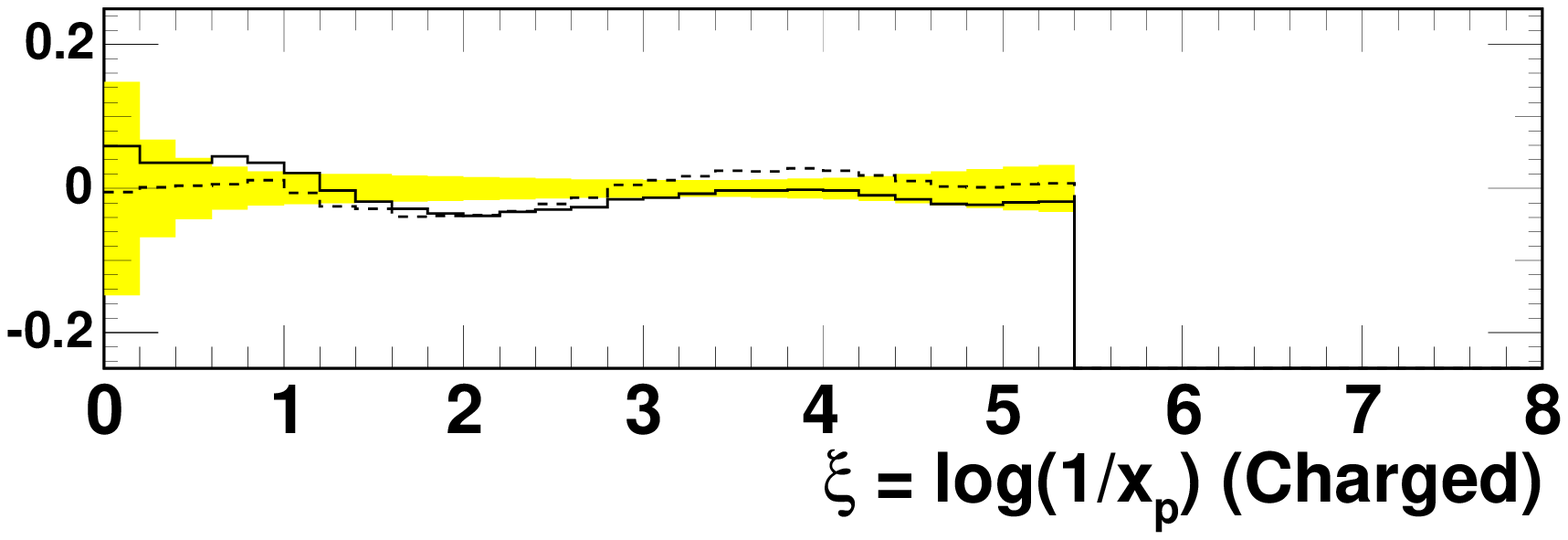}}
\end{picture}
\caption{\label{fig:lep_charged_multi} Charged multiplicity and scaled
  momentum at LEP 1. The left plot shows the distribution of the
  number of charged particles together with a measurement by Opal
  \cite{Acton:1991aa}. The mean charge multiplicity is also stated
  together with its PDG value \cite{Eidelman:2004wy}. 
  On the right hand side, a scaled momentum distribution
  is plotted against Delphi data \cite{Abreu:1996na}.
  The data are contrasted with results obtained through the parton
  shower alone (solid lines) with those obtained when the merging of
  matrix elements for up to five jets and the parton shower has been
  employed (dashed lines). In the latter case coloured lines indicate
  the contributions from individual matrix elements.}
\end{figure}

\noindent
Having validated the numerical correctness of the parton shower
implementation, the focus shifts from the investigation of the
behaviour of the Sudakov form factors to the study of experimental
observables. There, experimental data can be confronted with the
results obtained with the parton shower, eventually after a merging
with multijet matrix elements. As mention above, in the previous
section, the considered data are on the hadron level, consequently 
a hadronisation model has to be applied. The parameters of the
model were tuned, the perturbative input consisted of a sample of
multijet matrix elements for up to five jets, merged with the parton
shower of \apa. The quality of the tuning can be judged by considering
the multiplicity distribution of charged hadrons and of their scaled
momentum, cf.\ Fig.\ \ref{fig:lep_charged_multi}. In this plot as well
as in each of the following ones, data are confronted with the parton
shower implemented in \apa (solid lines) and with the merged multijet
matrix element plus parton shower (dashed lines), both after
hadronisation. For the merged results, contributions from different
jet multiplicities are indicated in different colours. Clearly, the
merged sample tends to produce a slightly larger fraction of events 
with harder jets, leading to higher parton and hadron multiplicities;
this is visible from the fact that the ``parton shower only'' sample
slightly undershoots the bins with comparably high hadron
multiplicity. For the momentum distribution, however, this minor trend
washes out. In both cases, the results obtained from the simulation
show excellent agreement with data, this is even more so when taking
into account that in all plots shown the statistical errors on the
Monte Carlo results are of the same order as the experimental errors,
indicated by the yellow error bands. 

\begin{figure}
\begin{picture}(400,240)
\put(-15,0){\includegraphics[width=8.0cm]{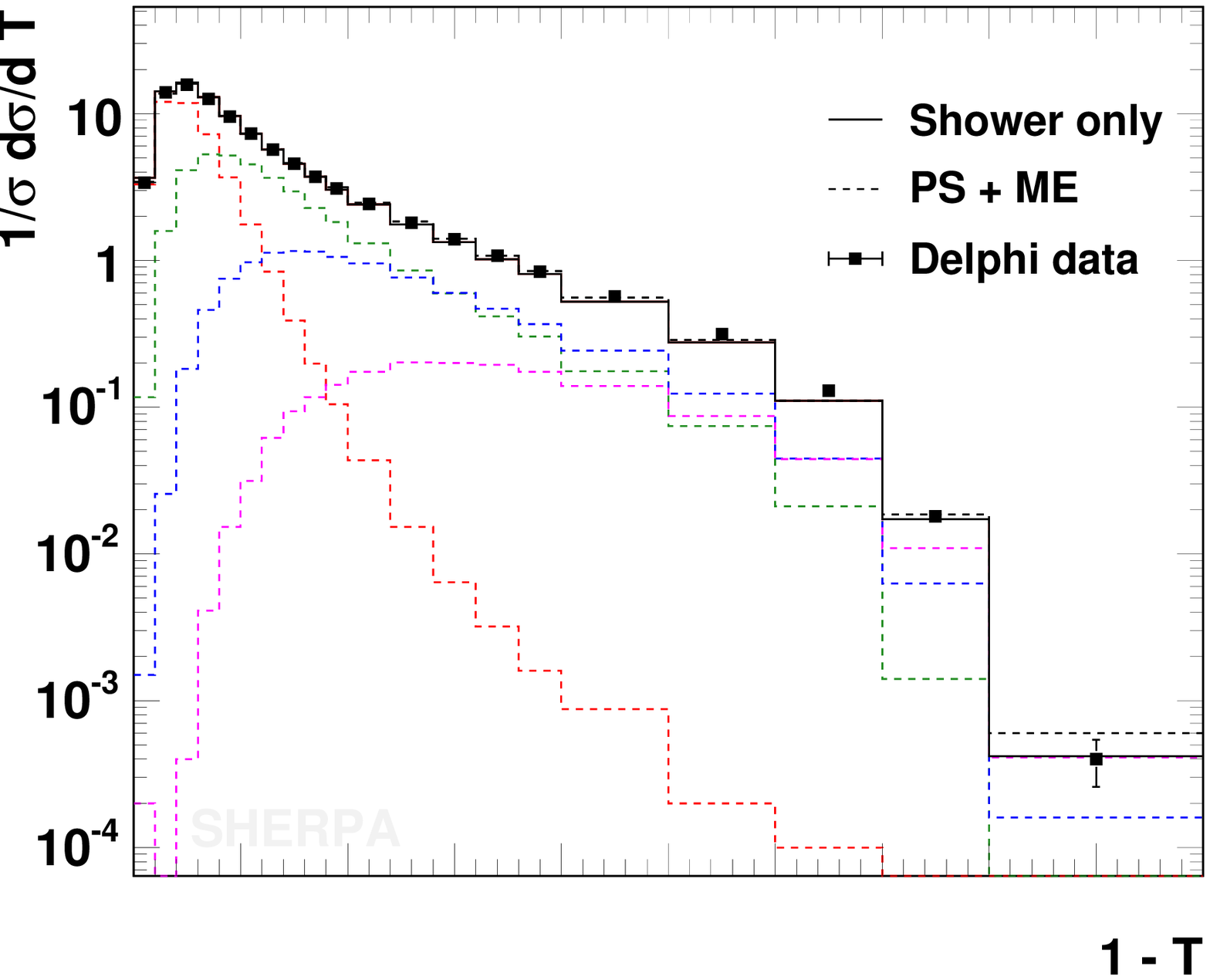}}
\put(-15,0){\includegraphics[width=8.0cm]{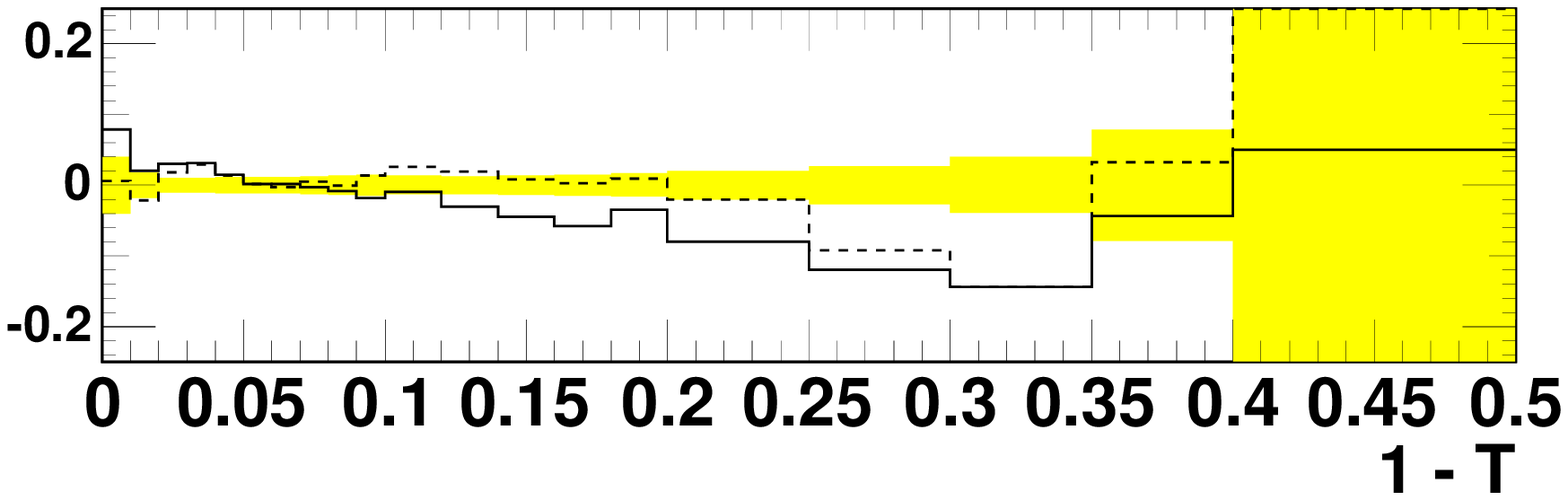}}
\put(200,0){\includegraphics[width=8.0cm]{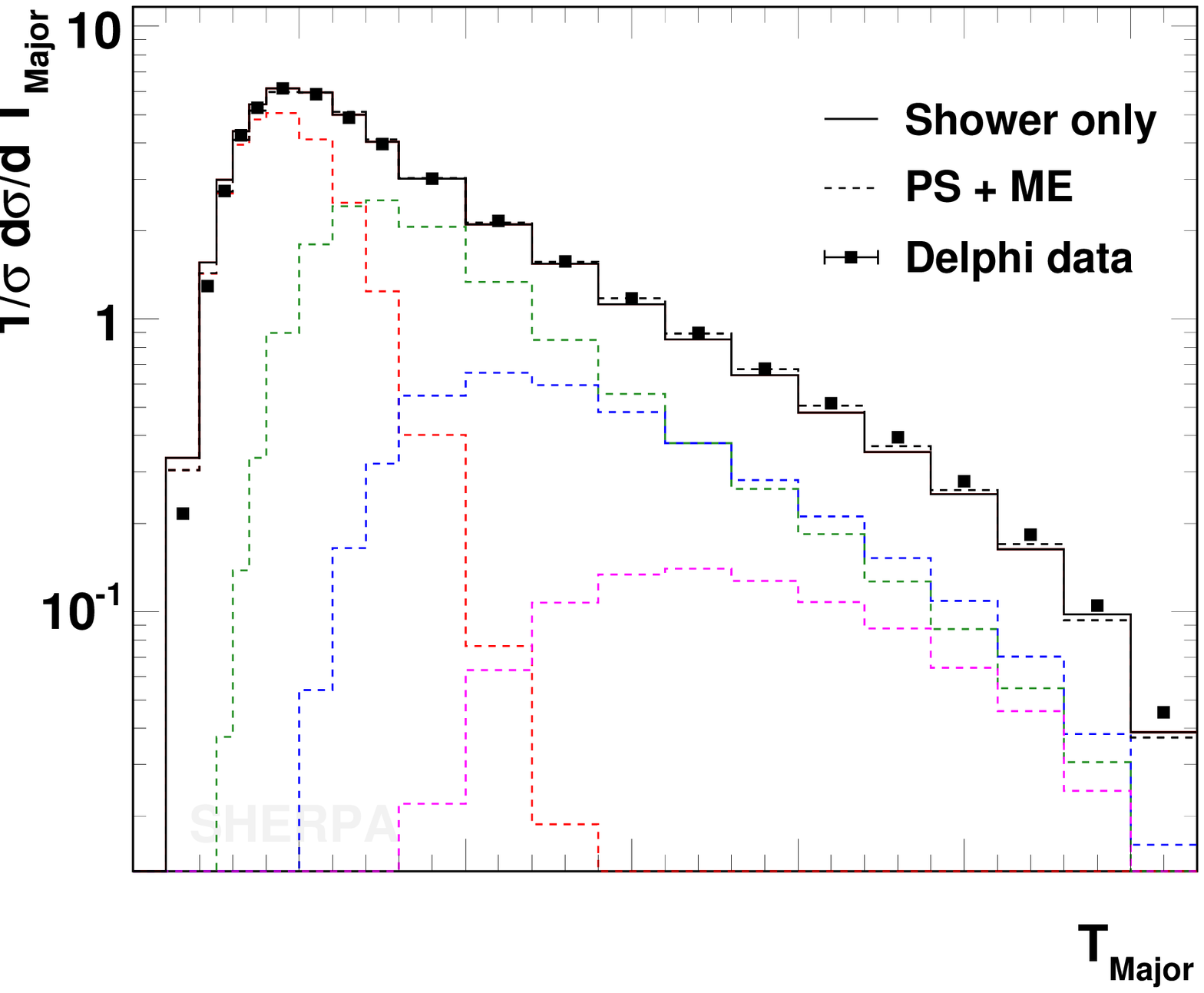}}
\put(200,0){\includegraphics[width=8.0cm]{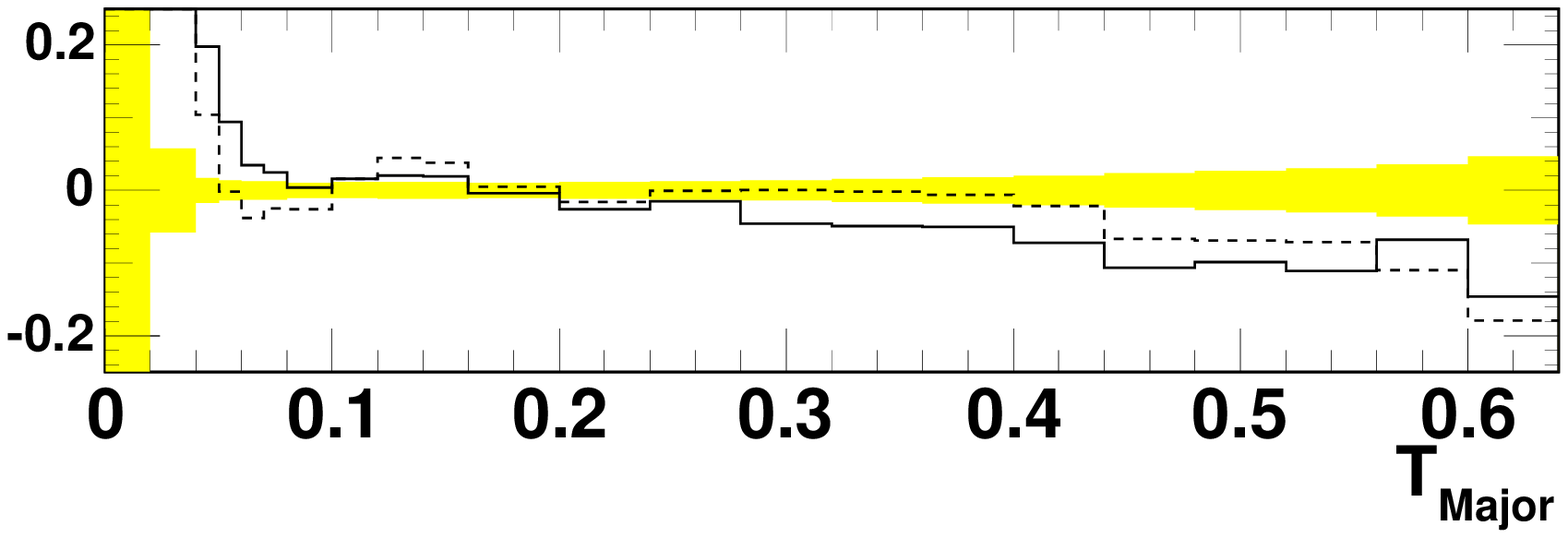}}
\end{picture}
\caption{\label{fig:lep_thrust} Thrust and Major at LEP 1. The hadron
  level result of {\tt SHERPA} is contrasted with measurement from the
  Delphi collaboration \cite{Abreu:1996na}. Line styles and colours
  are the same as in Fig.\ \ref{fig:lep_charged_multi}.}
\end{figure}

\begin{figure}
\begin{picture}(400,240)
\put(-15,0){\includegraphics[width=8.0cm]{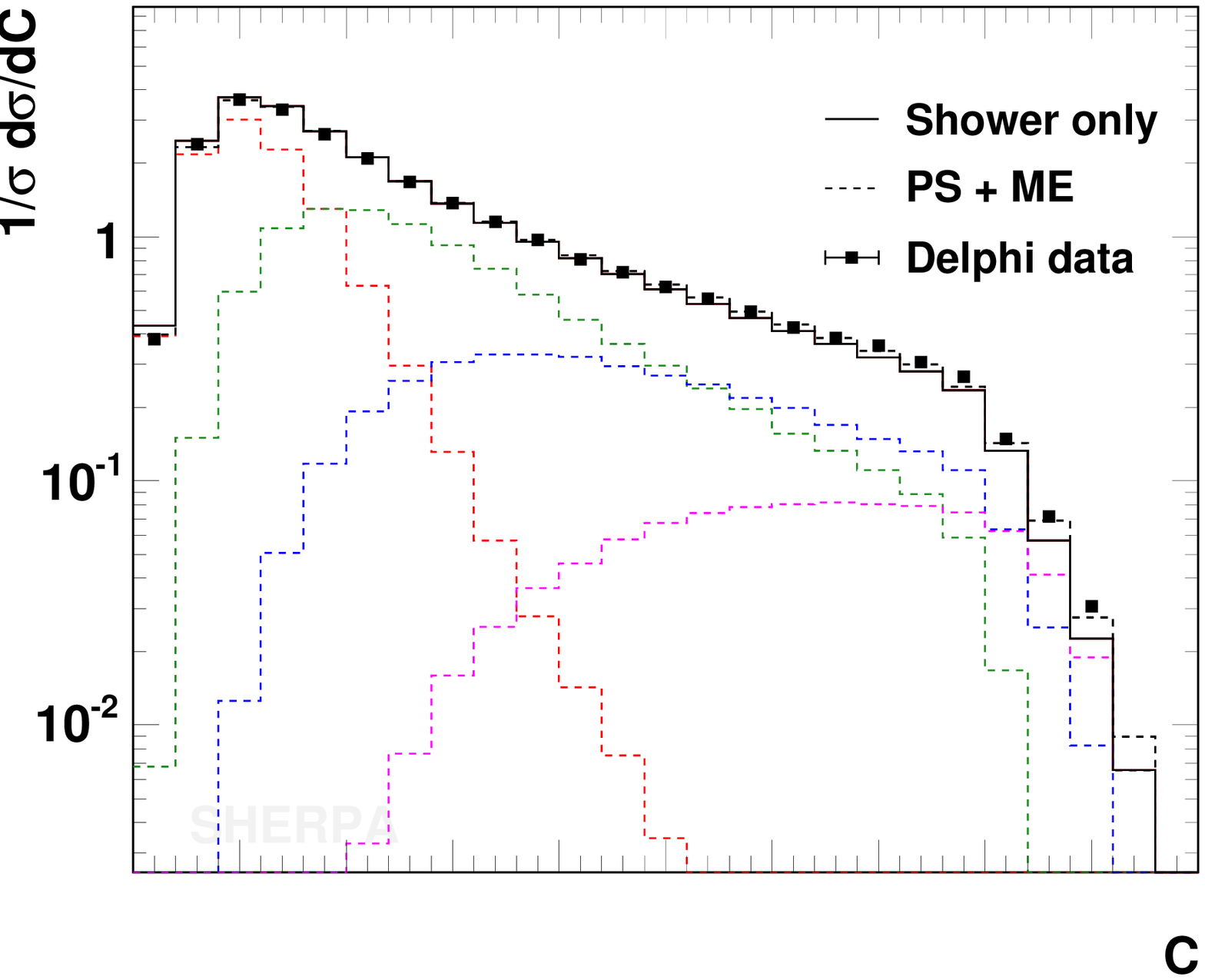}}
\put(-15,0){\includegraphics[width=8.0cm]{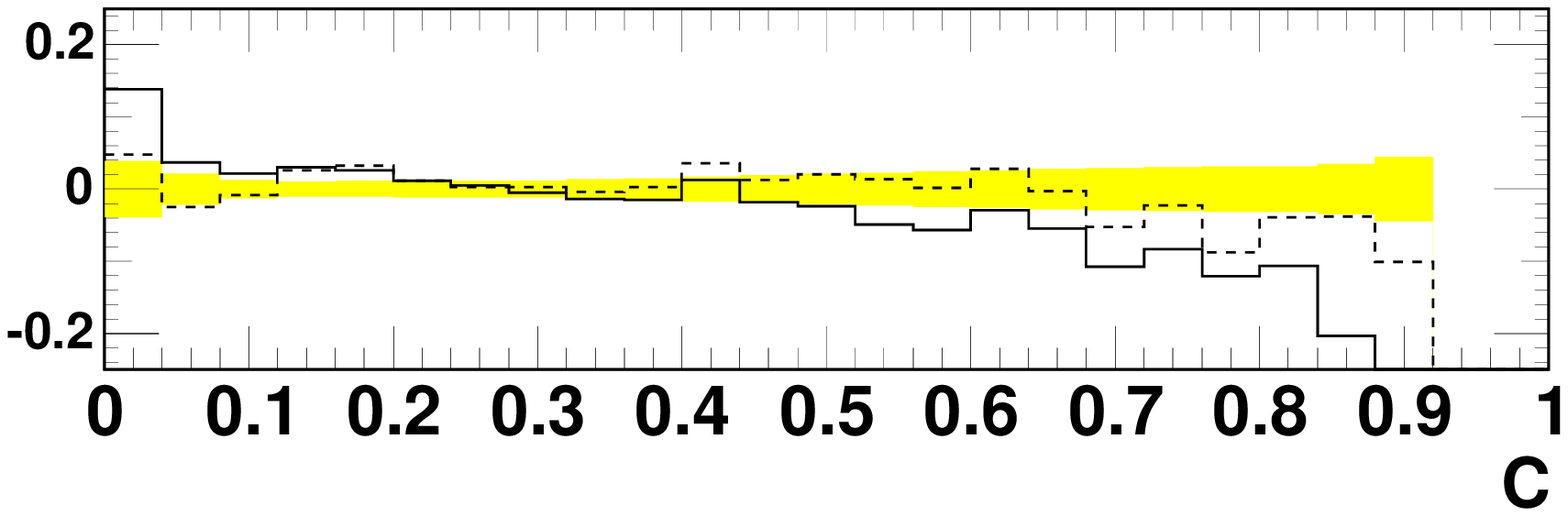}}
\put(200,0){\includegraphics[width=8.0cm]{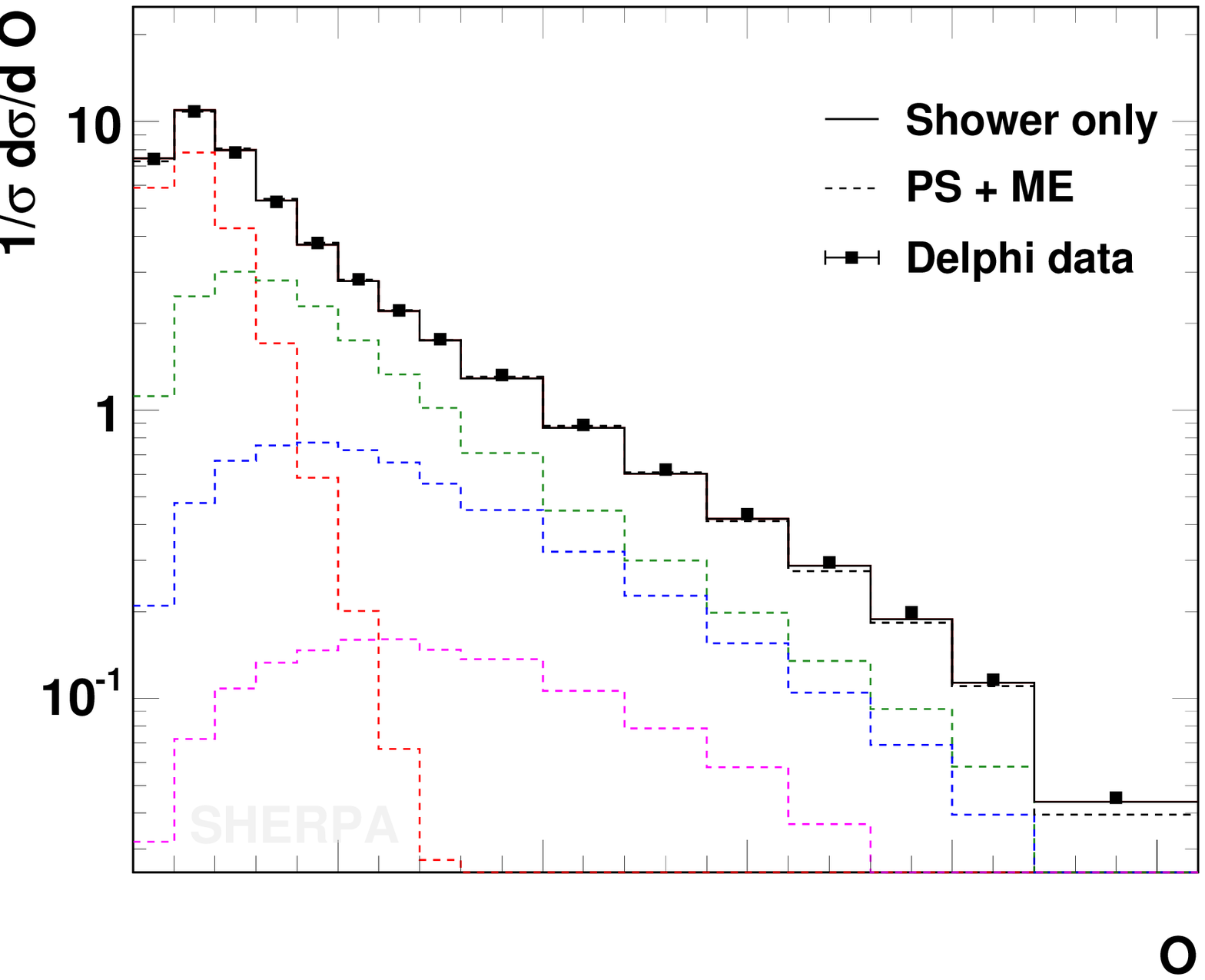}}
\put(200,0){\includegraphics[width=8.0cm]{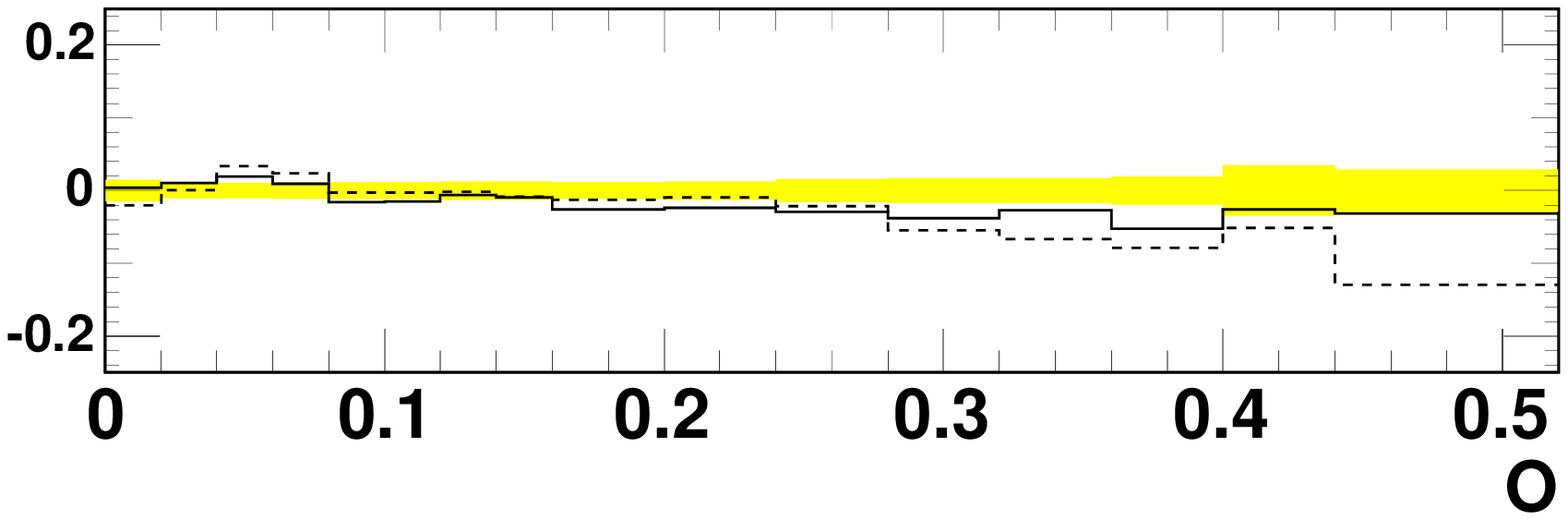}}
\end{picture}
\caption{\label{fig:lep_oblatness} C-Parameter and Oblateness at LEP
  1. This plot shows the event shape variables C-parameter and oblateness,
  together with Delphi data\cite{Abreu:1996na}. Line styles and
  colours are the same as in Fig.\ \ref{fig:lep_charged_multi}.}
\end{figure}

\noindent
The first real test of the parton shower performance is to check
whether it is able to reproduce event shape observables, such as
thrust, major, the C-parameter, or oblateness \footnote{
  For definitions, the reader is referred to appendix
  \ref{EvtShapes_Sec}.}, 
cf.\ Figs.\ \ref{fig:lep_thrust} and \ref{fig:lep_oblatness}. In all
cases, the agreement of the data with the generated events is
excellent. In the ``parton shower only'' sample the trend mentioned
above, namely of being a little bit softer than the merged sample, 
is continued.  

\begin{figure}
\begin{picture}(400,420)
\put(-15,280){\includegraphics[width=8.cm]{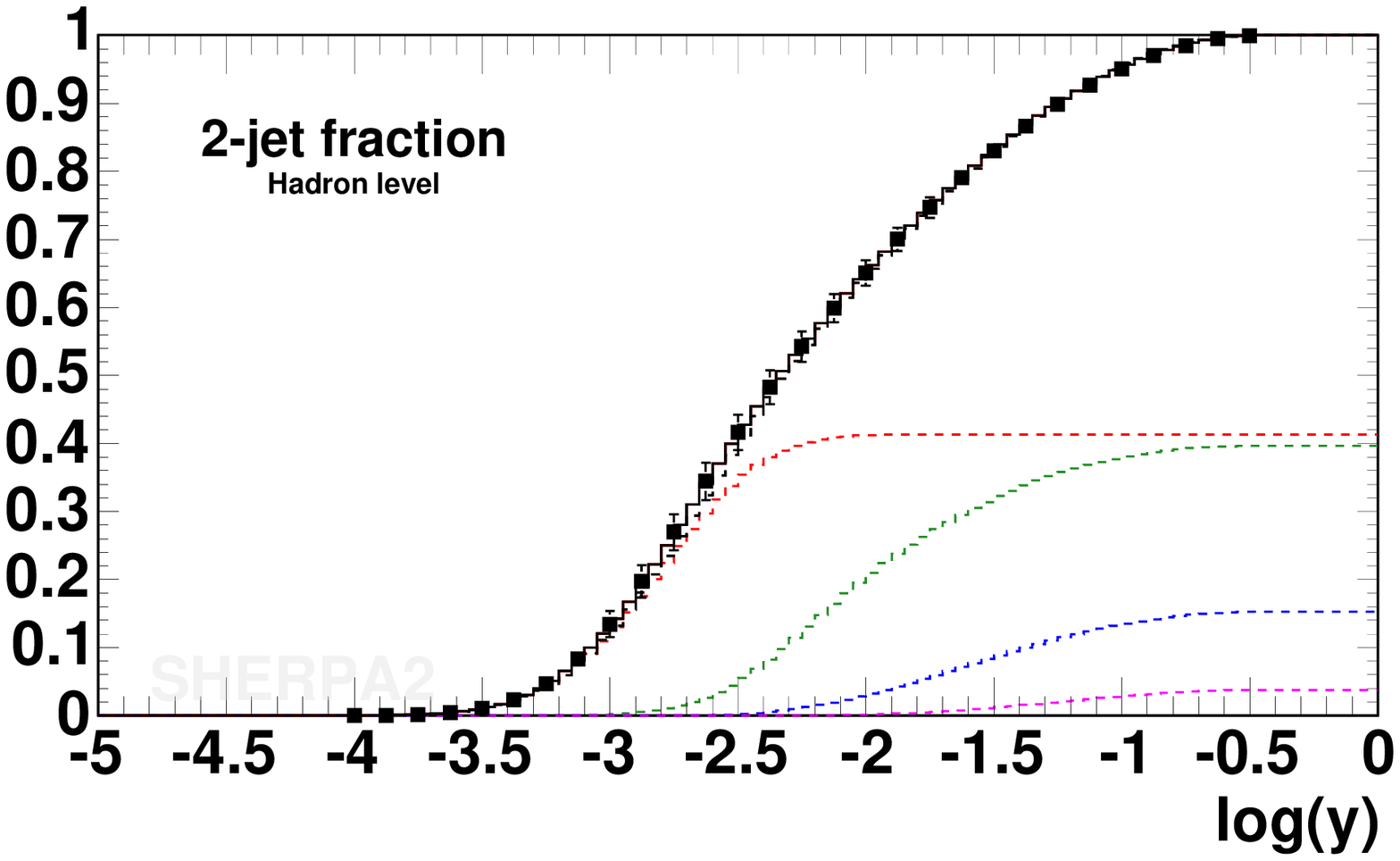}}
\put(200,280){\includegraphics[width=8.cm]{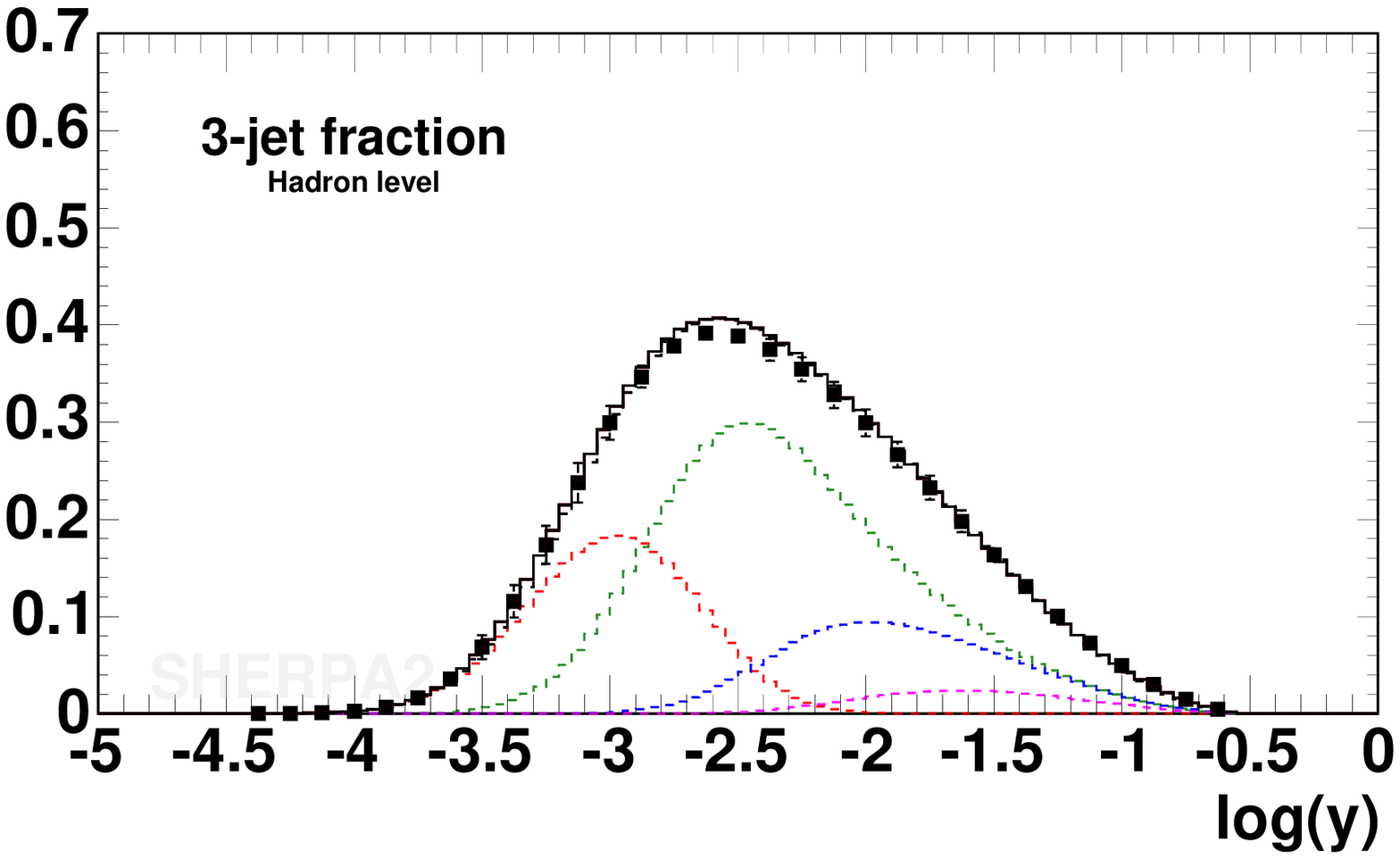}}
\put(-15,140){\includegraphics[width=8.cm]{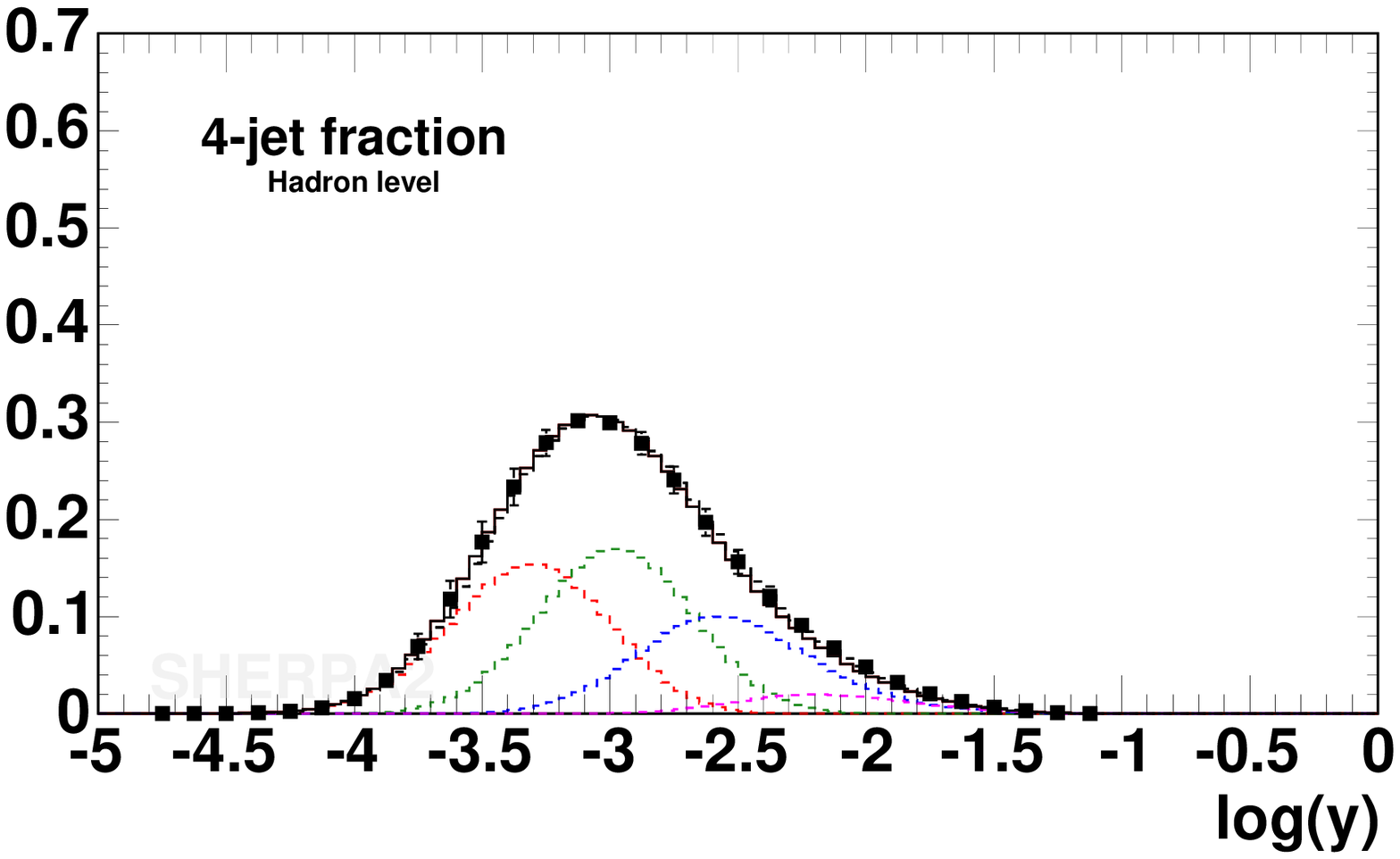}}
\put(200,140){\includegraphics[width=8.cm]{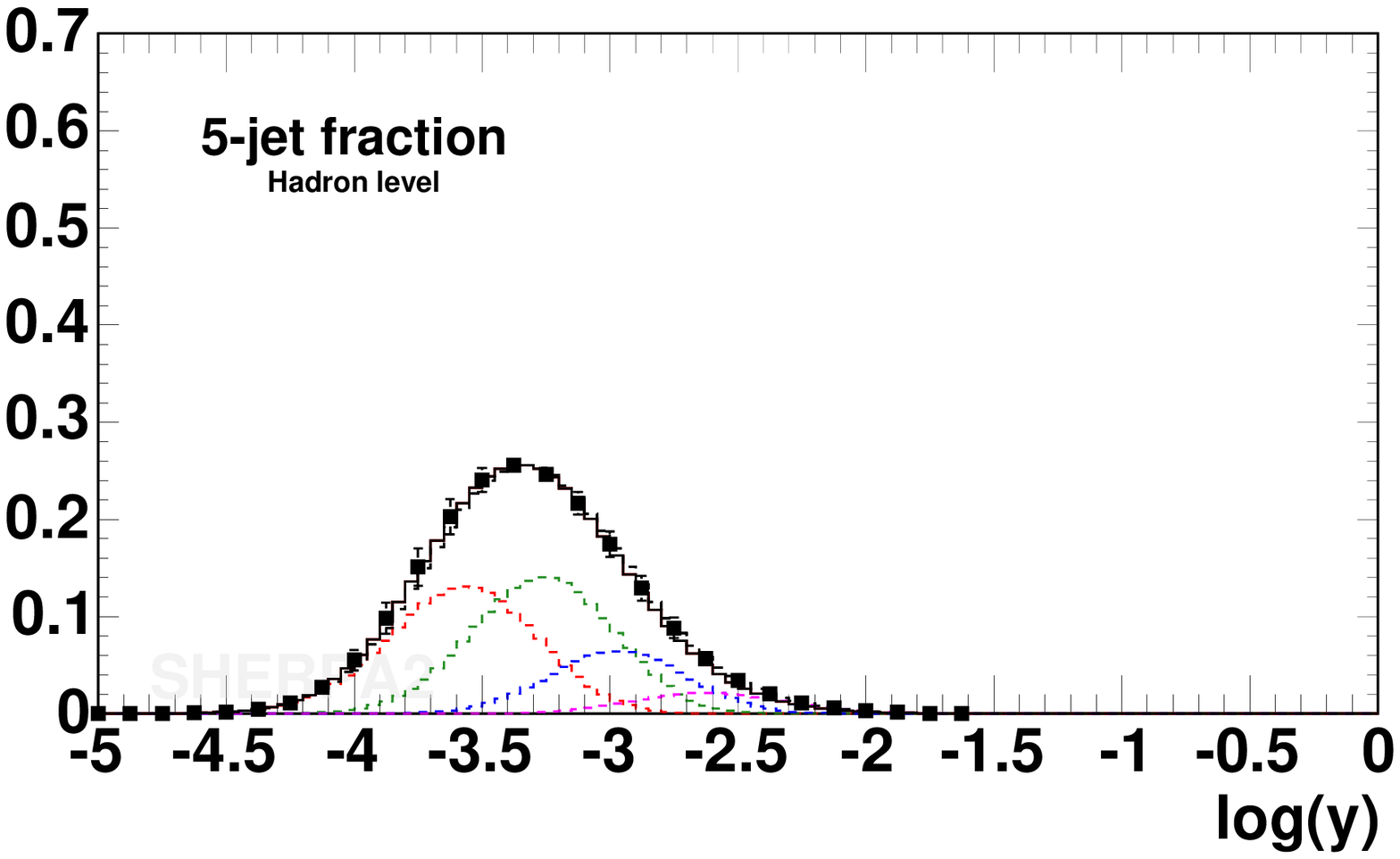}}
\put(93,0){\includegraphics[width=8.cm]{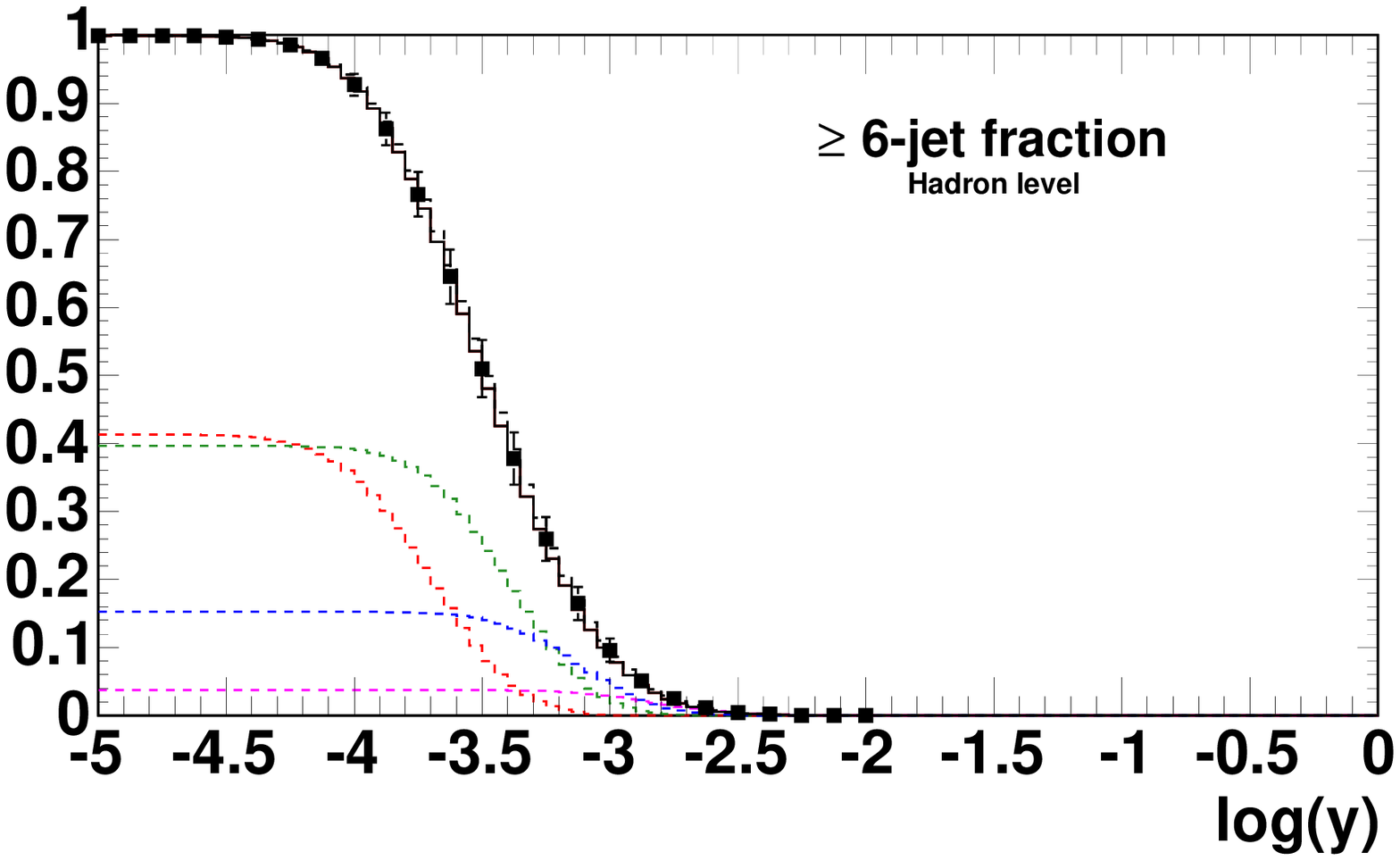}}
\end{picture}
\caption{\label{fig:lep_durham} Durham jet rates at LEP 1, 
  taken from \cite{Pfeifenschneider:1999rz}. Line styles and
  colours are the same as in Fig.\ \ref{fig:lep_charged_multi}.}
\end{figure}

\begin{figure}
\begin{picture}(400,260)
%\graphpaper(0,0)(400,260)
\put(-15,40){\includegraphics[width=8.cm]{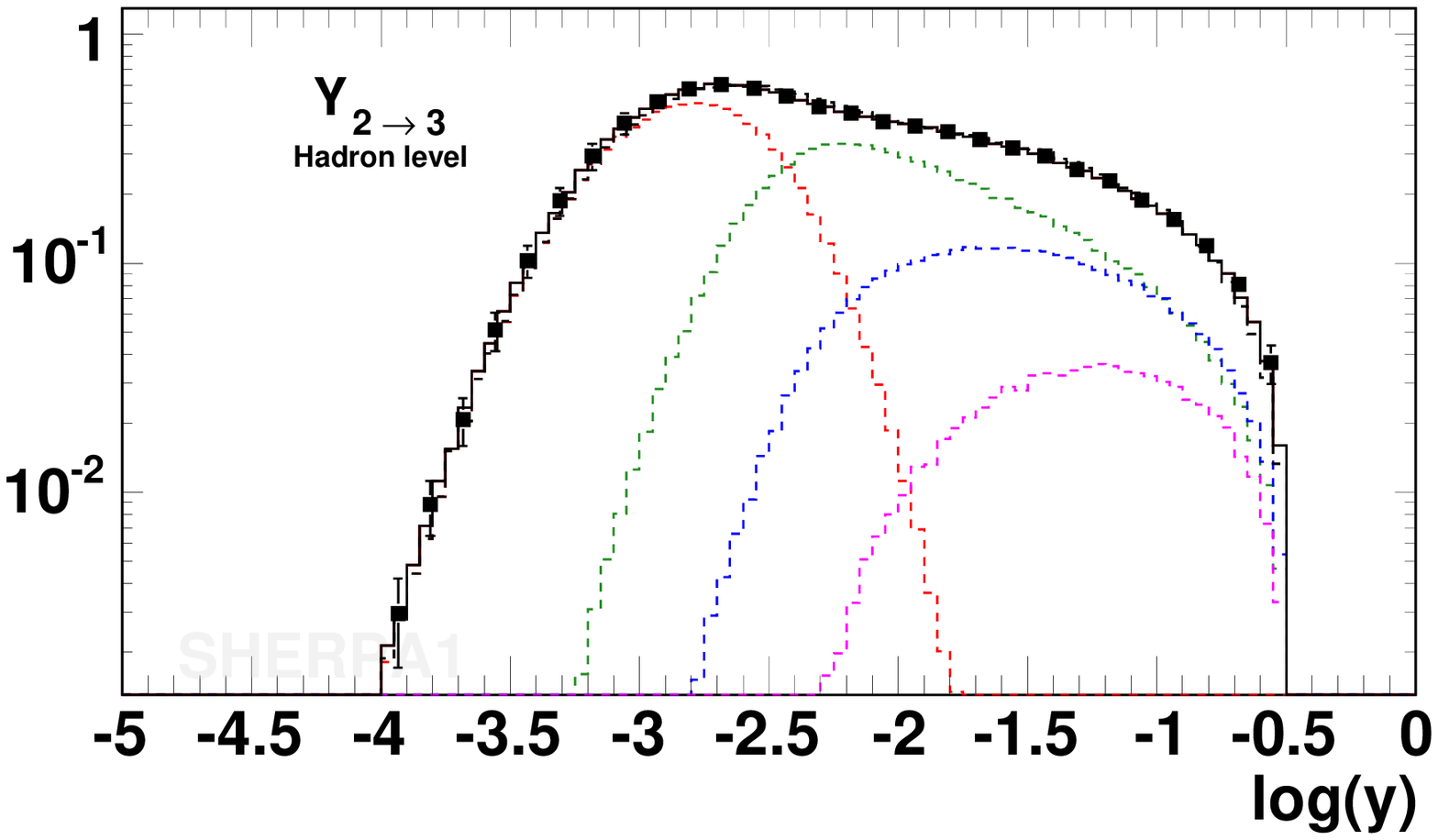}}
\put(200,40){\includegraphics[width=8.cm]{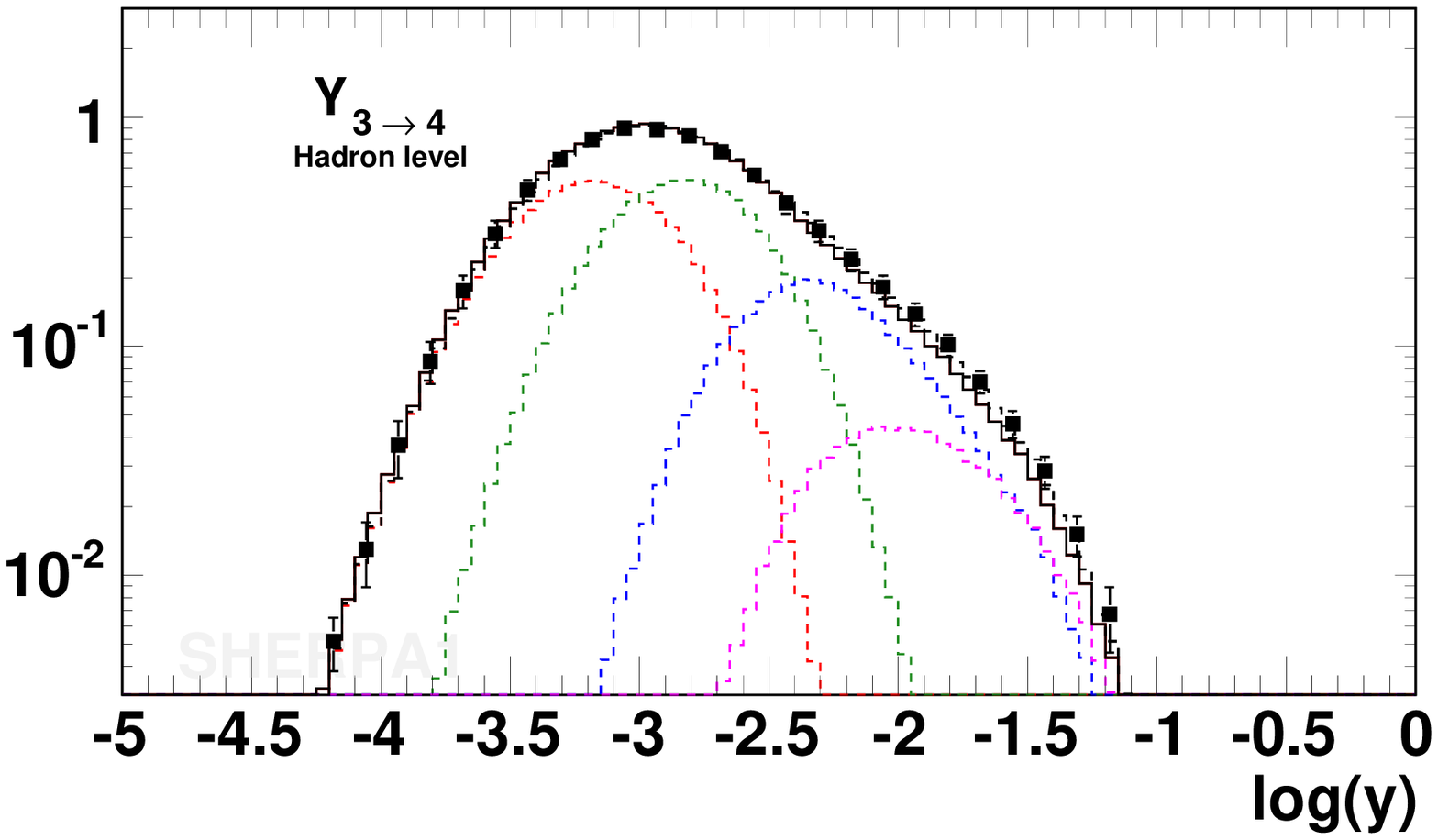}}
%\put(93,-100){\includegraphics[width=8.cm]{Analysed_KtJetrates_1_jet_1_1_3.dat_1.eps}}
\put(-15,-100){\includegraphics[width=8.cm]{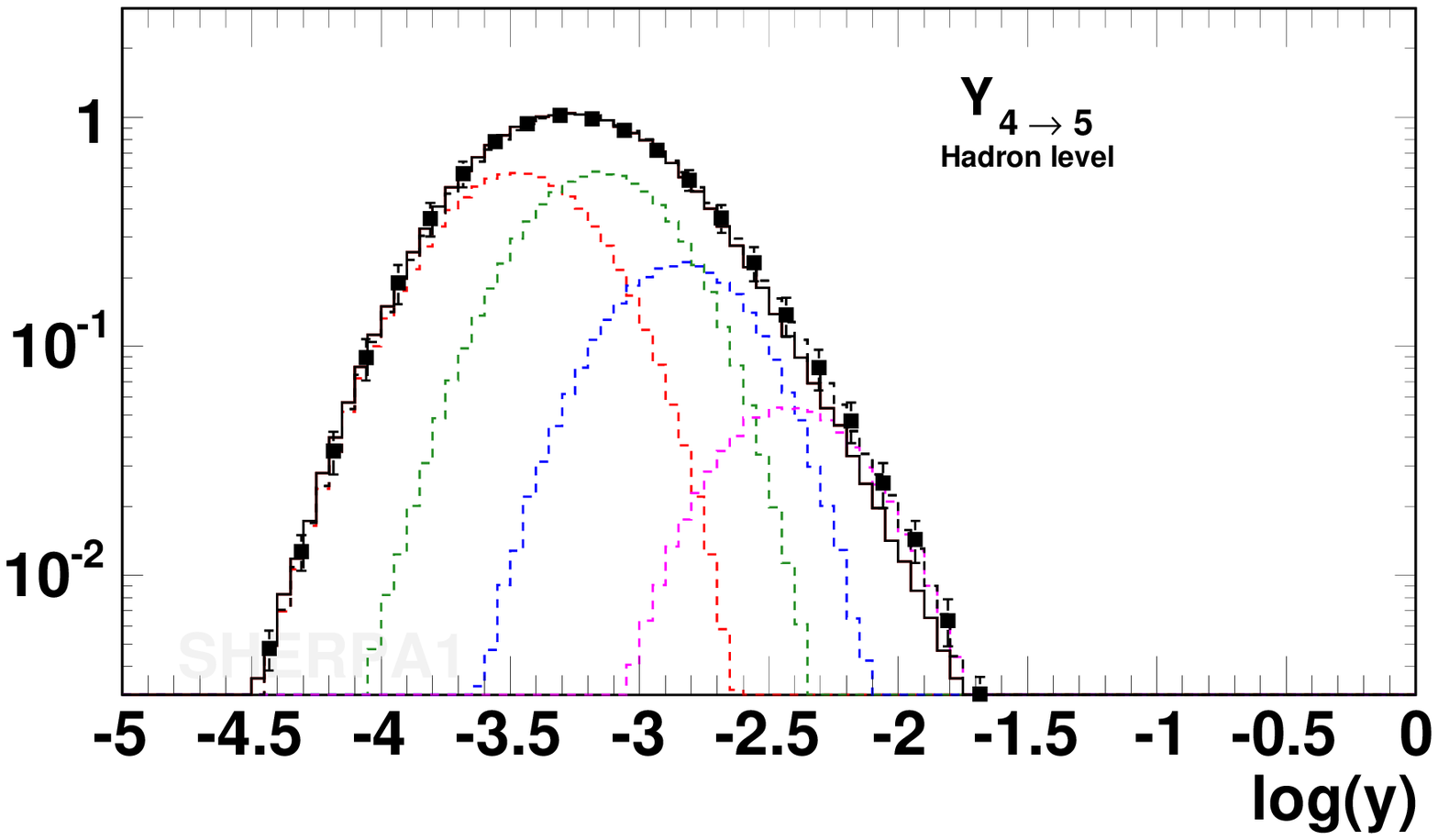}}
\put(200,-100){\includegraphics[width=8.cm]{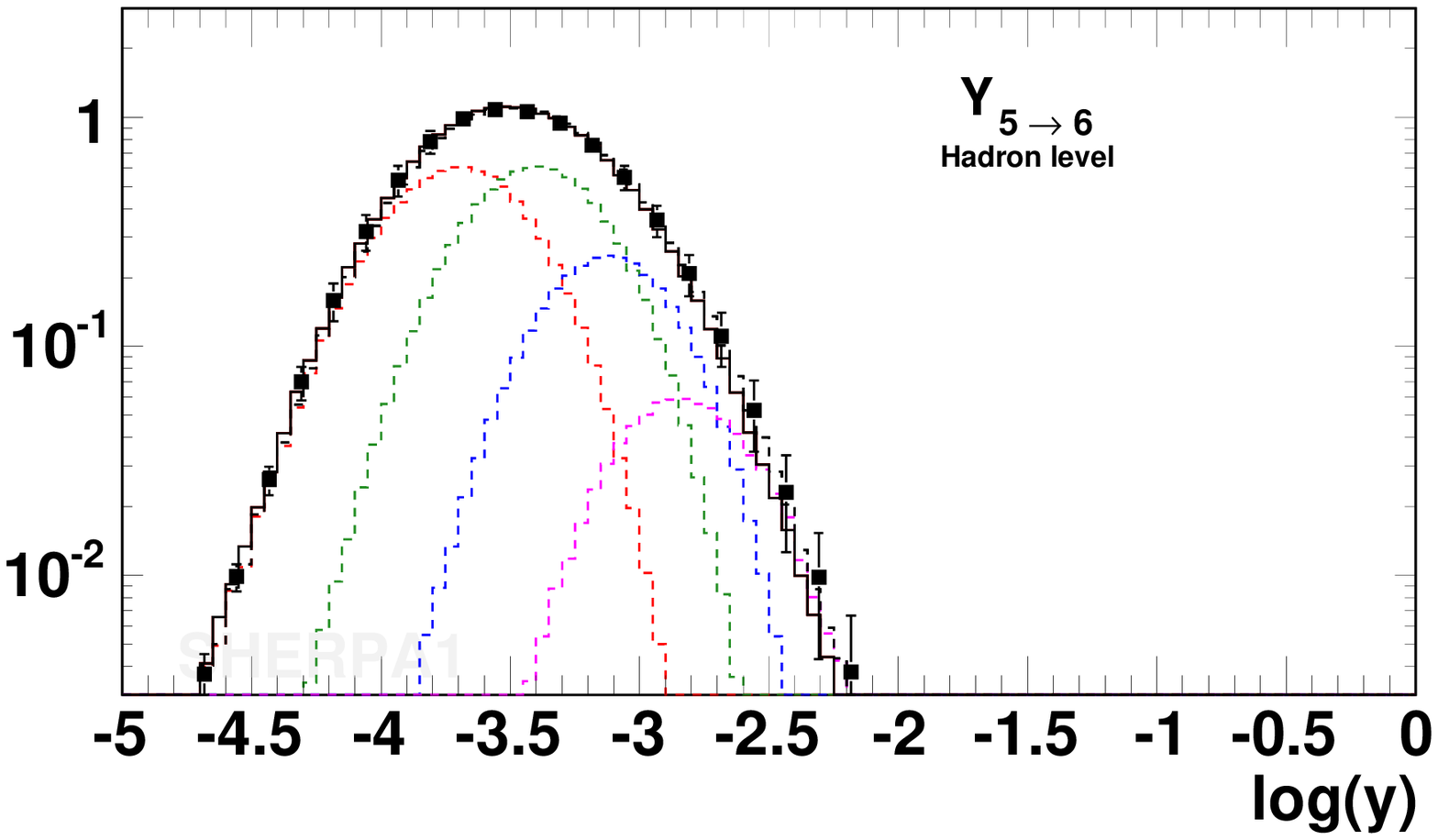}}
\end{picture}
\caption{\label{fig:lep_diffdurham} Differential jet rates in the Durham
  scheme at LEP 1, taken from \cite{Pfeifenschneider:1999rz}. Line styles and
  colours are the same as in Fig.\ \ref{fig:lep_charged_multi}.}
\end{figure}

\begin{figure}
\begin{picture}(400,240)
\put(-15,0){\includegraphics[width=8.0cm]{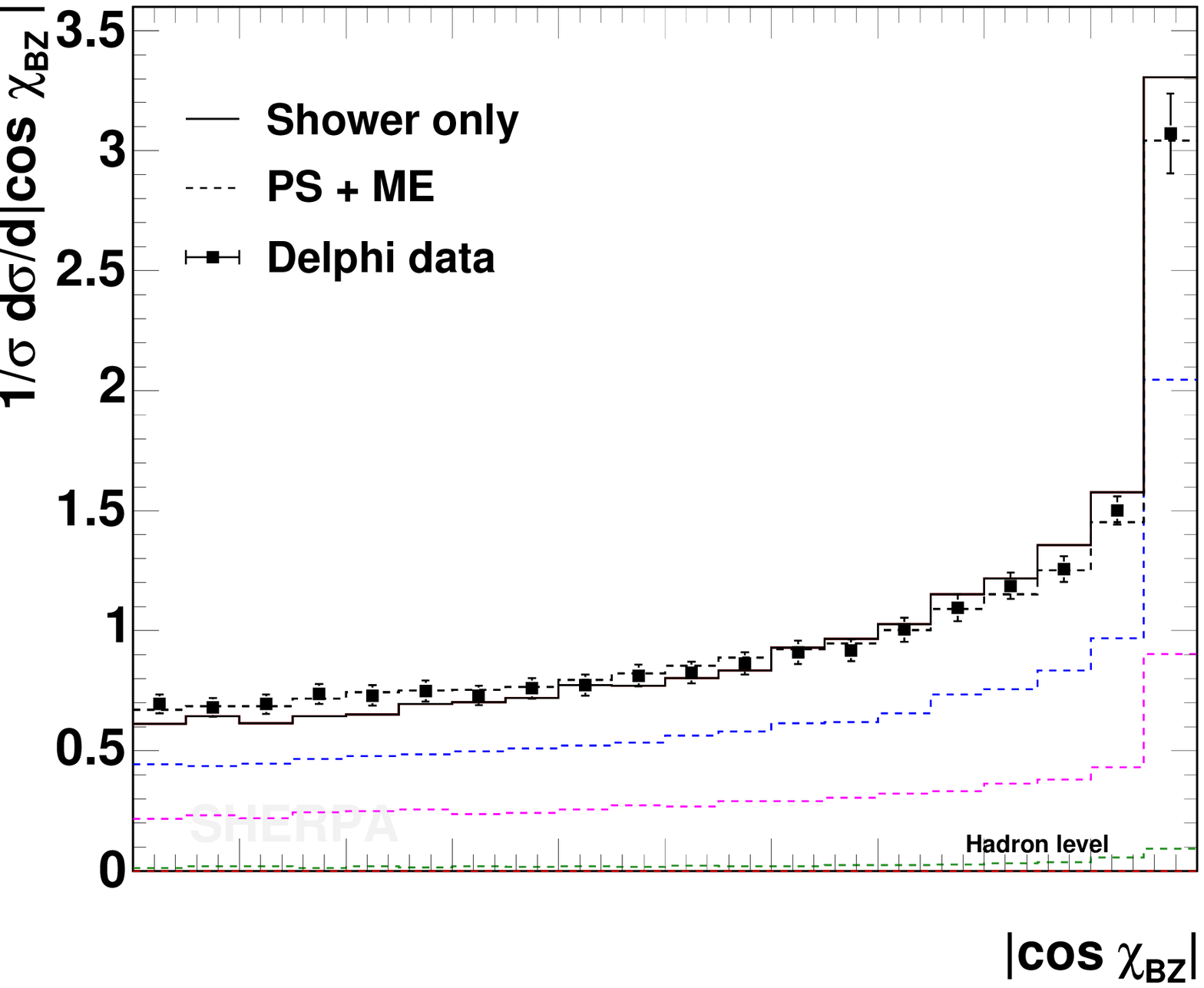}}
\put(-15,0){\includegraphics[width=8.0cm]{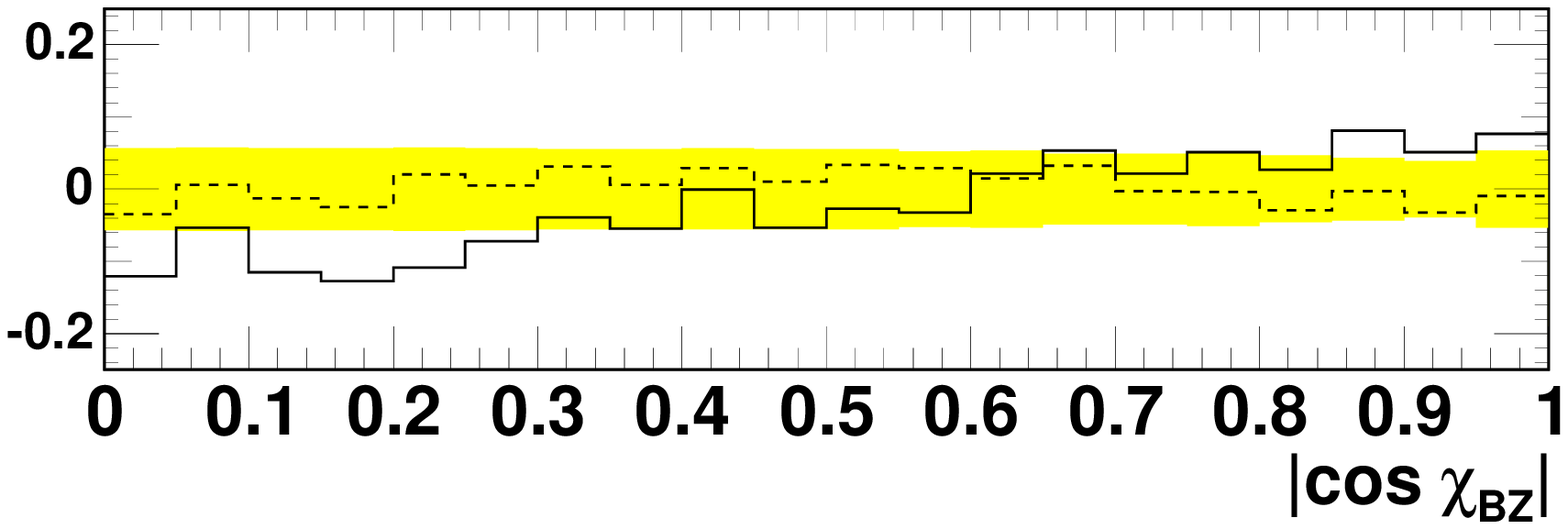}}
\put(200,0){\includegraphics[width=8.0cm]{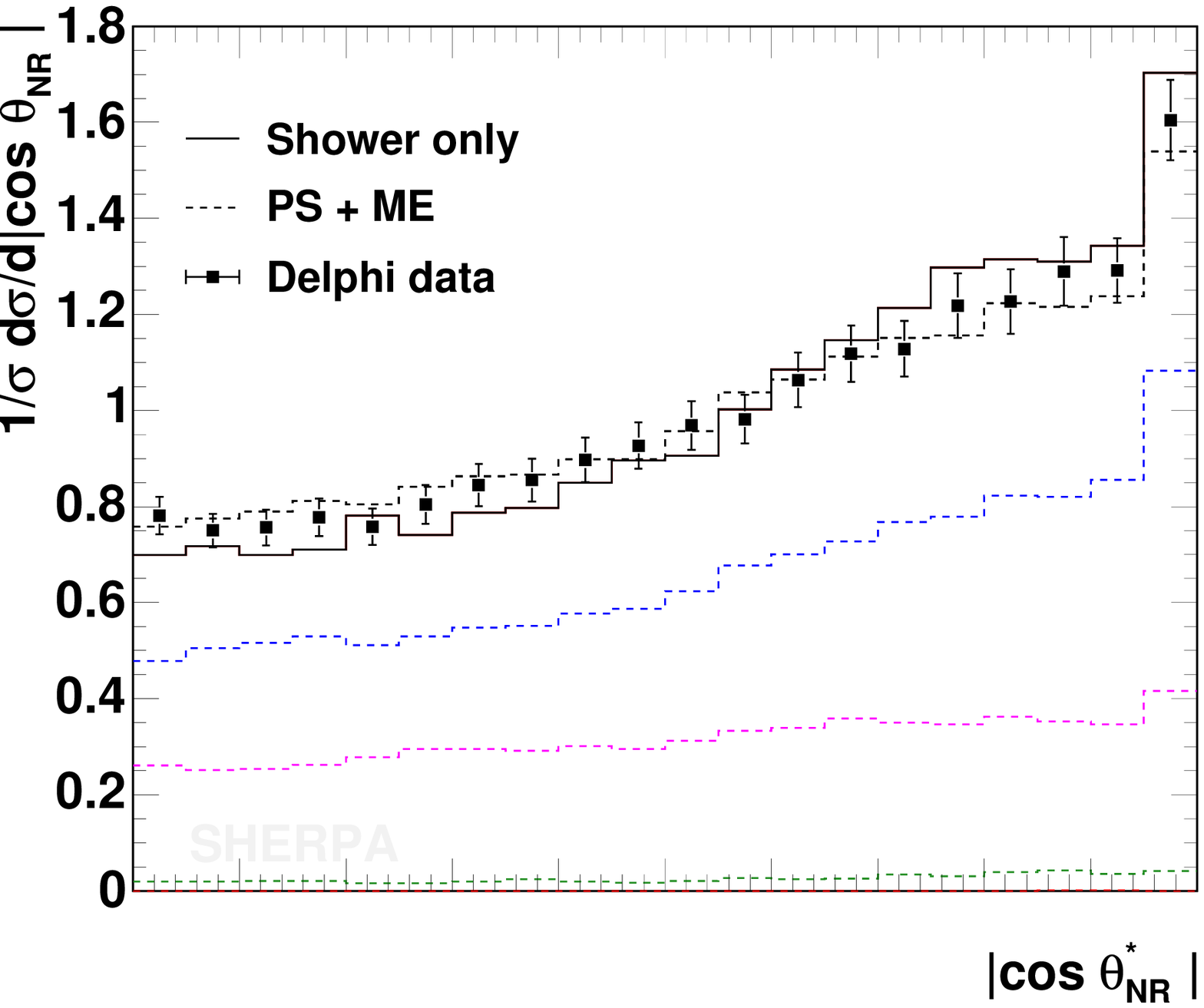}}
\put(200,0){\includegraphics[width=8.0cm]{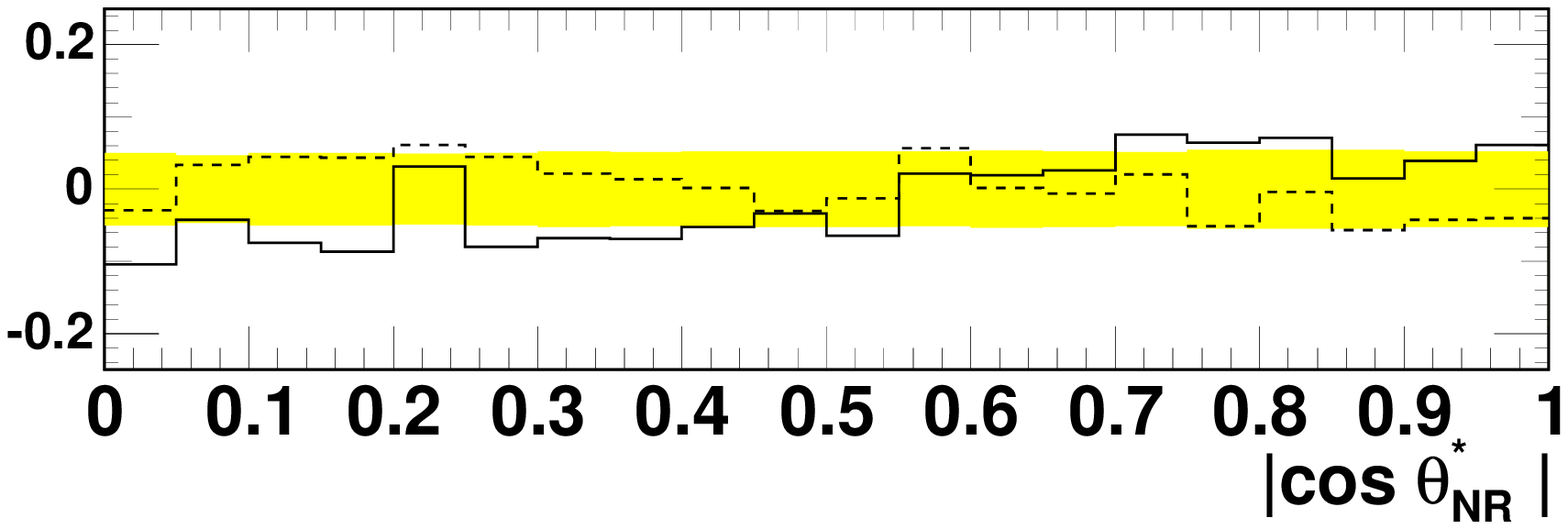}}
\end{picture}
\caption{\label{fig:lep_fourjet_angles} Four jet angle
  distributions. Shown are the Bengtsson-\-Zerwas angle (left) and the  
  modified Nachtmann-Reiter angle (right). The data points are from a DELPHI
  measurement \cite{Hendrikdipl}.}
\end{figure}

\noindent
Turning from the event shape observables to jet observables, the small
differences between the two samples vanish nearly completely. In Fig.\
\ref{fig:lep_durham}, the relative fractions of events with different
numbers of jets are exhibited in dependence on the jet resolution
parameter $y$. The mutual agreement of the two samples and their
agreement with data is excellent: The parton shower is perfectly
capable of describing jet multiplicities at LEP 1. This finding is 
repeated for the differential jet rates, viz Fig.\
\ref{fig:lep_diffdurham}. In Fig.\ \ref{fig:lep_fourjet_angles}, the
topological structure of four jet events is investigated. To this end,
the Bengtsson-\-Zerwas angle \cite{Bengtsson:1988qg} and the modified
Nachtmann-\-Reiter angle \cite{Nachtmann:1982xr} are
studied. Apparently, the merged sample is in perfect agreement with
the data, whereas the ``parton shower only'' sample exhibits a slight
shift away from them. This, however, is not a big surprise; after all,
these observables do depend on interferences between different
diagrams. To take this into account clearly is well beyond the
abilities of the parton shower.

\subsection*{Tevatron}

\begin{figure}[h]
\begin{center}
\includegraphics[width=8.5cm]{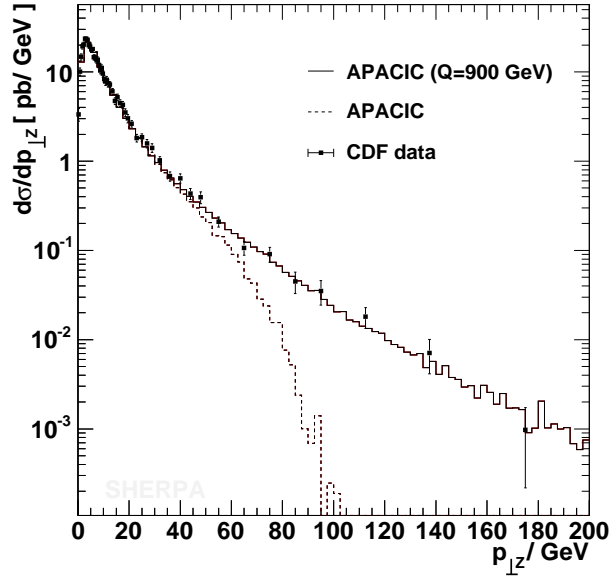}
\includegraphics[width=8.5cm]{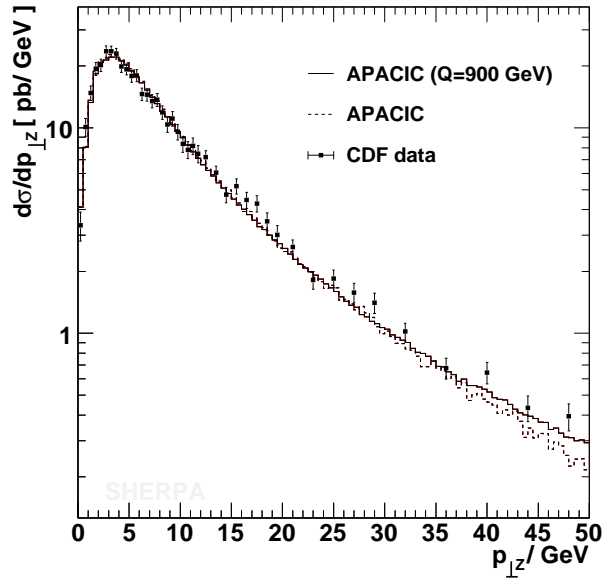}
\end{center}
\caption{\label{Z_data} The $p_\perp$ distribution of the $Z$-boson in
  comparison with data from CDF at the Tevatron, Run I
  \cite{Affolder:1999jh}. The bottom plot shows the same distribution
  as the top one, but with focus on the low momentum region.  
  The solid line indicates the shower result when using a fixed start
  scale of 1800 GeV, while the the dashed line is obtained when using
  $s'\approx M_{Z}^2$ as start scale. The {\tt SHERPA} results have
  been multiplied by a constant $K$-factor of 1.45 to match the data.
}
\end{figure}

\begin{figure}[h]
\begin{center}
\includegraphics[width=8.5cm]{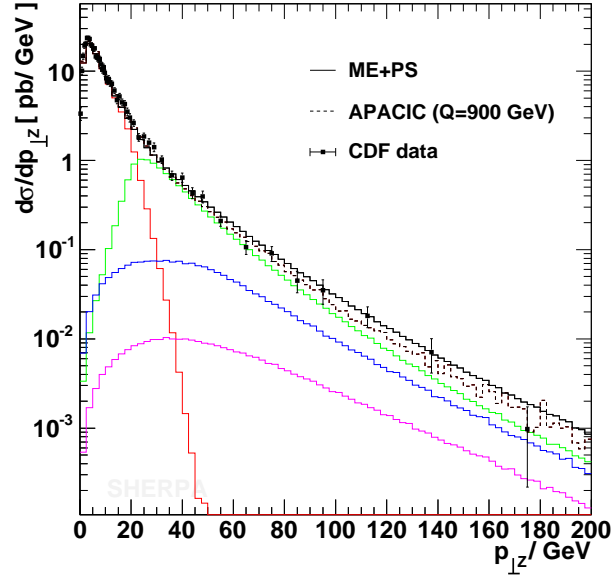}
\includegraphics[width=8.5cm]{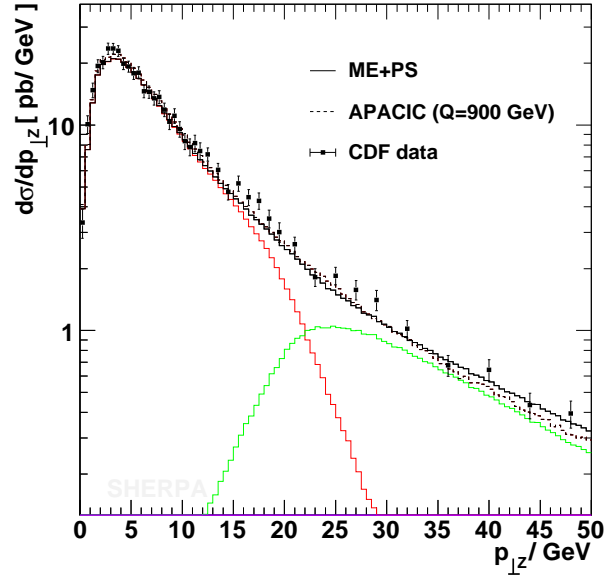}
\end{center}
\caption{\label{Z_data_meps} The $p_\perp$ distribution of the $Z$-boson in
  comparison with data from CDF at the Tevatron, Run I
  \cite{Affolder:1999jh}. This plot shows the same distributions as
  Fig. \ref{Z_data}, but this time the pure shower performance is
  compared with the result obtained when merging matrix elements with
  up to 3 extra jets. The jet scale $Q_{\rm cut}$ was fixed to 20 GeV.
  The coloured lines give the contributions from individual matrix
  elements. 
}
\end{figure}

\noindent
Having investigated the final state parton shower, the focus shifts
now on a study of the parton shower in the initial state. There, a
good observable to judge the performance of the parton shower is the
transverse momentum distribution of lepton pairs in Drell-\-Yan
scattering. In Fig.\ \ref{Z_data} the $p_\perp$ distribution of such 
pairs with masses at the $Z$-pole ($91 \pm 15$ GeV) is displayed. In
order to describe the left side of the distribution, the initial
partons have to be supplemented with intrinsic transverse momentum
w.r.t.\ the hadron they stem from. In {\tt SHERPA}, this intrinsic $k_\perp$
is distributed according to a Gaussian, with expectation value 
$k_\perp = 0.8$ GeV. In this plot, two different simulations runs are
confronted with data, both only use the parton shower for parton
radiation. They differ in the choice of the starting scale of the
shower evolution; one is starting at $t = M_{ll}^2 \approx M_Z^2$, the
other one is starting at the c.m.\ energy squared of the incoming
hadrons. Clearly, this leads to differences in the treatment of hard
radiation, and as expected in the former case, parton radiation ceases
to exist at scales of the order of the hard scale. The agreement of
both simulation runs with data at scales up to 40-50 GeV, however, is
excellent. In Fig.\ \ref{Z_data_meps}, the same observable is depicted
once more, this time, however, the data are confronted with the pure
parton shower, starting at the high scale, and with the merged result,
including up matrix elements for the production of the $Z$ accompanied
with up to three hard jets. The results again are in great agreement
with data. This shows that also the implementation of the initial
state parton shower in \apa\ has succeeded. 

\clearpage
\section{Conclusions}

\noindent
In this paper, the new version of the program \apa\ has been
presented. In contrast to its first version, \apa\ ceased to be a
stand-alone program; now, it is just the parton shower module of a
larger framework. This transformation resulted in a relocation of many
steering and service classes. On the other hand, the scope of \apa\
widened in such a way that in its present state it is also capable of
performing the parton shower in the initial state of hadronic
collisions. In addition, the merging procedure with the matrix elements
has been refined. On the technical side, some algorithms have been
changed, in particular, the look-up tables for the Sudakov form
factors have been replaced by an algorithm based on the hit-or-miss
method.

\noindent
Some exemplary results obtained with the new version of \apa\ have
been presented as well. The comparison with known analytical results,
that properly resum large logarithms, validates the physical and
numerical correctness of the implementation of the Sudakov form
factors. This is not too obvious, since the analytical results
contains logarithms in terms of transverse momentum scales, whereas
the parton shower in \apa\ has been formulated in terms of virtual
mass, supplemented with appropriate scale choices and an explicit
angular ordering. Both the parton shower alone and it being merged
with matrix elements results in an excellent agreement with a
collection of precise data. These results span up to five orders of
magnitude and the agreement is, especially for the LEP 1 data, on the
level of a few percent. This proves the validity and underlines the
abilities of the parton shower implemented in \apa.

\section*{Acknowledgements}

\noindent
F.K.\ and A.S.\ mourn the dead of Gerhard Soff, teacher and friend.

\noindent
F.K.\ and A.S.\ are indebted to S.~H{\"o}che, T.~Gleisberg, S.~Schumann
and J.~Winter for fruitful collaboration on {\tt SHERPA} and in
particular for many patient tests of \apa. A.S.\ and F.K.\ thank
K.~Hamacher and H.~Hoeth for many helpful discussions and for the
patient tuning of the parton shower.  F.K.\ would like to thank
S.~Catani, S.~Gieseke, M.~Mangano, P.~Richardson, and B.~Webber for
pleasant discussions concerning parton showers and their merging with
matrix elements. The authors acknowledge financial support by BMBF,
GSI and DFG. 

\newpage
\begin{appendix}

\section{\label{app:program} Brief program documentation}

\noindent
\apa\ is the module responsible for parton showers inside the \sherpa\
framework 

\noindent
Therefore, basic physics tools, like e.g.\ four-momentum, PDFs,
particle definitions, and the jet algorithm, are provided through the
overall framework. Furthermore, a number of features related to the
jet-veto are closely connected to the implementation of the merging
procedure in \sherpa. Nevertheless, \apa\ could be employed by the
framework of a different event generator, provided the basic physics
tools are made available.  

\noindent
This section gives a brief summary of the tasks each class in
\apa\ is responsible for. Where needed, some details on specific
implementation issues are presented that should, in principle, enable
the interested user to implement and test some of his or her own ideas. 
As stated above, in Sec.\ \ref{Impl_Sec}, \apa\ represents the
evolving parton shower in terms of binary {\tt Tree}s, consisting of
doubly linked {\tt Knot}s. Each of them represents one individual
parton splitting. \apa\ steers the shower evolution in the initial and
final through two different classes, {\tt Initial\_\-State\_\-Shower}
and {\tt Final\_\-State\_\-Shower}, respectively. These shower classes
fill the {\tt Knot}s through corresponding Sudakov form factors,
encoded in {\tt Timelike\_\-Sudakov} and {\tt Spacelike\_\-Sudakov},
where the latter carries a link to the appropriate PDF. Therefore,
there are two instances of {\tt Spacelike\_\-Sudakov} with potentially
different PDF in the {\tt Initial\_\-State\_\-Shower}. For
convenience, both Sudakov classes are derived from the class 
{\tt Splitting\_\-Group}; as such, they contain all relevant 
{\tt Splitting\_\-Function}s. However, both showers reconstruct the
branching kinematics of each splitting from quantities like the scale
$t$ and the energy splitting parameter $z$. This is achieved in the
two classes {\tt Timelike\_\-Kinematics} and 
{\tt Spacelike\_\-Kinematics}, respectively. 

\noindent

\subsection*{The interface with {\tt SHERPA}:
             the class {\tt Apacic}}

\noindent
This class defines the general interface to the shower package \apa. 
The parton shower evolution for a set of given partons is performed
along the following steps: 

\begin{itemize}
\item 
  The parton shower of each event is started after setting appropriate
  initial conditions, taken from the merging with the matrix
  element. These initial conditions - basically starting scales $t$
  and maximal angles for the coherent shower evolution - are brought
  directly into the {\tt Tree}s by filling the pseudo parton shower
  history of the matrix elements into corresponding {\tt Knot}s. 
  Pointers to the {\tt Tree}s needed for this operation are extracted
  from the showers and handed to the outside world through the methods
  {\tt FinTree()} and {\tt IniTrees()}.
\item 
  By calling {\tt PerformShowers()}, the corresponding methods in the
  class {\tt Final\_\-State\_\-Shower} and in the class 
  {\tt Initial\_\-State\_\-Shower} are triggered to perform the shower
  evolution. All necessary boosts at the beginning and at the end of
  the shower\footnote{
    For instance, the final state shower is always performed in the
    rest frame of the particles starting it.} 
  are carried out in here. The methods {\tt Set\-Jetveto\-Pt2()} and
  {\tt Set\-Facto\-risa\-tion\-Scale()} are used to pass settings relevant for
  the parton shower piece of the merging into the showers. The 
  check of whether a jet was lost during shower evolution is
  performed through {\tt Final\_\-State\_\-Shower::ExtraJetCheck()}.   
\item  
  The final result of the shower evolution can be transfered to the
  \sherpa\ frame work with the help of {\tt ExtractPartons()}.
\end{itemize}

\subsection*{Running the showers}

\subsubsection*{The class {\tt Final\_\-State\_\-Shower}}

\noindent
This is the central class of the final state shower implementation. It
controls the sequence of evolution steps\footnote{
  Note that the class {\tt Final\_\-State\_\-Shower} is also utilised
  for the (time-like) evolution of any parton emitted off the initial
  state shower.}. 
For the individual shower evolution of a single parton the class
{\tt Timelike\_\-Sudakov} is employed. Branching kinematics are
constructed after the evolution has finished, using the class  
{\tt Timelike\_\-Kinematics}. It is also responsible for
some kinematics checks during the shower evolution and for the jet
veto. 

\noindent
The final state shower evolution of a given jet ensemble is performed by
{\tt PerformShower()}, while the method {\tt FirstTimelikeFromSpacelike()}
is called for the final state (time-like) shower of a parton emitted
during the initial state (space-\-like) shower evolution.  

\noindent
The different methods of the class {\tt Final\_\-State\_\-Shower} are
responsible for the following tasks:
\begin{itemize}
\item
  The method {\tt PerformShower()} initiates the final state shower on
  a given tree, i.e.\ starting from its root {\tt Knot}. This is done
  in {\tt InitializeJets()}, the kinematics are constructed afterwards
  utilising {\tt Timelike\-Kinematics::Do\-Kinematics()}.
\item
  {\tt InitializeJets()} initialises the jet system emerging from a
  given (dummy) mother knot. The algorithm is performed recursively
  along the following steps:  
  \begin{enumerate}
  \item If one or both of the two daughters are allowed to decay, the
    mother {\tt Knot} is filled accordingly and the parton system
    produced by the decaying daughter(s) must be further
    evolved. These tasks are achieved by {\tt FillBranch()} and 
    {\tt EvolveJet()}, respectively.  
  \item If any the two daughters can not decay (internal lines of a
    ME), {\tt Initialize\-Jets()} is called again, with the daughter
    in question taking the role of the mother knot.
  \end{enumerate}
\item
  To fill a branch, its splitting scale $t$ and energy splitting
  variable $z$ have to be determined. In \apa, this is realised
  through {\tt FillBranch()}. There, a mother with parameters
  $\{t_a,\,z_a\}$ for a decay $a\to bc$ into two already specified
  massless daughters serves as input for the determination of their
  $\{t_i,\,z_i\}$ ($i=b,\,c$). 

  \noindent
  Step by step one of the daughters is chosen to obtain a new trial 
  virtuality $\bar t_i$ according to Eq.\ (\ref{eq:TL_No_Branch}),
  realised by {\tt Timelike\_\-Sudakov::Dice()}. In each step, the
  last $\bar t_i$ serves as starting condition for the determination
  of the new one. Only if the system of both daughters passes the
  kinematic constraints, this sequence of alternating reduction of the
  $\bar t_i$ is interrupted and they are accepted. Then, the mothers
  energy splitting variable $z_a$ must be modified according to Eqs.\
  (\ref{eq:TL_Shuffle_Z}, \ref{eq:TL_Shuffle_Z_ME}) implemented in
  {\tt Timelike\_\-Kinematics::Shuffle()} to compensate for the gain
  in virtuality of the daughters.  
\item
  {\tt EvolveJet()} evolves a given parton system, consisting of a
  mother knot with two daughters. The algorithm works recursively:
  Both daughters may branch further, therefore {\tt FillBranch()} is
  called first to determine both virtualities within the given
  kinematical constraints. In case neither of the daughters branches,
  the algorithm stops at once. Otherwise, {\tt EvolveJet()} is called
  for the corresponding daughter(s). 
\item
  {\tt FirstTimelikeFromSpacelike()} in contrast tries to initialise a
  new jet system emerging from a time-like particle emitted by a
  space-\-like shower. If, starting from the virtuality $t_a$ of the
  space-\-like branch $b\to ac$, the {\tt Timelike\_\-Sudakov::Dice()}
  yields a suitable virtuality $t_c>t_0$ respecting the constraint 
  Eq.\ (\ref{eq:MaxT}), daughters are initialised and the jet
  is evolved by {\tt EvolveJet()}. 
\item
  After the shower is completed, colours are set through the method
  {\tt Set\-All\-Colours()} according to the algorithm described
  above. Then, the method {\tt ExtractPartons()} extracts the partons 
  (instances of the class {\tt Particle} of the full framework) from 
  the outgoing {\tt Knot}s of the {\tt Tree} and fills them into the
  event record.
\end{itemize}

\subsubsection*{The class {\tt Initial\_\-State\_\-Shower}}

\noindent
The class {\tt Initial\_\-State\_\-Shower} governs the space-\-like shower
evolution. Starting with the particles entering the hard $2\to 2$
piece of the process, a backward shower is performed with the help of
the classes {\tt Spacelike\_\-Sudakov} (two objects, one for
each beam) and {\tt Spacelike\_\-Kinematics}. For the treatment of
time-like parton emissions the control is transfered to the class
{\tt Final\_\-State\_\-Shower}. The {\tt Initial\_\-State\_\-Shower}
object contains the following methods:

\begin{itemize}
\item 
  {\tt PerformShower()} is called in order to start the space-\-like
  shower evolution. Information one the matrix element kinematics is
  given in form of two partially filled trees, corresponding to the
  evolution  of the left and the right incoming parton, respectively.
  Similar to the final state case, the shower is initiated by
  employing {\tt InitializeSystem()}. After the evolution has finished
  some consistency checks are done, and the shower history is
  transfered back into the laboratory frame. 
\item
  {\tt InitializeSystem()} determines the initial system for the
  shower evolution. The algorithm starts with calling 
  {\tt FillBranch()} for each of the given matrix element partons in
  order to obtain the virtuality $t_i$ and energy fraction $z_i$ of
  the first splittings. Then, the off-shell momenta are constructed
  with the help of {\tt Spacelike\_\-Kinematics::\-InitKinematics()}.
\item
    Additional branchings are appended by {\tt EvolveSystem()}. This
    methods performs the evolution of a given system of two space-\-like
    partons, by recursively appending additional branchings. In each
    step a system consisting two partons of different incoming
    beams. The shower evolution is performed backward, i.e.\ from the
    hard interaction towards the beam particle. The parton with larger
    virtuality is supposed to be closer to the hard interaction and is
    selected for the next evolution step. There, the virtuality $t$
    and energy fraction $z$ is fixed by consecutive calls of 
    {\tt FillBranch()}, {\tt CalculateMaxT()}, and 
    {\tt First\-Timelike\-From\-Spacelike()}. After each step, the
    four-\-momenta of the participating partons are evaluated with 
    {\tt DoKinematics()}, leading to a new system of two space-\-like
    partons. 
\item
  After the shower evolution is finished, the created parton set can
  be accessed by calling {\tt ExtractPartons()}. It fills all final
  state particles connected to the two incoming {\tt Trees} into the
  event record.
\item
  Two configuration methods need to be mentioned, which are of special
  importance in the context of merging ME and PS, namely
  {\tt Set\-Jetveto\-Pt2()} and {\tt Set\-Facto\-risa\-tion\-Scale()}. In order to
  guarantee a clean separation between, the jet-veto is imposed
  on any trial emission inside the shower. Usually, the $p_\perp$ for
  any trial radiation is restricted by the jet resolution scale 
  $p_\perp^{\rm cut}$. For the matrix element with the highest
  multiplicity of jets, the separation cut is dynamically fixed by 
  the smallest transverse momentum present in the hard interaction, 
  owing to the fact that shower is not supposed
  to produce radiation harder than any QCD radiation present in the
  matrix element. In full analogy, merging ME and PS involves
  evaluating the PDFs at a specific factorisation scale $\mu_F$. The
  dependence on the $\mu_F$ is to be cancelled by a corresponding PDF
  evaluation during the first parton emission in the initial state
  shower. The factorisation scale $\mu_F$ does not necessary coincide
  with the jet-veto scale mentioned above $p_\perp^{\rm cut}$. For
  instance for (leptonically decaying) $W$ bosons produced in hadronic
  collisions, there is no jet-veto applied, but of course $\mu_F$ is
  not vanishing. In fact, in this example case, the factorisation scale
  is identified with the hard c.m energy $\sqrt{s'}$.
\end{itemize}

\subsubsection*{Splitting functions \& Sudakov form factors}
\begin{figure}
\begin{center}
\includegraphics[height=4cm]{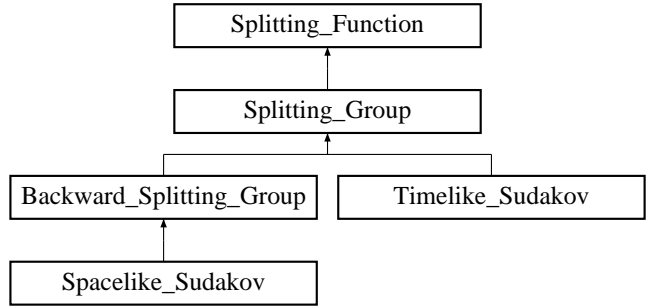}
\end{center}
\caption{\label{fig:splittinggroup}class hierarchy }
\end{figure}

\noindent
The class {\tt Splitting\_\-Function} is purely virtual. It defines a
common interface to the splitting functions $P_{a\to bc}(z)$. Among
others, this class and its specific instantiations include methods to
\begin{itemize}
\item access its flavours ({\tt GetFlA()}, {\tt GetFlB()}, and
  {\tt GetFlC()}) 
\item to determine $z$ according to an approximative distribution
  ({\tt GezT()}), and its integral in a given region
  ({\tt CrudeInt()}),  
\item a correction weight ({\tt GetWeight()}), and the exact
  functional form of the splitting function through its 
  {\tt operator()}) 
\item and some methods to manage {\tt Splitting\_Groups}.
\end{itemize}
All implemented splitting functions are derived from this class, cf.\ 
Tab. \ref{tab:splittingclasses}. This abstraction renders the
inclusion of further splitting functions an easy task.

\noindent
The class {\tt Splitting\_\-Group} is a container derived from the
abstract class {\tt Splitting\_\-Function}. It contains all splitting
functions of a given flavour, either in the forward splitting or for backward
splitting as a {\tt Backward\_\-Splitting\_\-Group} object. It is thus
responsible for the determination of a branching for one specific
flavour. The integration routine therefore returns the sum of the
integrals of all single {\tt Splitting\_\-Functions}. After the
integration the routine {\tt SelectOne()} can be called to choose one 
splitting mode out of the available options according to the integrals.
All subsequent calls to flavour access methods, dice routines or the
weight calculation then correspond to the selected branching. Note
that the {\tt Splitting\_\-Group} is also the base of any Sudakov form
factor determination, as depicted in Fig. \ref{fig:splittinggroup}. 

\noindent
Both Sudakov classes are derived from the corresponding {\tt Splitting\_\-Group}.
Of course, in the space-\-like case, the mother class is the 
{\tt Backward\_\-Splitting\_\-Group}. Following their names, the classes
{\tt Timelike\_\-Sudakov} and {\tt Spacelike\_\-Sudakov} govern
the time-\-like and space-\-like shower evolution, respectively.
The latter one must be supplemented with a link to a PDF
implementation, realised in {\tt SHERPA} through an object derived
from an abstract {\tt PDF\_\-Base}. However, both Sudakov classes
store appropriate {\tt Splitting\_\-Group}s for every flavour taking
part in the shower. In both classes, the method {\tt Dice()} is
responsible for the determination of virtualities ({\tt ProduceT()}),
daughter flavours, and energy fractions. The algorithm chosen 
is the hit-\-or-\-miss method, and both classes thus contain various
veto methods implementing cuts and correction weights. To exemplify
this, both classes have a method {\tt CplVeto()} incorporating the
correction weight $\alpha_s(Q)/\alpha_s(Q_{\rm min})$. Especially for
the incorporation of issues related with coupling constants, both have
a link to the class {\tt Sudakov\_\-Tools}.

\subsubsection*{Kinematics}

Kinematics are implemented in two classes, 
{\tt Timelike\_\-Kinematics} and {\tt Space\-like\_\-Kinematics} for the
determination of the kinematics in the final and initial state
showers, respectively. They include 
\begin{itemize}
\item checks whether a branching is kinematically possible 
  ({\tt KinCheck()}), 
\item a check whether an emission yields a jet that must be vetoed in
  the merging procedure ({\tt JetVeto()}), and
\item a method to construct kinematics ({\tt DoKinematics()}).
\end{itemize}
In addition, {\tt Timelike\_\-Kinematics} provides a method to redefine
the energy splitting $z$ according to Eqs.\ (\ref{eq:TL_Shuffle_Z},
\ref{eq:TL_Shuffle_Z_ME}) ({\tt Shuffle()}), and 
{\tt Spacelike\_\-Kinematics} has a method to determine the maximal
kinematically allowed $t$ for the initialisation of a final state
shower off an initial state splitting ({\tt CalculateMaxT()}), cf.\
Eq.\ (\ref{eq:MaxT})

\begin{figure}
\begin{center}
\begin{tabular}{|r|l|}
% * class name
% * coupling
% * formula
\multicolumn{2}{l}{QED splitting functions} \\
\hline
{\tt Fermion\_To\_Fermion\_Photon} & 
  $P_{f\to f \gamma}(z)$ cf. Eq. \eqref{eq:pff} \\
{\tt Photon\_Fermion\_Fermion} & 
  $P_{\gamma\to f \bar f}(z)$ cf. Eq. \eqref{eq:ppf} \\ 
\hline
% --- QCD ---
\multicolumn{2}{l}{QCD splitting functions} \\
\hline
{\tt Quark\_To\_Quark\_Gluon} & 
  $P_{q\to q g}(z)$ cf. Eq. \eqref{eq:pqq} \\
{\tt Gluon\_To\_Quark\_Quark} & 
  $P_{g\to q \bar q}(z)$ cf. Eq. \eqref{eq:pgq} \\
{\tt Gluon\_To\_Gluon\_Gluon} & 
  $P_{g\to g g}(z)$ cf. Eq. \eqref{eq:pgg} \\
\hline
% --- SUSY ---
\multicolumn{2}{l}{SUSY QCD splitting functions} \\
\hline
{\tt Gluino\_To\_Gluino\_Gluon} & 
$P_{\tg\to \tg g}(z)$
                                cf. Eq. \eqref{eq:ptgtg} \\
{\tt Gluon\_To\_Gluino\_Gluino} & $P_{g\to \tg \tg}(z)$
                                cf. Eq. \eqref{eq:pgtg} \\
\hline
{\tt SQuark\_To\_SQuark\_Gluon} & $P_{\tq\to \tq g}(z)$
                                cf. Eq. \eqref{eq:ptqtq} \\
{\tt Gluon\_To\_SQuark\_SQuark} & $P_{g\to \tq \btq}(z)$
                                cf. Eq. \eqref{eq:pgtq} \\
\hline
\end{tabular}
\end{center}
\caption{\label{tab:splittingclasses}A summary of implemented splitting
functions and their \class\ names in \sherpa.}
\end{figure}

\subsection*{Basic structures}

\noindent
{\tt Tree}s are used as representations of parton shower histories in
terms of interconnected binary splittings. It provides the basic
structure and all necessary routines to handle operations, like
boosts, on the whole or parts of a tree. Three trees form a complete
parton shower history: Two trees correspond to the initial state
shower evolution of the right and the left incoming particle,
respectively, and the other tree holds information of the final state
shower of the outgoing particles, cf.\ Sec.\ \ref{Impl_Sec}.

\noindent
The {\tt Knot}s are the basic elements forming a binary 
{\tt Tree}. They store all properties of a single branching, like
flavour, momentum, virtuality, energy component, etc..

\section{\label{app:splittings}Altarelli--Parisi splitting functions}

In this appendix all splitting functions, relevant for the shower
evolution of \apa\ are listed, cf.\
\cite{Catani:2000ef,Beenakker:1996ch}. The splitting functions are 
obtained after averaging over the azimuthal angle. The mass terms in
each case are parametrised by the variable
$\mu_{ij}^2=(m_i^2+m_j^2)/[(p_i+p_j)^2-m_{(ij)}^2]$. The expressions
for splittings with the decay products exchanged fulfil the obvious
symmetry relation $P_{a\to bc}(z)=P_{a\to cb}(1-z)$. 

{\bf Fermions and photons:}
\begin{align}
\label{eq:pff}
{ P_{f\to f\gamma}(z;\mu_{f\gamma}^2) } &= e_f^2\,
\left[ \frac{1+z^2}{1-z} - 2\mu_{f\gamma}^2\right] \;,\\
\label{eq:ppf}
{ P_{\gamma \to f \bar{f}}(z;\mu_{f{\bar f}}^2) } &= e_f^2\,
\left[ z^2 + (1-z)^2 + \mu_{f{\bar f}}^2\right] \;.
\end{align}
{\bf Quarks and gluons:}
\begin{align}
\label{eq:pqq}
{ P_{q\to qg}(z;\mu_{qg}^2) } &= C_F\,
\left[ \frac{1+z^2}{1-z} - 2\mu_{qg}^2\right] \;,
\\ 
%\label{eq:pqg}
%{ P_{q \to gq}(z;\mu_{qg}^2) } &= C_F\,
%\left[ \frac{2(1-z)}{z} + z
%     - 2\mu_{qg}^2 \right] \;,
%\\ 
\label{eq:pgq}
{ P_{g\to q\bar{q}}(z;\mu_{q{\bar q}}^2) } &= T_R\,
\left[ z^2 + (1-z)^2 - \mu_{q{\bar q}}^2\right] \;,
\\ 
\label{eq:pgg}
{ P_{g\to gg}(z) } &= C_A\,
\left[ \frac{z}{1-z} + \frac{1-z}{z} +  z(1-z) \right] \;.
\end{align}
In the massless limit ($\mu\to0$), these splitting functions reduce to
the well-known form
\begin{align}
\label{pqq1}
{ P_{qq}(z) } &= C_F\, \frac{1+z^2}{1-z}  \;,
\\ 
\label{pgq1}
{ P_{gq}(z) } &= T_R\,
\left[ z^2 + (1-z)^2\right] \;,
\\ 
\label{pgg1}
{ P_{gg}(z) } &= C_A\,
\left[ \frac{z}{1-z} + \frac{1-z}{z} +  z(1-z) \right] \;.
\end{align}

{\bf Gluinos and gluons:}
\begin{align}
\label{eq:ptgtg}
{ P_{\tg\to\tg g}(z;\mu_{\tg g}^2) } &= C_A\,
\left[ \frac{1+z^2}{1-z} - 2\mu_{\tg g}^2\right] \;,
\\ 
%\label{eq:ptgg}
%{ P_{\tg\to g\tg}(z;\mu_{\tg g}^2) } &= C_A\,
%\left[ \frac{2(1-z)}{z} + z
%     - 2\mu_{\tg g}^2 \right] \;,
%\\
\label{eq:pgtg}
{ P_{g\to\tg\tg}(z;\mu_{\tg\tg}^2) } &= C_A\,
\left[ z^2 + (1-z)^2 + \mu_{\tg\tg}^2\right] \;.
\end{align}

{\bf Squarks and gluons:}
\begin{align}
\label{eq:ptqtq}
{ P_{\tq\to\tq g}(z;\mu_{\tq g}^2) } &= C_F\,
\left[ \frac{2z}{1-z} - 2\mu_{\tq g}^2\right] \;,
\\ 
%\label{eq:ptqg}
%{ P_{\tq\to g\tq}(z;\mu_{\tq g}^2) } &= C_F\,
%\left[ \frac{2(1-z)}{z} - 2\mu_{\tq g}^2 \right] \;,
%\\ 
\label{eq:pgtq}
{ P_{g\to\tq\bar{\tq}}(z;\mu_{\tq{\bar{\tq}}}^2) } &= T_R\,
\frac{1}{2}\left[2z(1-z) - \mu_{\tq{\bar{\tq}}}^2 \right] \;.
\end{align}

\section{\label{EvtShapes_Sec}Definitions of event shapes}

The global properties of hadronic events may be characterised by set
of observables, usually called event shapes. In section
\ref{Results_Sec} the following shape observables have been
considered. 
\begin{itemize}
 \item Thrust $T$ : \\
   The thrust axis $\vec{n}_T$ maximises the
   following quantity
  \begin{equation}
    T = \max_{\vec{n}_T} \left( \frac{\sum_i | \vec{p}_i \cdot
        \vec{n}_T |}{\sum_i | \vec{p}_i |} \right) \,,
  \end{equation}
   where the sum extends over all particles in the event.
   The thrust $T$ tends to 1 for events that has two thin back-to-back 
   jets (``pencil-like'' event), and it tends towards $1/2$ for
   perfectly isotropic events. 
 \item Thrust Major $T_{\rm Major}$ :  \\
   The thrust major vector $\vec{n}_{\rm Major}$ is defined in the same way as
   the thrust vector, but with the additional condition that
   $\vec{n}_{\rm Major}$ must lie in the plane perpendicular to $\vec{n}_T$:
  \begin{equation}
    T_{\rm Major} = \max_{\vec{n}_{\rm Major} 
      \perp \vec{n}_T } \left( \frac{\sum_i | \vec{p}_i \cdot
        \vec{n}_{\rm Major} |}{\sum_i | \vec{p}_i |} \right) \,.
  \end{equation}
\item Thrust Minor $T_{\rm Minor}$ : \\
  The minor axis is perpendicular to both the thrust axis and
  the major axis, 
  $\vec{n}_{\rm Minor} = \vec{n}_T \times \vec{n}_{\rm Major}$. The 
  value of thrust minor is then given by
  \begin{equation}
    T_{\rm Minor} = \frac{\sum_i | \vec{p}_i \cdot
        \vec{n}_{\rm Minor} |}{\sum_i | \vec{p}_i |} \,.
  \end{equation}
\item Oblateness $O$ : \\
  The oblateness is defined as the difference between thrust major
  $T_{\rm Major}$ and  thrust minor $T_{\rm Minor}$ :
  \begin{equation}
    O = T_{\rm Major} - T_{\rm Minor}
  \end{equation}

\item C-parameter $C$ : \\
  The C-parameter is derived from the eigenvalues of the linearised
  momentum tensor $\Theta^{\alpha\beta}$ , defined by
  \begin{equation}
    \Theta^{\alpha\beta} = \frac{1}{\sum_i |\vec{p}_i|} \sum_i
    \frac{p_i^\alpha p_i^\beta}{|\vec{p}_i|} \,, 
    \; \alpha,\beta = \{x,y,z\} \;.
  \end{equation}
  The three eigenvalues $\lambda_i$ of this tensor define $C$ with
  \begin{equation}
    C = 3\, (\lambda_1 \lambda_2 + \lambda_2 \lambda_3 + \lambda_1
    \lambda_3) \,.
  \end{equation}
\end{itemize}

\end{appendix}

\newpage

\include{biblio}
%\end{fmffile}

\end{document}

%% file: mysettings.tex
\newcommand{\sherpa}{{\tt SHERPA}}
\newcommand{\apa}{{\tt APACIC++}}
\newcommand{\ame}{{\tt AMEGIC++}}
\newcommand{\class}{{\tt class}}

\newcommand{\dt}{{\rm d}t}
\newcommand{\dz}{{\rm d}z}

\newcommand{\be}{\begin{eqnarray}}
\newcommand{\ee}{\end{eqnarray}}
\newcommand{\bea}{\begin{eqnarray}}
\newcommand{\eea}{\end{eqnarray}}
\newcommand{\barl}{\begin{array}{rl}}
\newcommand{\ba}{\begin{array}}
\newcommand{\ea}{\end{array}}

\newcommand{\intl}{\int\limits}

\def\tq{{\tilde q}}
\def\btq{{\bar{\tilde q}}}
\def\tg{{\tilde g}}